\documentclass[reprint,natbib209,
 amsmath,amssymb,
 aps,aa
floatfix,
]{revtex4-1}
\usepackage{graphicx}
\usepackage{dcolumn}
\usepackage{bm}
\usepackage{hyperref}
\usepackage{multirow}
\usepackage{wasysym}

\newlength{\dslashwidth}

\newcommand{\beq}{\begin{equation}} 
\newcommand{\eeq}{\end{equation}}
\newcommand{\beqa}{\begin{eqnarray}} 
\newcommand{\eeqa}{\end{eqnarray}}
\newcommand{\newc}{\newcommand}
\newcommand{\bq}{\begin{equation}}
\newcommand{\eq}{\end{equation}}
\newcommand{\ba}{\begin{array}}
\newcommand{\ea}{\end{array}}
\newcommand{\bqa}{\begin{eqnarray}}
\newcommand{\eqa}{\end{eqnarray}}

\newcommand{\lnf}{{\ifmmode \Lambda^{(N_f)} \else $\Lambda^{(N_f)}$\fi}}
\newcommand{\ms}{{\ifmmode \overline{MS} \else $\overline{MS}$\fi}}
\newcommand{\dr}{{\ifmmode \overline{DR} \else $\overline{DR}$\fi}}
\newcommand{\lms}{{\ifmmode \Lambda^{(5)}_{\overline{MS}} \else $\Lambda^{(5)}_{\overline{MS}}$\fi}}
\newcommand{\lam}{{\ifmmode \Lambda \else $\Lambda$\fi}}
\newcommand{\mev}{{\ifmmode {\rm MeV} \else ${\rm MeV}$\fi}}
\newcommand{\gev}{{\ifmmode {\rm GeV} \else ${\rm GeV}$\fi}}
\newcommand{\gevc}{{\ifmmode {\rm GeV/c^2} \else ${\rm GeV/c^2}$\fi}}
\newcommand{\tev}{{\ifmmode {\rm TeV} \else ${\rm TeV}$\fi}}
\newcommand{\tevc}{{\ifmmode {\rm TeV/c^2} \else ${\rm TeV/c^2}$\fi}}
\newcommand{\lp}{{\ifmmode L^+  \else $L^+$\fi}}
\newcommand{\lm}{{\ifmmode L^-  \else $L^-$\fi}}
\newcommand{\mlp}{{\ifmmode M(L^-) \else $M(L^-)$\fi}}
\newcommand{\mlz}{{\ifmmode M(L^0) \else $M(L^0)$\fi}}
\newcommand{\lz}{{\ifmmode L^0 \else $L^0$\fi}}
\newcommand{\ev}{{\ifmmode GeV/c^2 \else $GeV/c^2$\fi}}
\newcommand{\tri}{{\ifmmode \triangleup \else $\triangleup$\fi}}
\newcommand{\unl}{{\ifmmode U_{lL^0} \else $U_{lL^0}$\fi}}\newcommand{\gL}{{\ifmmode g_L \else $g_{L}$\fi}}
\newcommand{\gR}{{\ifmmode g_R  \else $g_{R}$\fi}}
\newcommand{\gumu}{{\ifmmode \gamma^{\mu} \else $\gamma^{\mu}$\fi}}
\newcommand{\gunu}{{\ifmmode \gamma^{\nu} \else $\gamma^{\nu}$\fi}}
\newcommand{\gdmu}{{\ifmmode \gamma_{\mu} \else $\gamma_{\mu}$\fi}}
\newcommand{\gdnu}{{\ifmmode \gamma_{\nu} \else $\gamma_{\nu}$\fi}}
\newcommand{\stw}{{\ifmmode\sin^2\theta_W \else $\sin^{2}\theta_{W}$ \fi}}
\newcommand{\sws}{{\ifmmode \;\sin^2\theta_W  \else $\;\sin^{2}\theta_{W}$ \fi}}
\newcommand{\cws}{{\ifmmode \;\cos^2\theta_W  \else $\;\cos^{2}\theta_{W}$ \fi}}
\newcommand{\sw}{{\ifmmode \;\sin\theta_W  \else $\sin\theta_{W}$ \fi}}
\newcommand{\cw}{{\ifmmode \;\cos\theta_W  \else $\;\cos\theta_{W}$ \fi}}
\newcommand{\tw}{{\ifmmode \;\tan\theta_W  \else $\;\tan\theta_{W}$ \fi}}
\newcommand{\qq}{{\ifmmode q\overline{q} \else $q\overline{q}$\fi}}
\newcommand{\lR}{{\ifmmode l_R \else $l_R$\fi}}
\newcommand{\lL}{{\ifmmode l_L \else $l_L$\fi}}
\newcommand{\nt}{{\ifmmode \nu_{\tau} \else $\nu_{\tau}$\fi}}
\newcommand{\nuR}{{\ifmmode \nu_R  \else $\nu_R$\fi}}
\newcommand{\nuL}{{\ifmmode \nu_L  \else $\nu_L$\fi}}
\newcommand{\qR}{{\ifmmode g_R  \else $q_R$\fi}}
\newcommand{\qL}{{\ifmmode q_L  \else $q_L$\fi}}
\newcommand{\qRp}{{\ifmmode q_R'  \else $q_{R}$'\fi}}
\newcommand{\qLp}{{\ifmmode q_L'  \else $q_{L}$'\fi}}
\newcommand{\est}{{\ifmmode e^{\bf \ast} \else $e^{\bf \ast}$\fi}}
\newcommand{\lst}{{\ifmmode l^{\bf \ast} \else $l^{\bf \ast}$\fi}}
\newcommand{\must}{{\ifmmode \mu^{\bf \ast} \else $\mu^{\bf \ast}$\fi}}
\newcommand{\taust}{{\ifmmode \tau^{\bf \ast} \else $\tau^{\bf \ast}$ \fi}}
\newcommand{\pperp}{{\ifmmode p_t  \else $p_t$\fi}}
\newcommand{\et}{{\ifmmode E_t  \else $E_t$\fi}}
\newcommand{\xt}{{\ifmmode x_t  \else $x_t$\fi}}
\newcommand{\smumu}{{\ifmmode \sigma_{\mu\mu}  \else $\sigma_{\mu\mu}$ \fi}}
\newcommand{\eg}{{\ifmmode e\gamma  \else $e\gamma$\fi}}
\newcommand{\epem}{{\ifmmode e^+e^-  \else $e^+e^-$\fi}}
\newcommand{\lplm}{{\ifmmode L^+L^-  \else $L^+L^-$\fi}}
\newcommand{\pp}{{\ifmmode p\overline p  \else $p\overline p$\fi}}
\newcommand{\llz}{{\ifmmode L^0\overline{L}^0 \else $L^0\overline{L}^0$\fi}}
\newcommand{\epemt}{{\ifmmode e^+e^- \to  \else $e^+e^- \to$\fi}}
\newcommand{\eb}{{\ifmmode E_{beam}  \else $E_{beam}$\fi}}
\newcommand{\ip}{{\ifmmode pb^{-1}  \else $pb^{-1}$\fi}}
\newcommand{\upm}{{\ifmmode ^{\pm}  \else $^{\pm}$\fi}}
\newcommand{\de}{{\ifmmode ^{\circ}  \else $^{\circ}$ \fi}}
\newcommand{\appr}{{\ifmmode \sim \else $\sim$ \fi}}
\newcommand{\corresp}{{\ifmmode \stackrel{\wedge}{=} \else $\stackrel{\wedge}{=}$ \fi}}
\newcommand{\sqrts}{{\ifmmode \sqrt{s} \else $\sqrt{s}$\fi}}
\newcommand{\zz}{{\ifmmode Z^0  \else $Z^0$\fi}}
\newcommand{\mz}{{\ifmmode M_{Z}  \else $M_{Z}$\fi}}
\newcommand{\mzs}{{\ifmmode M_{Z}^2  \else $M_{Z}^2$\fi}}
\newcommand{\mw}{{\ifmmode M_{W}  \else $M_{W}$\fi}}
\newcommand{\mws}{{\ifmmode M_{W}^2  \else $M_{W}^2$\fi}}
\newcommand{\mh}{{\ifmmode M_{Higgs}  \else $M_{Higgs}$\fi}}
\newcommand{\msusy}{{\ifmmode M_{SUSY}  \else $M_{SUSY}$\fi}}
\newcommand{\msusys}{{\ifmmode M_{SUSY}^2  \else $M_{SUSY}^2$\fi}}
\newcommand{\su}{{\ifmmode SU(3)_C\otimes\- SU(2)_L\otimes\- U(1)_Y \else $SU(3)_C\otimes\A0SU(2)_L\otimes U(1)_Y$\fi}}
\newcommand{\suthree}{{\ifmmode SU(3)_C  \else $SU(3)_C$\fi}}
\newcommand{\sutwo}{{\ifmmode  SU(2)_L\otimes U(1)_Y \else $SU(2)_L\otimes U(1)_Y$\fi}}
\newcommand{\taup}{{\ifmmode \tau_{proton} \else $\tau_{proton}$\fi}}
\newcommand{\as}{{\ifmmode \alpha_{s}  \else $\alpha_{s}$\fi}}
\newcommand{\mgut}{{\ifmmode M_{GUT}  \else $M_{GUT}$\fi}}
\newcommand{\mguts}{{\ifmmode M_{GUT}^2  \else $M_{GUT}^2$\fi}}
\newcommand{\mzero}{{\ifmmode m_0        \else $m_0$\fi}}
\newcommand{\mhalf}{{\ifmmode m_{1/2}    \else $m_{1/2}$\fi}}
\newcommand{\sq}{{\ifmmode \tilde{q}    \else $\tilde{q}$\fi}}
\newcommand{\gl}{{\ifmmode \tilde{g}    \else $\tilde{g}$\fi}}
\newcommand{\mb}{{\ifmmode m_{b}    \else $m_{b}$\fi}}
\newcommand{\mt}{{\ifmmode m_{t}    \else $m_{t}$\fi}}
\newcommand{\mts}{{\ifmmode m_{t}^2    \else $m_{t}^2$\fi}}

\newcommand{\mtau}{{\ifmmode m_{\tau}  \else $m_{\tau}$\fi}}
\newcommand{\dpp}{{\ifmmode \delta_{pert} \else $\delta_{pert}$\fi}}
\newcommand{\dnp}{{\ifmmode\delta_{non-pert}\else$\delta_{non-pert}$\fi}}
\newcommand{\dew}{{\ifmmode \delta_{\rm EW}\else $\delta_{\rm EW}$\fi}}
\newcommand{\rt}{{\ifmmode R_{\tau}  \else $R_{\tau} $\fi}}
\newcommand{\rz}{{\ifmmode R_{Z}  \else $R_{Z} $\fi}}

\newcommand{\swb}{{\ifmmode \sin^2\theta_{\overline{MS}} \else $\sin^2\theta_{\overline{MS}}$\fi}}
\newcommand{\cwb}{{\ifmmode \cos^2\theta_{\overline{MS}} \else $\cos^2\theta_{\overline{MS}}$\fi}}

\newc\AIPCP[3] {{\em AIP Conf. Proc.} {\bf #1} (#2) #3}
\newc\AJ[3] {{\em Astrophys. J.} {\bf #1} (#2) #3}
\newc\AMS[3] {{\em Ann. Math. Statist.} {\bf #1} (#2) #3}
\newc\AP[3] {{\em Ann. Phys.} {\bf #1} (#2) #3}
\newc\APJ[3] {{\em Astropart. J.} {\bf #1} (#2) #3}
\newc\APP[3] {{\em Astropart. Phys.} {\bf #1} (#2) #3}
\newc\APS[3] {{\em Astrophys. J. Suppl.} {\bf #1} (#2) #3}
\newc\ARNPS[3] {{\em Ann. Rev. Nucl. Part. Sci.} {\bf C#1} (#2) #3}
\newc\BA[3] {{\em Bayesian Anal.} {\bf C#1} (#2) #3}
\newc\CPC[3] {{\em Comput. Phys. Commun.} {\bf C#1} (#2) #3}
\newc\CP[3] {{\em Contemp. Phys.} {\bf #1} (#2) #3}
\newc\EPJ[3] {{\em Euro. Phys. Journ.} {\bf C#1} (#2) #3}
\newc\JCAP[3] {{\em JCAP} {\bf #1} (#2) #3}
\newc\JHEP[3] {{\em JHEP} {\bf #1} (#2) #3}
\newc\JPG[3] {{\em J. Phys.} {\bf G #1} (#2) #3}
\newc\IJMP[3] {{\em Int. J. Mod. Phys.} {\bf A #1} (#2) #3}
\newc\MNRAS[3] {{\em Mon. Not. Roy. Astron. Soc.} {\bf #1} (#2) #3}
\newc\MPL[3] {{\em Mod. Phys. Lett.} {\bf A #1} (#2) #3}
\newc\NAR[3] {{\em New Astron. Rev.} {\bf #1} (#2) #3}
\newc\NCA[3] {{\em Nuovo Cimento} {\bf #1} (#2) #3}
\newc\NIM[3] {{\em Nucl. Instrum. Methods} {\bf #1} (#2) #3}
\newc\NIMA[3] {{\em Nucl. Instrum. Methods} {\bf A #1} (#2) #3}
\newc\NAT[3] {{\em Nature} {\bf #1} (#2) #3}
\newc\NPB[3] {{\em Nucl. Phys.} {\bf B #1} (#2) #3}
\newc\NPA[3] {{\em Nucl. Phys.} {\bf A #1} (#2) #3}
\newc\NPPS[3] {{\em Nucl. Phys. Proc. Suppl.} {\bf #1} (#2) #3}
\newc\PLB[3] {{\em Phys. Lett.} {\bf B #1} (#2) #3}
\newc\PR[3] {{\em Phys. Rep.} {\bf #1} (#2) #3}
\newc\PRL[3] {{\em Phys. Rev. Lett.} {\bf #1} (#2) #3}
\newc\PRD[3] {{\em Phys. Rev.} {\bf D #1} (#2) #3}
\newc\PRC[3] {{\em Phys. Rev.} {\bf C #1} (#2) #3}
\newc\PTP[3] {{\em Prog. Theor. Phys.} {\bf #1} (#2) #3}
\newc\RMP[3] {{\em Rev. Mod. Phys.} {\bf #1} (#2) #3 }
\newc\RPP[3] {{\em Rept. Prog. Phys.} {\bf #1} (#2) #3 }
\newc\SC[3] {{\em Science} {\bf #1} (#2) #3 }
\newc\ZPC[3] {{\em Z. Phys.} {\bf C #1} (#2) #3}
\newc\Err[3] {{\em Erratum-ibid.} {\bf #1} (#2) #3 }

\begin{document}

\preprint{Effectively Scanning the NMSSM  parameter space}

\title{An effective scanning method of the NMSSM parameter space
}

\author{Conny Beskidt}
\email{Conny Beskidt@kit.edu}
\author{Wim de Boer}
\email{Wim.de.Boer@kit.edu}
\affiliation{Dept. of Phys., Karlsruhe Inst. for Technology KIT, Karlsruhe, Germany
}%

\begin{abstract}

The next-to-minimal supersymmetric standard model (NMSSM) naturally provides a 125 GeV Higgs boson without the need for large loop corrections from multi-TeV stop quarks. Furthermore, the NMSSM provides an electroweak scale dark matter candidate consistent with all experimental data, like relic density and non-observation of direct  dark matter signals with the present experimental sensitivity. However, more free para\-meters are introduced in the NMSSM, which are strongly correlated.  A simple parameter scan without knowing the correlation matrix is not efficient and can  miss significant regions of the parameter space.  We introduce a new technique to sample the NMSSM parameter space, which takes into account the correlations. For this we project  the 7D NMSSM parameter space onto the 3D Higgs boson mass parameter space. The reduced dimensionality  allows for a non-random sampling  and therefore a complete coverage of the allowed NMSSM parameters. In addition, the parameter correlations and possible deviations of the signal strengths of the observed 125 Higgs boson from the SM values are easily predicted.

\end{abstract}

\maketitle

\tableofcontents
\onecolumngrid

\section{Introduction}
\label{Introduction}

The next-to-minimal supersymmetric standard model (NMSSM) distinguishes itself from the minimal supersymmetric standard model (MSSM) by a Higgs singlet in addition to the two Higgs doublets of the MSSM. This has  three advantages: 

i) The singlet solves the $\mu$ problem bagging the question why the dimension-full Higgs mixing parameter $\mu$ in the Lagrangian, which could take any value up to the GUT scale,  is required by radiative electroweak symmetry breaking to be at the electroweak scale. In the NMSSM  the $\mu$ parameter is related to the vev of the singlet, so it is naturally of the order of the electroweak scale,  see e.g. Refs. \cite{Miller:2003ay,Ellwanger:2009dp}; 

ii) The Higgs boson mass at tree level has contributions from the mixing with the singlet, so it is not restricted to be below the Z-mass at tree level, as is the case in the MSSM. Therefore, the NMSSM does not need the large loop corrections from stop quarks to bridge the gap between the Z-mass and the observed Higgs boson with a mass of 125 GeV.\cite{Aad:2012tfa,Chatrchyan:2012xdj} Bridging this gap requires multi-TeV stop masses in the MSSM, see e.g. \cite{Beskidt:2012sk,Buchmueller:2013rsa,Fowlie:2012im,Bechtle:2013mda} and references therein. However, in the NMSSM stop quark masses  can be of the order of the TeV scale, see e.g.  Ref. \cite{Beskidt:2013gia}, so the quadratic divergencies to the Higgs mass are effectively canceled by a not-too-large mass difference between top- and stop quark masses, thus avoiding the fine-tuning problem \cite{Haber:1984rc,deBoer:1994dg,Martin:1997ns}; 

iii) The dark matter candidate in the NMSSM is usually singlino-like with a mass at the electroweak scale, which  fulfills all experimental constraints, especially it has direct scattering cross sections with nuclei not yet excluded by experiments, see e.g. Ref. \cite{Beskidt:2017xsd} and references therein.

The introduction of an additional Higgs singlet  in the NMSSM yields more parameters for the interactions between the singlet and the Higgs doublets and the singlet self interaction. One usually performs random scans of the parameters to investigate experimental signatures in the parameter region allowed by the experimental constraints of the observed 125 GeV Higgs boson and its SM-like couplings and/or constraints from the dark matter sector.\cite{Dermisek:2008uu,King:2012is,Cao:2012fz,Ellwanger:2012ke,Gunion:2012zd,Cao:2013gba,Badziak:2013bda,Barbieri:2013nka,King:2014xwa,Bernon:2014nxa,Guchait:2015owa,Potter:2015wsa,Bandyopadhyay:2015tva,Bomark:2015hia,Cao:2016uwt,
Muhlleitner:2017dkd,Das:2016eob,Mariotti:2017vtv,Baum:2017gbj} In  random scans of highly correlated parameters it is difficult to reach all parameter combinations (''complete coverage''), since  the correlations require simultaneously specific  values of several  parameters.  Such combinations can be found efficiently in random scans only, if a correlation matrix is used to tell in which direction one has to step for parameter n1, if parameters n2 to nx take specific values. Incomplete coverage  can lead to wrong predictions, e.g. of the allowed cross sections for spin-dependent and spin-independent direct dark matter searches \cite{Beskidt:2017xsd}.  To cope with the large NMSSM parameter space of the Higgs sector, and especially the large correlations between these parameters, we describe in this paper in detail a novel sampling technique to obtain the allowed range of the NMSSM parameters for the constraints from the observed Higgs boson mass and its couplings. This method was previously used for the analysis of the heavy Higgs boson \cite{Beskidt:2016egy}, a light singlet-like Higgs boson \cite{Beskidt:2017dil} and dark matter constraints \cite{Beskidt:2017xsd}, but never described in detail.

After a short summary of the Higgs sector in the NMSSM in Sect. \ref{higgs}, we present the novel sampling technique to sample the NMSSM parameter space efficiently in Sect. \ref{sampling}.  In Sect. \ref{results} some applications of the novel sampling technique are presented, like determining the optimal values of the couplings and possible deviations of the signal strenths of the observed 125 GeV Higgs boson from the expected SM-like signal strengths. SM-like signals are only expected  for cross sections without loops, since in the loops of gluon fusion and the Higgs boson decays into photons, SUSY contributions from e.g.  stop loops may contribute. So only the signal strengths from diagrams without loops are required to be SM-like, while the possible deviations from diagrams including loops are studied as function of the stop mass. We restrict the analysis to the well-motivated subspace of the SUSY parameters using the unification of masses and couplings at the GUT scale and allowing for radiative electroweak symmetry breaking. Present limits on the SUSY masses indicate they are rather heavy and for heavy SUSY masses the Higgs and the SUSY sector largely decouple, except for the stop sector, which influences the 125 GeV Higgs mass and the  signal strengths   for loop-induced processes.  The restriction of the SUSY particle masses by GUT scale parameters  does not really restrict the validity of the analysis, since it does not matter if one parametrizes the stop mass with the common mass parameters at the GUT scale or chooses the stop mass directly. However, the GUT scale definition of the parameters has the  advantage that the fixed point solutions of the trilinear couplings are taken into account, thus avoiding values not allowed  by  solutions of the renormalization group equations (RGE), as will be discussed later. 

\section{The Higgs sector in the semi-constrained NMSSM}
\label{higgs}

 Within the NMSSM the Higgs fields consist of the usual two Higgs doublets ($H_u, H_d$) with an additional complex Higgs singlet $S$.  The latter singlet distinguishes the NMSSM from the MSSM.
The neutral components of the two Higgs doublets and singlet mix to form three physical CP-even scalar  bosons and two physical CP-odd pseudo-scalar  bosons.

The mass eigenstates of the neutral Higgs bosons are determined by the diagonalization of the mass matrix, see e.g. \cite{Miller:2003ay}, so the scalar Higgs bosons $H_i$ are mixtures of the CP-even weak eigenstates $H_d, H_u$ and the singlet $S$:
\begin{eqnarray}\label{eq1}
H_i=S_{i1}  H_d  + S_{i2}  H_u  + S_{i3}  S, 
\end{eqnarray}
where the index $i$ increases with increasing mass of the Higgs boson $H_i$ and  $S_{ij}$ with $i,j=1,2,3$ are the elements of the 3x3  Higgs mixing matrix. 

As mentioned before, the analysis is restricted to the well-motivated subspace of the SUSY parameters using the unification of masses and couplings at the GUT scale. In contrast to the constrained MSSM the Higgs mixing parameter $\mu$ is not fixed by radiative electroweak symmetry breaking, but is related to the vev of the Higgs singlet and is considered to be a free parameter $\mu_{eff}$.    In total, this semi-constrained NMSSM has nine free parameters: 

\begin{equation}
 \mzero,~ \mhalf,~A_0,~ \tan\beta,  ~ \lambda, ~\kappa,  ~A_\lambda, ~A_\kappa, ~\mu_{eff}.
\label{parameter}
\end{equation}

The latter six parameters in Eq. \ref{parameter} enter the Higgs mixing matrix,  thus forming the 6D parameter space of the NMSSM Higgs sector. 
Here $\tan\beta$ corresponds to the ratio of the vacuum expectation values (vev's) of the Higgs doublets, i.e. $\tan\beta=v_u/v_d$. The coupling $\lambda$ represents the coupling between the Higgs singlet and doublets, while $\kappa$ corresponds to the self-coupling of the singlet. $A_{\lambda}$ and $A_{\kappa}$ are the corresponding trilinear soft breaking terms. $\mu_{eff}$  is related to the vev of the singlet $s$ via the coupling $\lambda$, i.e. $\mu_{eff}=\lambda\cdot s$. Therefore, $\mu_{eff}$ is naturally of the order of the
electroweak scale, thus avoiding the $\mu$-problem, see e.g. Ref. \cite{Ellwanger:2009dp}. 

In addition, we have the GUT scale parameters of the constrained minimal supersymmetric standard model (CMSSM) $m_0$, $m_{1/2}$ and $A_0$, where $m_0$ and $m_{1/2}$ are the common mass scales of the spin 0 and 1/2 SUSY particles at the GUT scale. $A_0$ is the trilinear coupling of the CMSSM  at the GUT scale. The trilinear coupling $A_0$ is highly correlated with $A_{\lambda}$ and $A_{\kappa}$ in the semi-constrained NMSSM, so fixing this parameter would restrict the range of $A_{\lambda}$ and $A_{\kappa}$ severely. Therefore, $A_0$ is allowed to vary, which leads  to 7 free parameters in total and thus a 7D NMSSM parameter space. From the free parameters in Eq. \ref{parameter} the complete SUSY spectrum and all Higgs boson masses can be calculated using the publicly available code NMSSMTools.\cite{Das:2011dg} 
The values of $m_0$ and $m_{1/2}$ can be fixed to values, which are consistent with the current LHC limits \cite{Aad:2015iea}. In the following we use $m_0=m_{1/2}=$1 TeV. The impact of higher common SUSY masses will be discussed in Sect. \ref{results}. 

One of the lightest NMSSM Higgs bosons should be SM-like. The cross section errors have a significant theoretical error from the  dependence on the renormalization - and factorization scales, see e.g. Ref. \cite{Djouadi:2005gi} and references therein, 
which can be reduced by using cross section ratios of the NMSSM and SM cross sections, the so-called reduced cross sections. If multiplied by the branching ratios one obtains the signal strength $\mu_j^i$  defined as: 
\begin{align}
\mu_j^i=\frac{\sigma_i \times BR_j}{(\sigma_i \times BR_j)_{SM}}=c_i^2 \cdot \frac{BR_j}{(BR_j)_{SM}},\label{coupling4} 
\end{align}
where the  reduced coupling $c_i$ squared equals the reduced cross section for production mode $i$, which is  multiplied by the corresponding ratio of branching ratios for the decay $j$. The reduced couplings $c_i$ depend only on the Higgs mixing matrix elements and $\tan\beta$ for processes without loops. The reduced couplings  {\it including} loops can have additional contributions from SUSY particles in the loops, preferentially from particles with large couplings to the Higgs boson, like the stop particles.  These modify  the reduced couplings of Higgs bosons to gluons $c_{gluon}$ and gammas $c_\gamma$, which are parametrized as effective couplings  within NMSSMTools.

\begin{figure}
\begin{center}
\begin{minipage}{\textwidth}
\begin{tabular}{cccc}
\multirow{7}{*}{\includegraphics[width=0.23\textwidth]{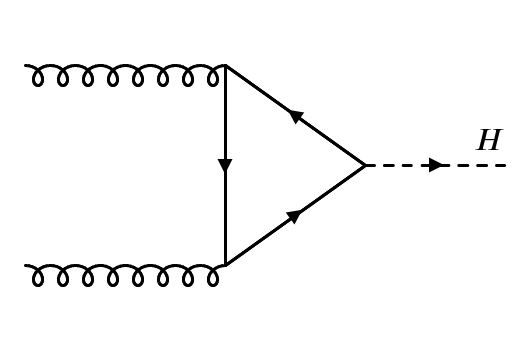}} & \multirow{7}{*}{\includegraphics[width=0.23\textwidth]{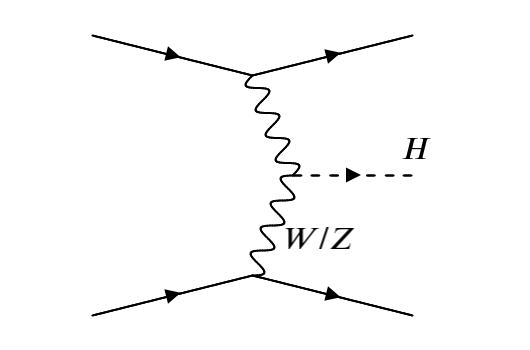}} &  \multirow{7}{*}{\includegraphics[width=0.23\textwidth]{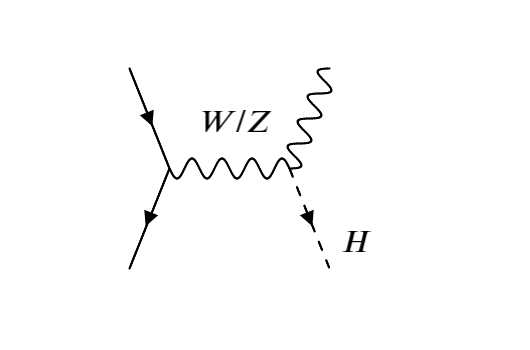}} &  \multirow{7}{*}{\includegraphics[width=0.23\textwidth]{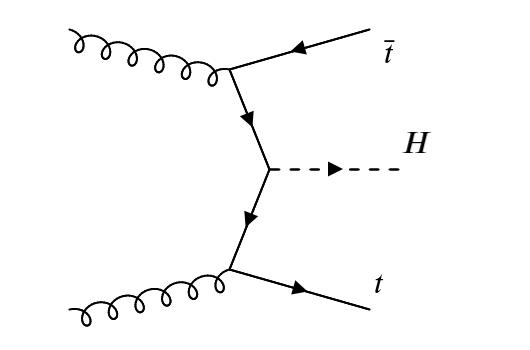}} \\
&  &  & \\
&  &  & \\
&  &  & \\
&  &  & \\
 & &  & \\
 & &  & \\
 & &  & \\
 \small{$\sim c_{gluon}^2$} &  \small{$\sim c_{W/Z}^2$}&  \small{$\sim c_{W/Z}^2$}&  \small{$\sim c_{u}^2$}\\
\end{tabular}
\end{minipage}
\caption[]{ Summary for the Higgs production channels at the LHC and the corresponding reduced couplings.  From left to right: the effective reduced gluon coupling $c_{gluon}$ for gluon fusion (ggf), $c_{W/Z}$ for vector boson fusion (VBF) and Higgs Strahlung (VH) and $c_u$ for top fusion (tth).
}
\label{f1}
\end{center}
\end{figure}

 The diagrams for the used reduced couplings have been summarized  in Fig. \ref{f1}, which shows  from left to right: the effective reduced gluon coupling $c_{gluon}$ for gluon fusion (ggf), $c_{W/Z}$ for vector boson fusion (VBF) and Higgs Strahlung (VH) and $c_u$ for top fusion (tth). We consider in our analysis two different fermionic final states (b-quarks and $\tau$-leptons) and two different bosonic final states ($W/Z$ and $\gamma$)  for  two different production modes (gluon fusion (ggf) and vector boson fusion (VBF)) leading to a total of 8 reduced cross sections (all calculated in NMSSMTools), which can be divided into  signal strengths without (with) effective couplings $\mu_{no-loop}(\mu_{loop})$:   
 
\begin{align}
\mu_{no-loop}&:&\mu_{\tau\tau}^{VBF/VH}&, &\mu_{ZZ/WW}^{VBF/VH}&, &\mu_{bb}^{VBF/VH}&, &\mu_{bb}^{tth},\notag\\
\mu_{loop}&:&\mu_{\tau\tau}^{ggf}&, &\mu_{ZZ/WW}^{ggf}&, &\mu_{\gamma\gamma}^{VBF/VH}&, &\mu_{\gamma\gamma}^{ggf}.\label{coupling5}
\end{align}

The  signal strengths $\mu_{loop}$ include loop diagrams at lowest order, see e.g. the left diagram in Fig. \ref{f1}, so  SUSY particles in the loop can lead  to deviations from the SM prediction. Therefore, it is reasonable to constrain only the signal strengths without loop contributions $\mu_{no-loop}$ to its SM-like expectation, i.e.  $\mu_{no-loop}$=1, while $\mu_{loop}$ is allowed to deviate from 1.  So we impose the following four constraints:  $\mu_{\tau\tau}^{VBF/VH}=1, \mu_{bb}^{VBF/VH}=1, \mu_{ZZ/WW}^{VBF/VH}=1$ and $\mu_{bb}^{tth}=1$.  The signal strengths  $\mu_{loop}$ can be calculated from the fitted NMSSM parameters using the constraint  $\mu_{no-loop}$=1. The fit will be discussed in the next section.

\section{Sampling technique}
\label{sampling}

\begin{figure}
\begin{center}
\includegraphics[width=0.6\textwidth]{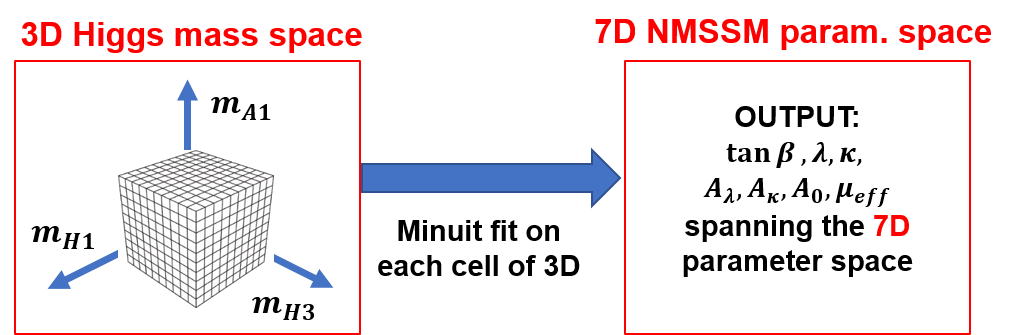}\\[2mm]
\caption[]{ 
Sketch of the sampling technique to determine the allowed NMSSM parameter space: 
Perform a Minuit fit on each cell of the 3D Higgs mass space  (left box) to determine the corresponding 7 free NMSSM parameters (right box). The relation between the NMSSM parameters and the Higgs masses is encoded in NMSSMTools. The second-lightest Higgs boson is chosen to be the 125 GeV Higgs, but we repeat the fit in case $m_{H1}$=125 GeV. Then the $m_{H1}$  becomes an $m_{H2}$ axis in the grid on the left.
}
\label{f2}
\end{center}
\end{figure}

The NMSSM parameters in Eq. \ref{parameter} completely determine the masses of the 6 Higgs bosons: 3 scalar Higgs masses $m_{H_i}$, 2 pseudo-scalar Higgs masses $m_{A_i}$ and the charged Higgs boson mass $m_{H^\pm}$. The masses of $A_2$, $H_3$ and $H^\pm$ are approximately equal in the decoupling limit, i.e.  the mass region with heavy Higgs masses  much larger than the Z-boson mass.\cite{Haber:1995be,Djouadi:2005gj}
Then only one of the heavy Higgs masses is independent which leads in total to a 4D Higgs mass space. Furthermore, one of the masses has to be 125 GeV, so only 3 Higgs masses are free in the decoupling limit, which can be chosen to be: $m_{A_1}$, $m_{H_1}$ and $m_{H_3}$. The Higgs masses can be calculated from the NMSSM parameters in Eq. \ref{parameter}, but vice versa  each combination of the 3 masses $m_{A_1}$, $m_{H_1}$ and $m_{H_3}$ uniquely determines the 7 NMSSM parameters assuming the decoupling limit. The parameters  can be obtained for each combination of Higgs masses in the 3D  space of Higgs masses from a fit to the masses with the NMSSM parameters as free parameters. Such a fit one would perform, if all Higgs boson masses would have been measured and we show later that the fit leads to unique solutions. In the fit we assume either $m_{H_1}$ or $m_{H_2}$ to be 125 GeV, leading to two independent fits for each Higgs mass combination.
 The procedure is illustrated in Fig. \ref{f2}. The fit can be performed for each cell of the 3D Higgs mass space on the left, thus providing for each Higgs mass combination the 7 NMSSM parameters on the right panel of Fig. \ref{f2}.  For each set of these parameters the  Higgs mixing matrix is fully specified and hence the Higgs sector including masses, couplings, branching ratios and cross sections can be calculated from these 7 parameters.

Note that with the low dimensionality of the 3D grid spanning the Higgs mass space one can perform the fit  for each cell in the 3D Higgs mass space, thus providing a complete coverage of the Higgs sector and the corresponding NMSSM parameters without having to resort to a random scan. 

As statistic for the fit determining the NMSSM parameters from the Higgs masses we choose the $\chi^2$ function, which  can be minimized  by Minuit.\cite{James:1975dr}  The following contributions are included in the $\chi^2$ function:
\beq\label{eq5}
\chi^2_{tot}=\chi^2_{H_S}+\chi^2_{H_3}+\chi^2_{A_1}+\chi^2_{H_{125}}+\chi^2_{\mu_{125}}+\chi^2_{LEP}+\chi^2_{LHC},
\eeq 
which are defined as:
\begin{itemize} 
\item $\chi^2_{H_S}=(m_{H_S} - m_{grid,H_S})^2/\sigma^2_{H_S}$: 
The term $\chi^2_{H_S}$ requires the NMSSM parameters to be adjusted such that the mass of the singlet-like light Higgs boson mass $m_{H_S}$ agrees with the chosen point in the 3D mass space $m_{grid,H_S}$. 
The value of $\sigma_{H_S}$ is set to $1\permil$ of $m_{grid,H_S}$.  A small error on the chosen Higgs mass   avoids a smearing in the 3D Higgs mass space and a corresponding smearing by the projection onto the 7D parameter space.
\item $\chi^2_{H_3}=(m_{H_3} - m_{grid,H_3})^2/\sigma^2_{H_3}$: as $\chi^2_{H_S}$, but for the heavy scalar Higgs boson $H_3$.
\item $\chi^2_{A_1}=(m_{A_1} - m_{grid,A_1})^2/\sigma^2_{A_1}$: as $\chi^2_{H_S}$, but for the light pseudo-scalar Higgs boson $A_1$.
\item $\chi^2_{H_{125}}=(m_{H_125} - m_{obs})^2/\sigma^2_{125}$: This term is analogous to the term for $m_{H_S}$, except that the  Higgs  mass $m_{H_125 }$ is required to agree with the observed Higgs boson mass, so $m_{obs}$ is set to $125.2$ GeV. The corresponding uncertainty $\sigma_{125}$ was set to $1\permil$ of $m_{obs}$. Note that the much larger error on the mass of the observed 125 GeV boson is not taken into account, since we want to determine the NMSSM parameters for a precise scan of the Higgs mass parameter space. Once the NMSSM parameters have been determined one can look for the region, where the predicted Higgs mass is within the experimental errors of the observed Higgs mass.
\item $\chi^2_{\mu_{125}}=\sum_i (\mu^i_{H_{125}} - \mu_{theo})^2/\sigma^2_{\mu}$: This term requires the Higgs boson $H_{125}$ to have SM-like couplings for  
 the four signal strengths, labeled as $\mu_{no-loop}$ in Eq. \ref{coupling5},  so $\mu_{theo}=1$ and $\sigma_{\mu}$ was chosen to be 0.005.  
The remaining 4 reduced cross sections $\mu_{loop}$, which include either gluons and/or gammas, can be calculated from the fitted NMSSM parameters and can deviate from one because of the  SUSY contributions, which depend on the choice of $m_0,m_{1/2}$. 
\item $\chi^2_{LEP}$ includes the LEP constraints on the couplings of a light Higgs boson below 115 GeV and the limit on the chargino mass. These constraints include upper limits on the decay of light Higgs bosons into b-quark pairs, which are particular important for the singlet Higgs, if it is the lightest one. The LEP constraints are in principle implemented in NMSSMTools, but small corrections were applied, as discussed in Ref. \cite{Beskidt:2014kon}.      
\item $\chi^2_{LHC}$ includes constraints from the LHC, as implemented in NMSSMTools, concerning light scalar and pseudo-scalar Higgs bosons. \cite{Khachatryan:2015nba,Aad:2015oqa,Khachatryan:2015wka}
\end{itemize}
We assume that the constraints on SUSY mass limits from LEP and LHC as well as the Higgs masses are uncorrelated. Note that we either assume the lightest Higgs $H_1$ to be the singlet-like Higgs boson $H_S$ and the second lightest Higgs boson $H_2$ to be the 125 GeV SM Higgs boson $H_{SM}$ or vice versa. In the first case, the singlet-like Higgs boson has a mass below 125 GeV while for the second case the mass is above 125 GeV. 
There are also solutions where $m_{H_1}=125$ GeV, $m_{H_2}>125$ GeV and $m_{H_3}$ is singlet-like, but we focus on the scenarios where the heavy Higgs bosons are MSSM-like, which leads naturally to $m_{A_2} \approx m_{H_3}$ and is a reasonable assumption.
We restrict the range of the Higgs masses in the 3D mass space as follows:
\begin{align}
5~\mathrm{GeV} &<  m_{H_S} < 500~\mathrm{GeV},\notag\\
125~\mathrm{GeV} & < m_{H_3} < 2~\mathrm{TeV},\label{range}\\
5~\mathrm{GeV} & < m_{A_1} < 500~\mathrm{GeV}.\notag
\end{align}
Although heavier Higgs boson are not forbidden, they are not relevant for LHC physics, so we did not investigate them here.

\section{Exemplary results}
\label{results}
\begin{figure}[ht]
\vspace{-0.7cm}
\begin{center}
\begin{minipage}{\textwidth}
\begin{tabular}{cccc}
& {\boldmath$\chi^2$} &{\boldmath$\mu$}& { \boldmath$m_{\tilde{t}}$} \\
&   \multirow{7}{*}{\includegraphics[height=0.2\textwidth]{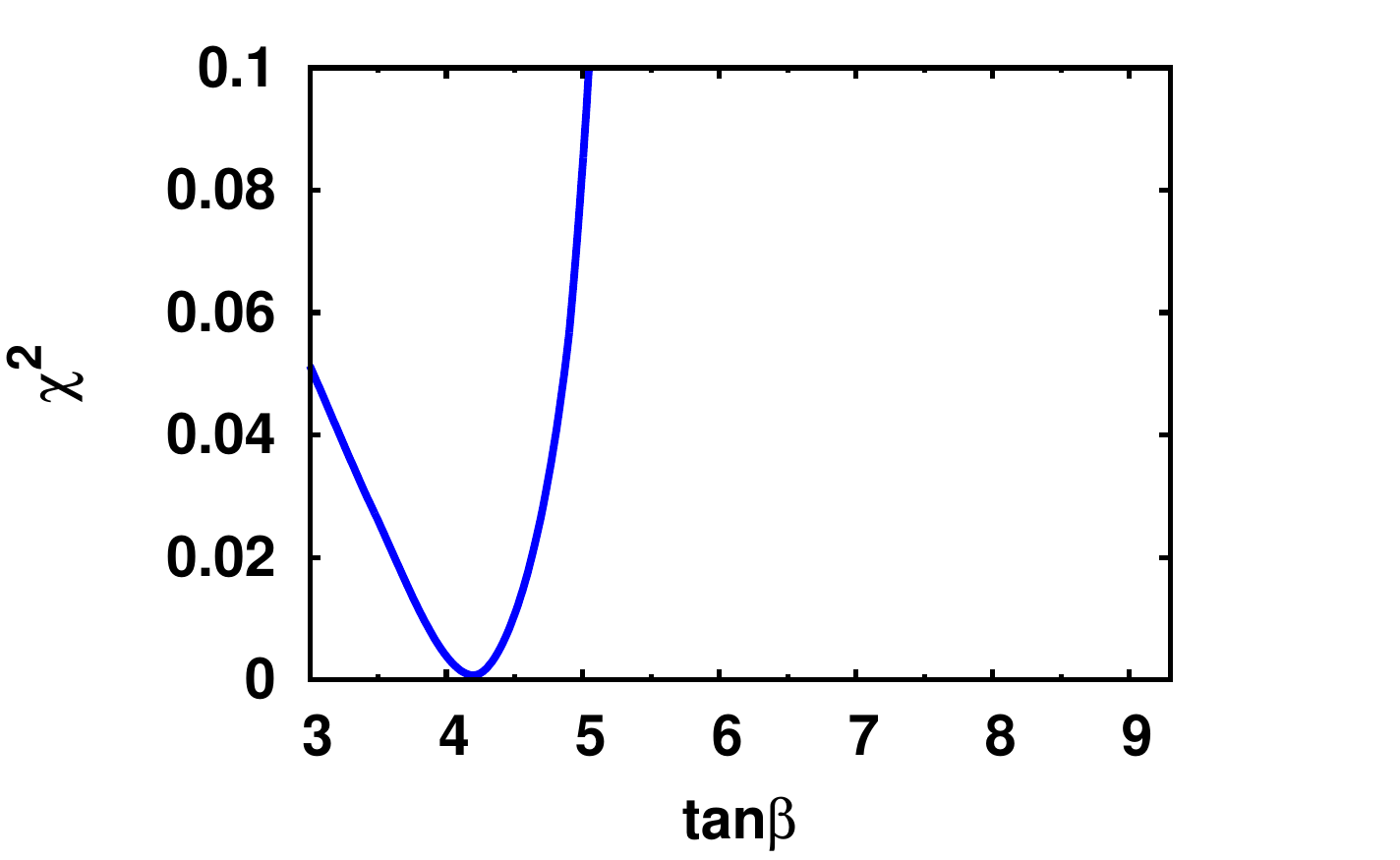}} 
& \multirow{7}{*}{\includegraphics[height=0.2\textwidth]{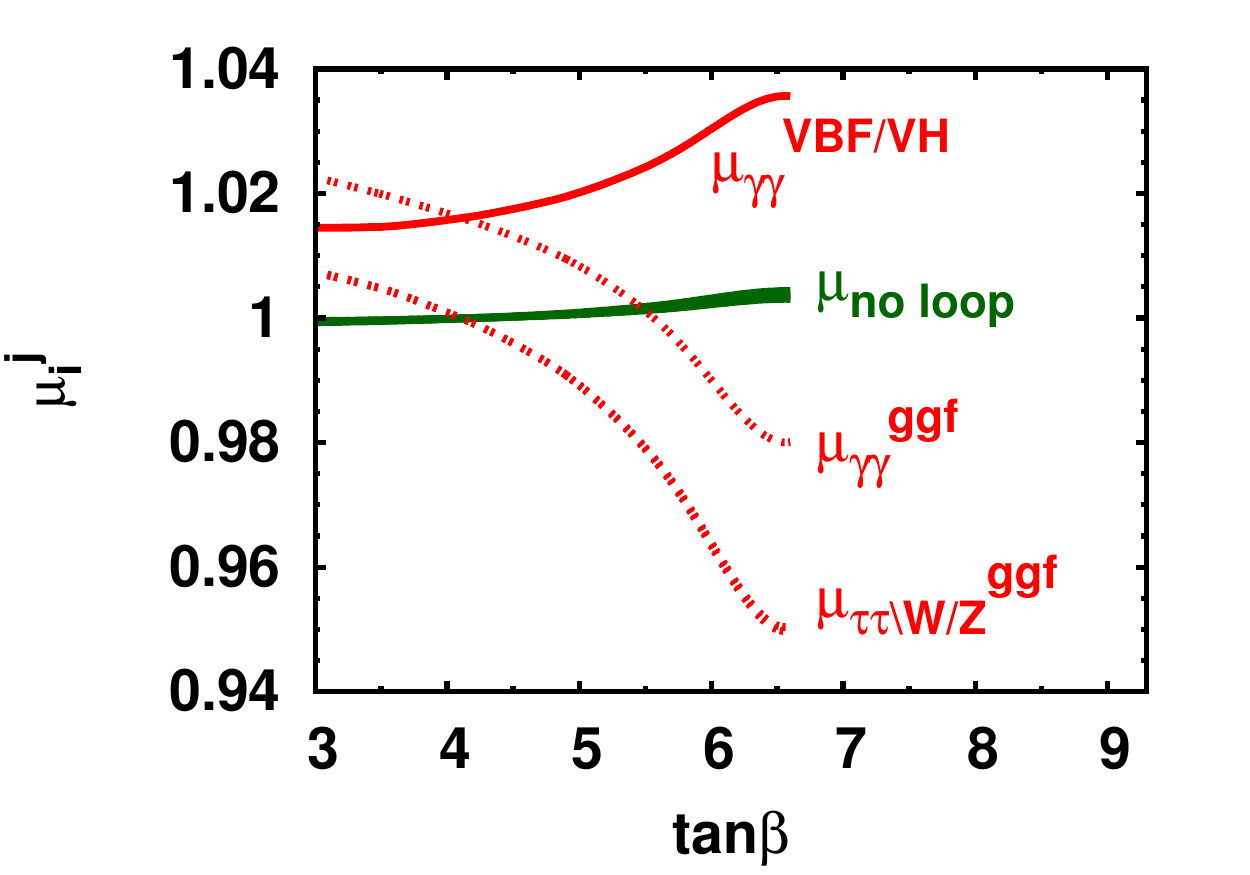}} 
&  \multirow{7}{*}{\includegraphics[height=0.2\textwidth]{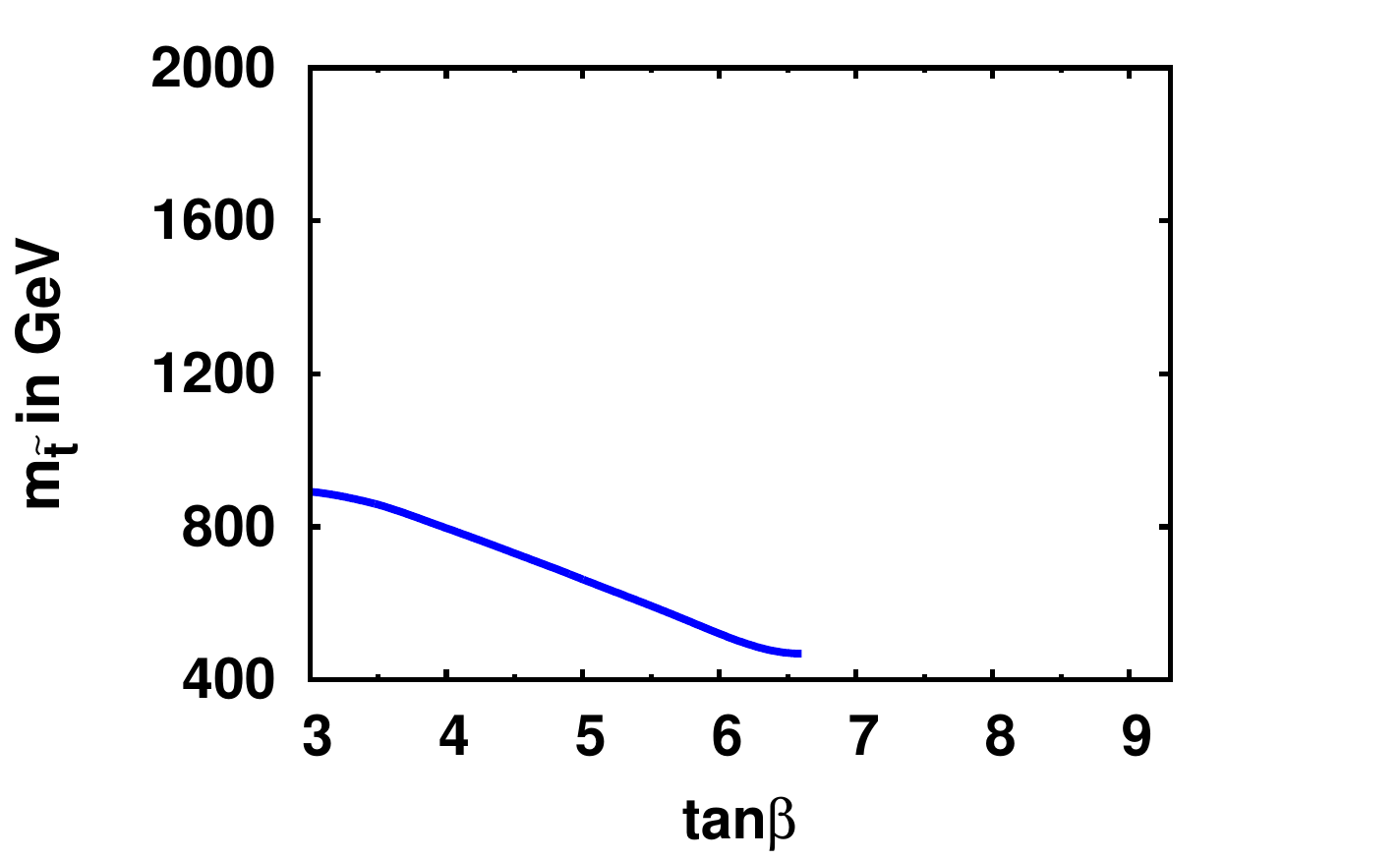}} \\[10mm]
 & &  & \\
 \small{\boldmath$m_0=m_{1/2}$} & & & \\
\small{\boldmath$=0.7$ \textbf{TeV}} & &  & \\
 & &  & \\
& &  & \\
 & &  & \\
&   \multirow{7}{*}{\includegraphics[height=0.2\textwidth]{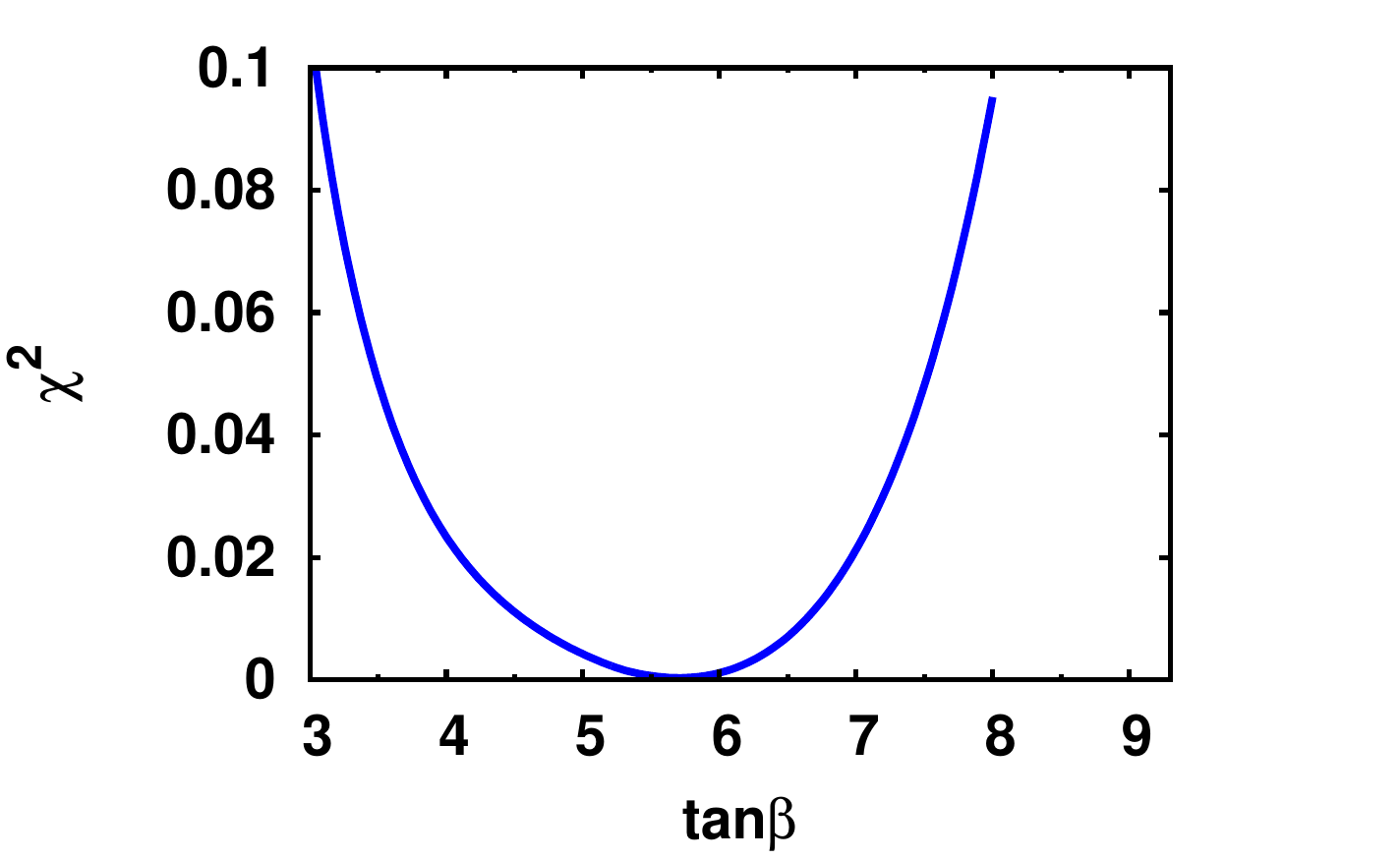}} 
& \multirow{7}{*}{\includegraphics[height=0.2\textwidth]{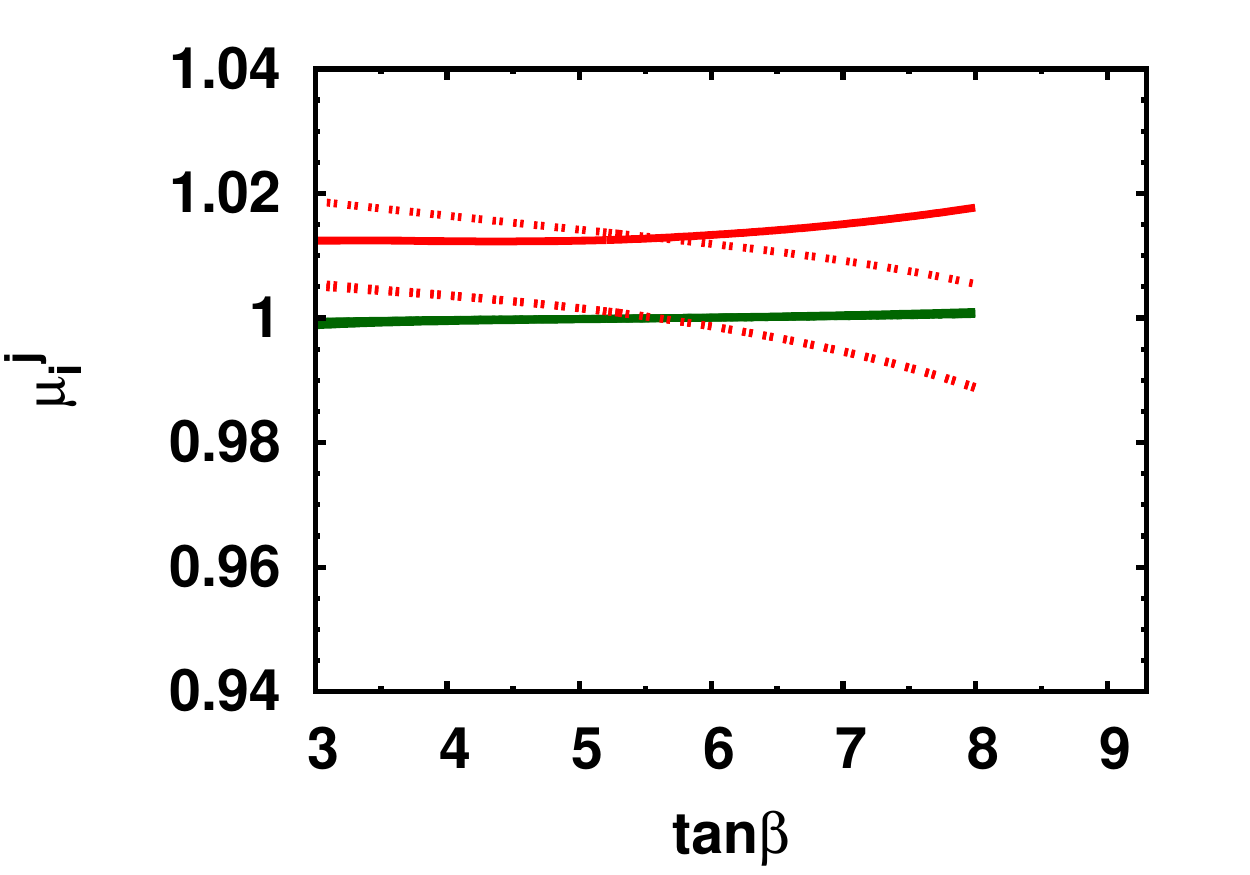}} 
&  \multirow{7}{*}{\includegraphics[height=0.2\textwidth]{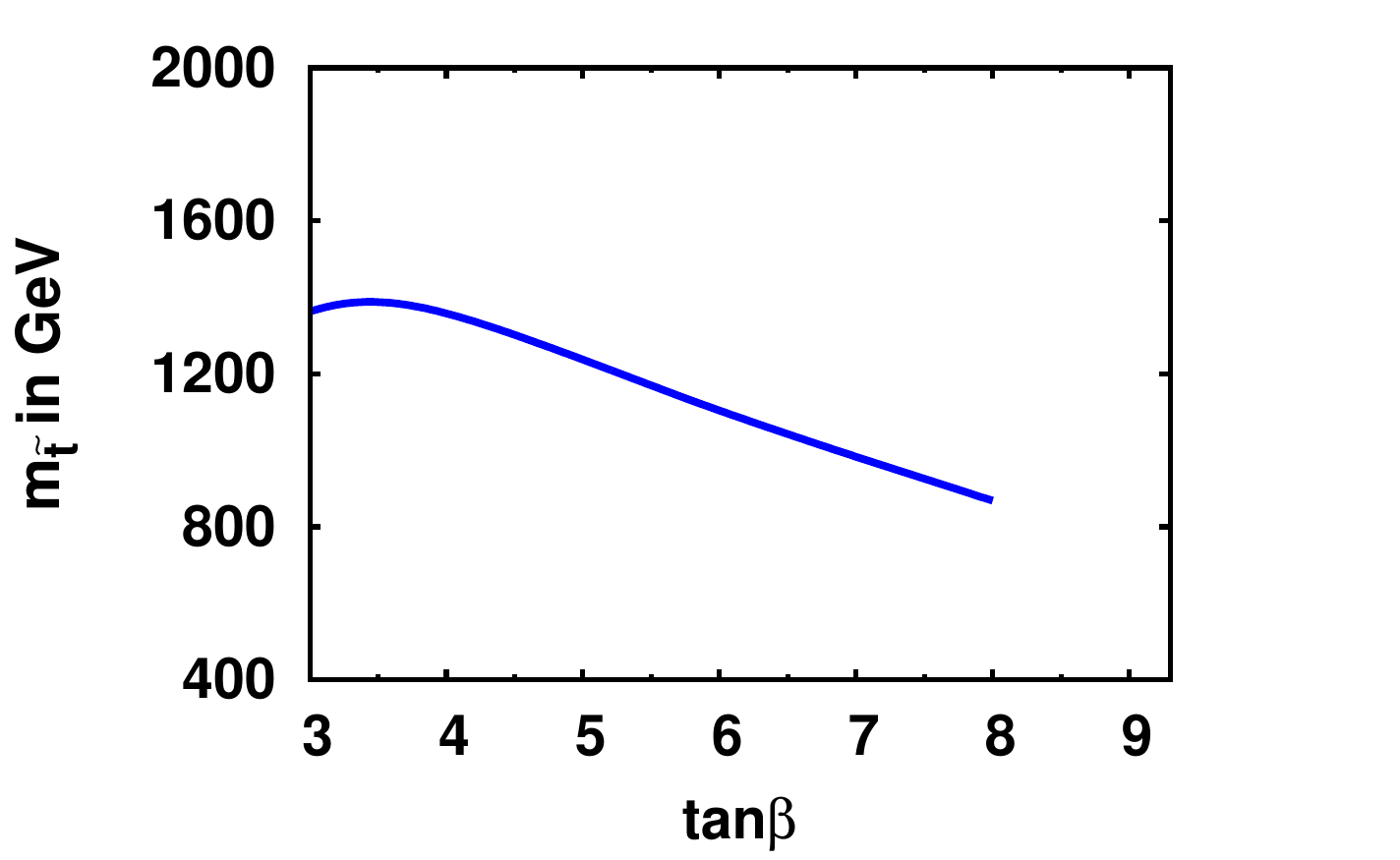}} \\[10mm]
 & &  & \\
 \small{\boldmath$m_0=m_{1/2}$} & & & \\
\small{\boldmath$=1.0$ \textbf{TeV}} & &  & \\
&  &  & \\
 & &  & \\
& &  & \\
&   \multirow{7}{*}{\includegraphics[height=0.2\textwidth]{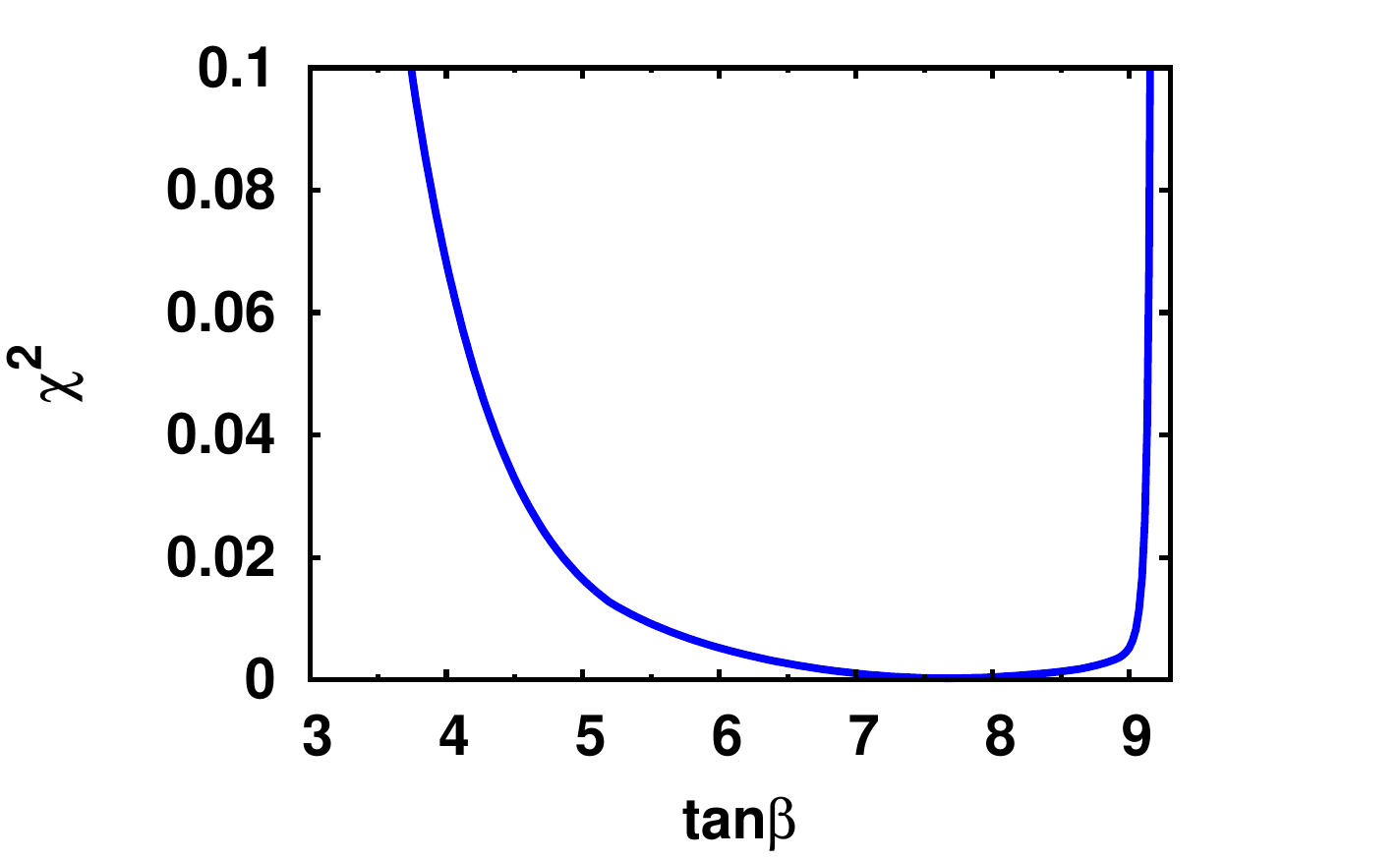}} 
& \multirow{7}{*}{\includegraphics[height=0.2\textwidth]{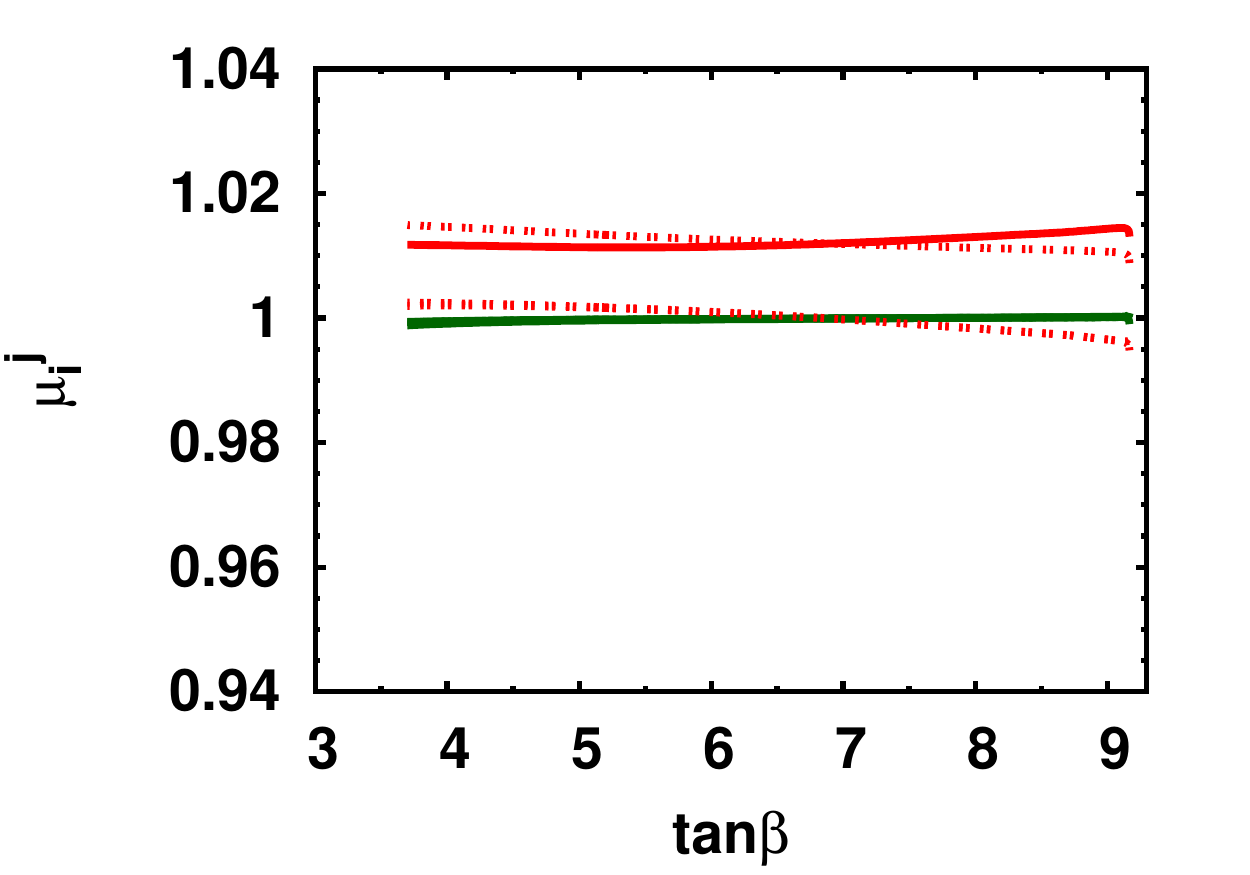}} 
&  \multirow{7}{*}{\includegraphics[height=0.2\textwidth]{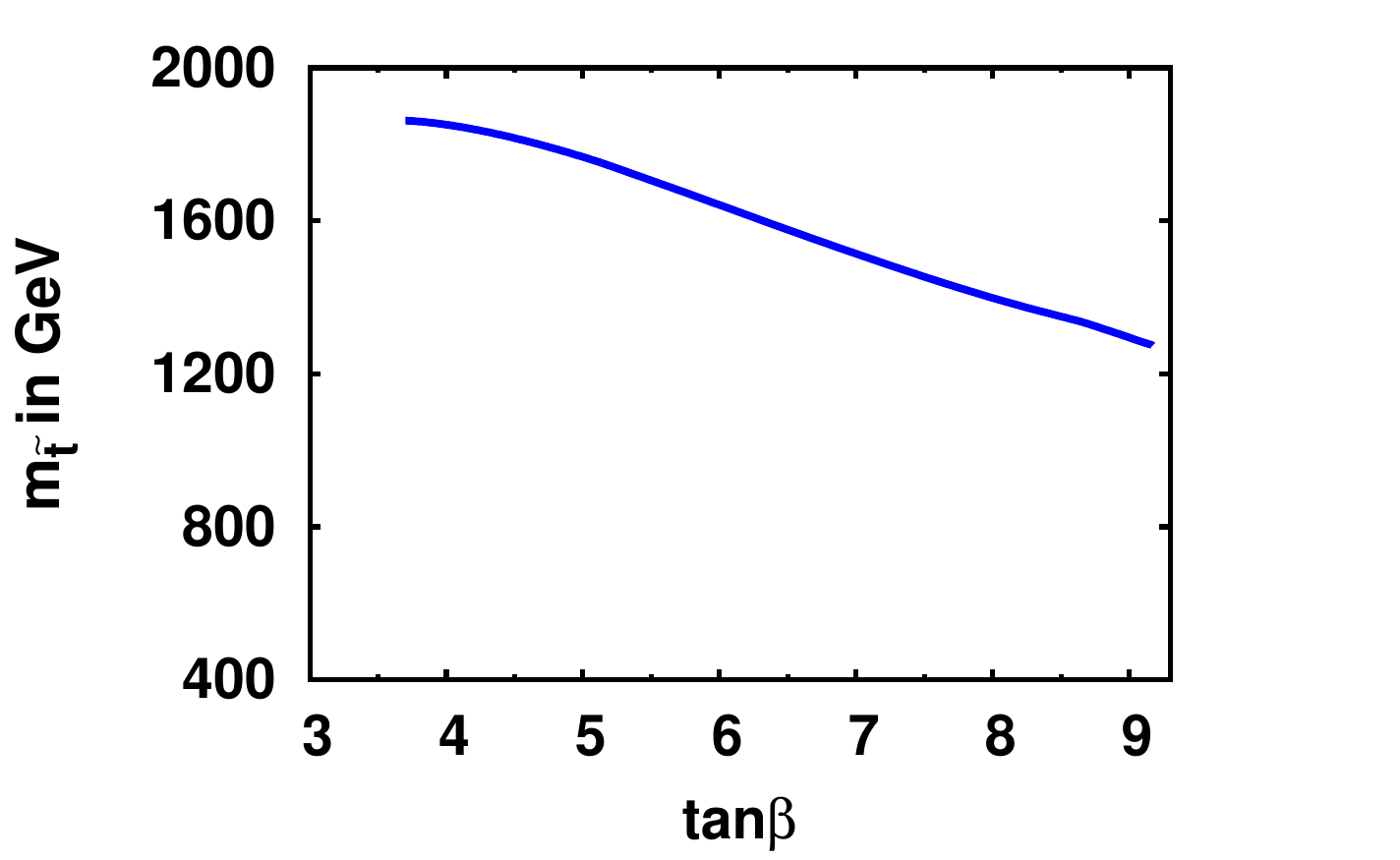}} \\[10mm]
 & &  & \\
 \small{\boldmath$m_0=m_{1/2}$} & & & \\
\small{\boldmath$=1.3$ \textbf{TeV}} & &  & \\
 & &  & \\
& &  & \\
&  &  & \\
\end{tabular}
\end{minipage}
\end{center}
\caption[]{
The left side shows the $\chi^2$ distribution as function of $\tan\beta$ for $m_0=m_{1/2}=$ 0.7/1.0/1.3 TeV from top to bottom, respectively. The fits are for an exemplary  Higgs mass combination  of $m_{H1}=90$ GeV, $m_{H3}=1000$ GeV, $m_{A1}=200$ GeV. The main contributions to the $\chi^2$ are coming from the signal strengths, which are shown in the middle panels for $m_0=m_{1/2}=$ 0.7/1.0/1.3 TeV, respectively. The signal strengths including gammas and/or gluons deviate from the SM prediction, since the corresponding reduced couplings, and hence the signal strengths, are sensitive to SUSY contributions. These contributions vary with the mass of the SUSY particles, which vary strongly from top to bottom, as can be seen from the right panels for $m_0=m_{1/2}=$ 0.7/1.0/1.3 TeV.  
}
\label{f3}
\end{figure}
The fit  for all Higgs mass combinations in the left box of Fig. \ref{f2} leads to the allowed region of the NMSSM parameters in the right box. 
The left side of Fig. \ref{f3}  shows the $\chi^2$ distribution as function of $\tan\beta$ for $m_0=m_{1/2}=$ 0.7/1.0/1.3 TeV from top to bottom, respectively for a given cell  in the left panel of Fig. \ref{f2} (in this case    $m_{H1}=90$ GeV, $m_{H3}=1000$ GeV, $m_{A1}=200$ GeV). The main contribution to the $\chi^2$ is coming from the signal strengths, which are close to the SM expectation of 1 for a large range of $\tan\beta$ for processes without loops, as shown on the middle panels of Fig. \ref{f3}. However, the signal strengths {\it including} loop contributions deviate from 1 because of the  SUSY contributions. The deviations are strongest for light stop masses (top row):  the gluon fusion reduced signal strenths decrease by $\sim 7$\% if the stop mass decreases from 800 to 400 GeV for $\tan\beta$ increasing from 3-6. The signal strength including the decay into gammas increases at the same time by $\sim 2$\%.  The stop masses as function of $\tan\beta$) are shown on the panels on the right for the different choices of  $m_0,m_{1/2}=$. 
Larger SUSY masses decrease the contribution from the stop loops, thus reducing the signal strength dependence on $\tan\beta$, as can be seen from the panels in the middle of Fig. \ref{f3}.

    The shift in the minimum of the $\chi^2$ function to higher $\tan\beta$  in the middle and bottom row is  caused by the increase in the stop mass, since  the reduced stop  corrections to the 125 GeV Higgs boson mass can be compensated largely by an increase in $\tan\beta$, although  other parameters  change somewhat as well for the minimum $\chi^2$ value.

Not only a shift but also a broader $\chi^2$ distribution is shown for $m_0,m_{1/2}$ varying from 0.7 to 1.3 TeV. This originates mainly from the 125 GeV mass constraint of the observed boson and can be understood as follows:
 For  the lowest stop mass  ($m_0,m_{1/2}$=0.7 TeV) one has the smallest loop corrections to the 125 GeV Higgs mass and one needs a large mass correction from NMSSM mixing between the singlet and other Higgs bosons, which requires a precise tuning of the NMSSM parameters  to reach the value of 125 GeV. For heavier stop mass the loop corrections to the 125 GeV Higgs boson are larger and the contribution from mixing  with the singlet can be smaller, which leads  to more freedom in the NMSSM parameters and thus a broader $\chi^2$ distribution. We have checked this by imposing a smaller value below 125 GeV for the observed  Higgs boson mass, in which case the freedom in the NMSSM parameters is larger for a given stop mass, thus leading to a broader $\chi^2$ distribution. This is demonstrated in   Appendix \ref{SUSYmasses}, where we used hypothetical values of 123, 124 and 125 GeV for the observed Higgs boson.

\begin{figure}
\begin{center}
\includegraphics[width=0.32\textwidth]{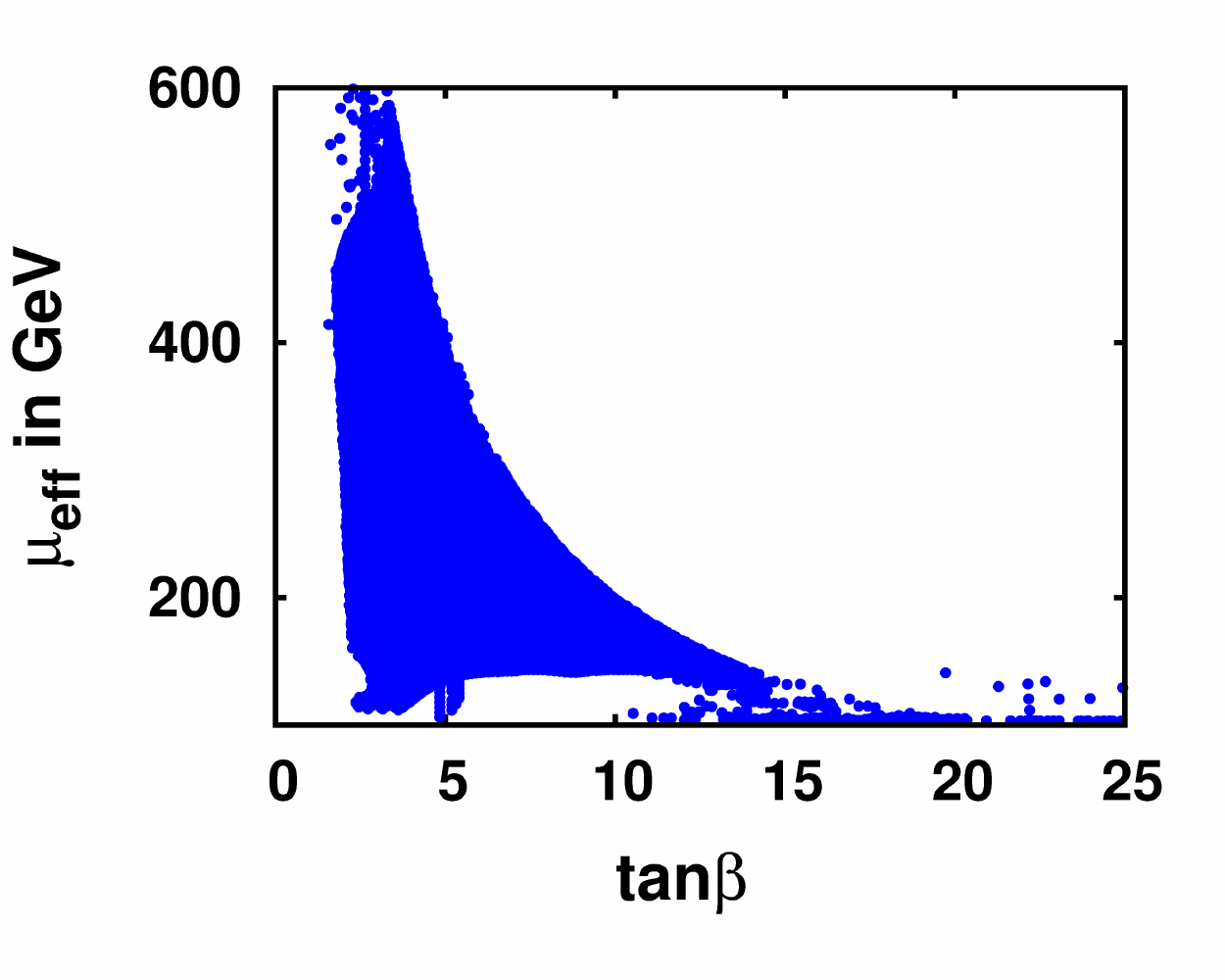}
\includegraphics[width=0.32\textwidth]{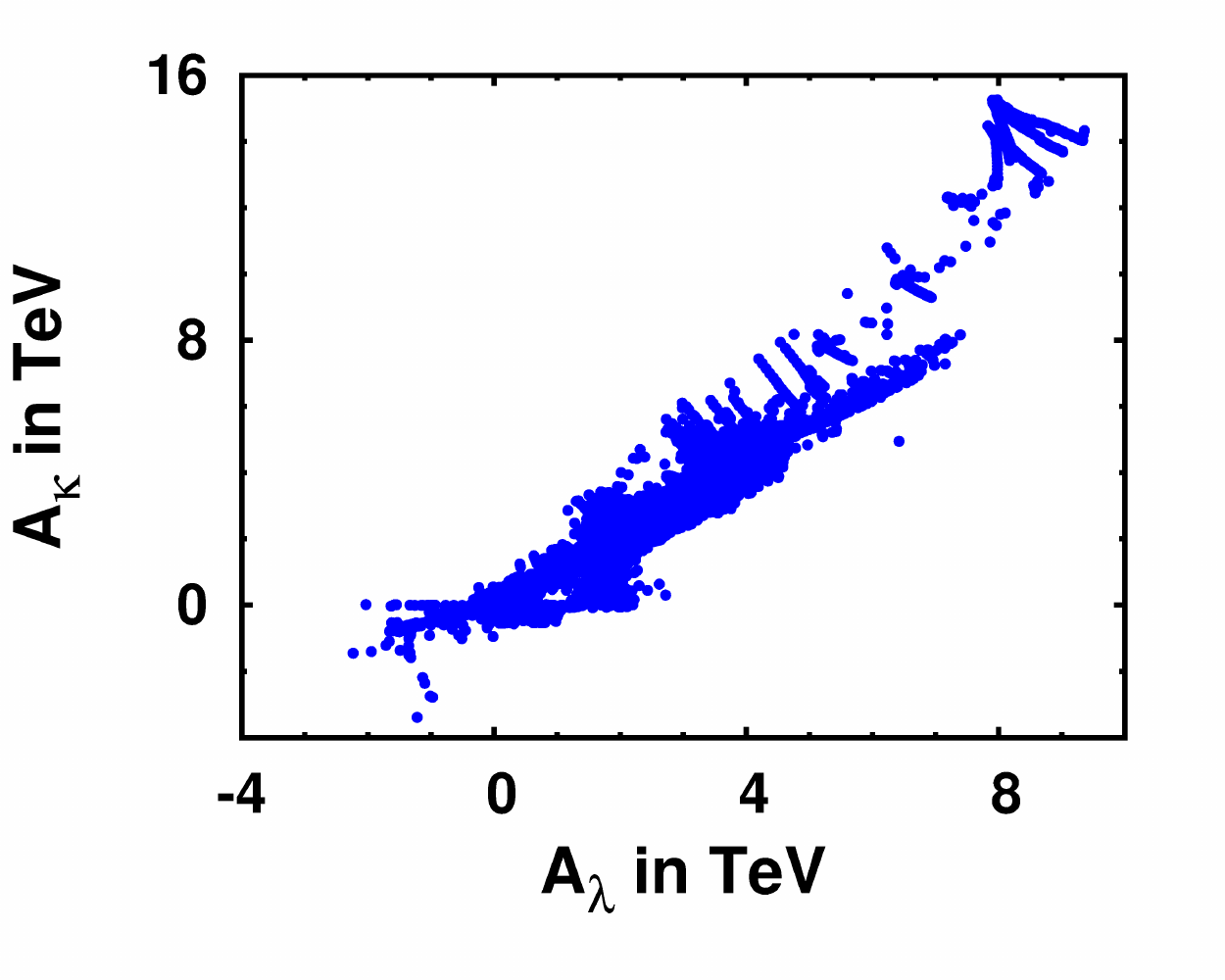}
\caption[]{
Allowed ranges for a few NMSSM parameters in the planes  $\tan\beta-\mu_{eff}$ (left) and  $A_\kappa-A_\lambda$ (right). 
}
\label{f4}
\end{center}
\end{figure}

The NMSSM parameters are correlated, but they are not degenerate, as can be observed from the correlations between the parameters  in the Minuit output. The maximum correlation occurs between the parameters $\mu_{eff}$ and $\kappa$, but the correlation is still below 0.76 and other correlations are  smaller, as shown for a given mass combination  in  Appendix   \ref{matrix}. After fitting all cells in the 3D Higgs mass space (left side of Fig. \ref{f2}) one obtains the allowed regions of the parameters, which are shown for a few parameters in the two dimensional plots  in Fig. \ref{f4}.
The parameters $\tan\beta$ and $\mu_{eff}$ show a strong negative correlation, while the trilinear couplings show a strong positive correlation for the GUT scale input parameters. 
The values of the trilinear couplings at the SUSY scale are much more restricted than their values at the GUT scale,  because of the fixed point solutions of the RGE for $A_\lambda$ and $A_\kappa$, which means that the SUSY scales are largely independent of the GUT scale values. However, the SUSY scale values depend on the chosen Higgs masses, so the ranges of the SUSY scale values of $A_\lambda$ and $A_\kappa$ are still appreciable, as demonstrated in  Appendix  \ref{running}. 

\begin{figure}
\begin{center}
\includegraphics[width=0.32\textwidth]{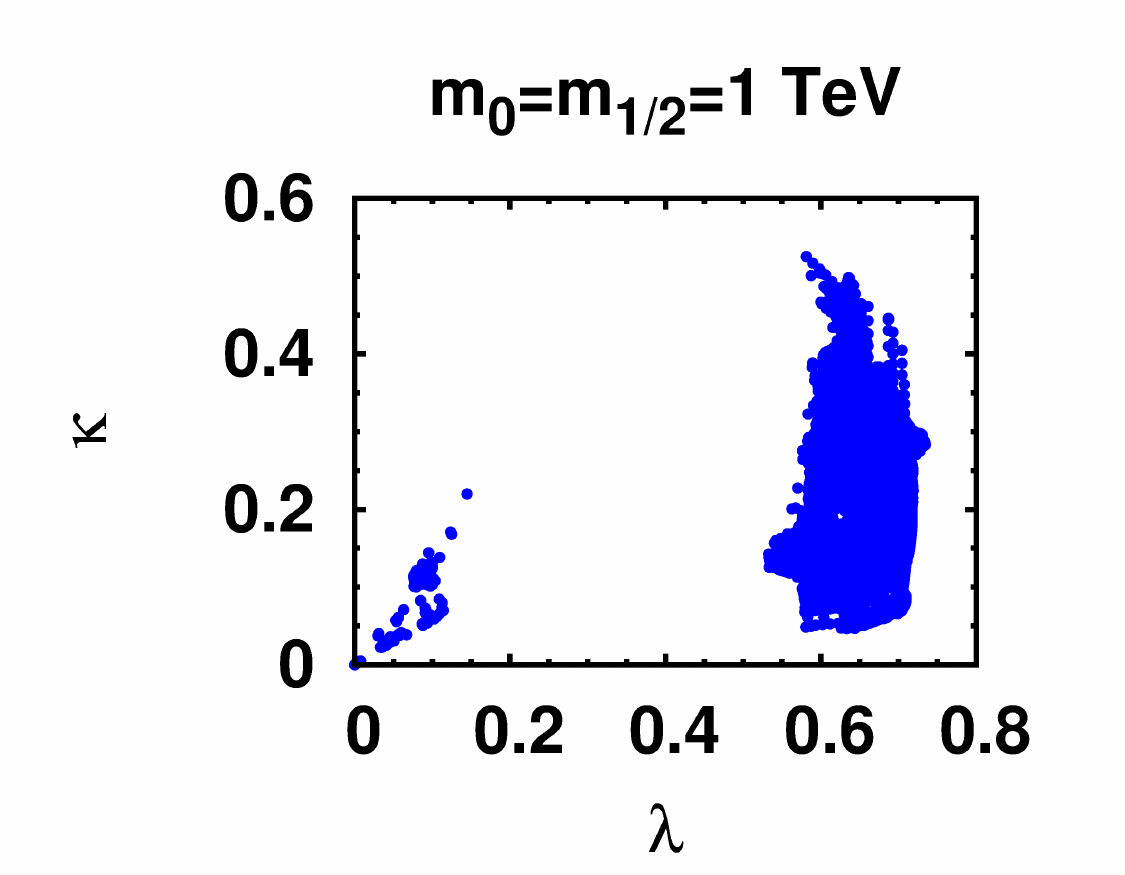}
\includegraphics[width=0.32\textwidth]{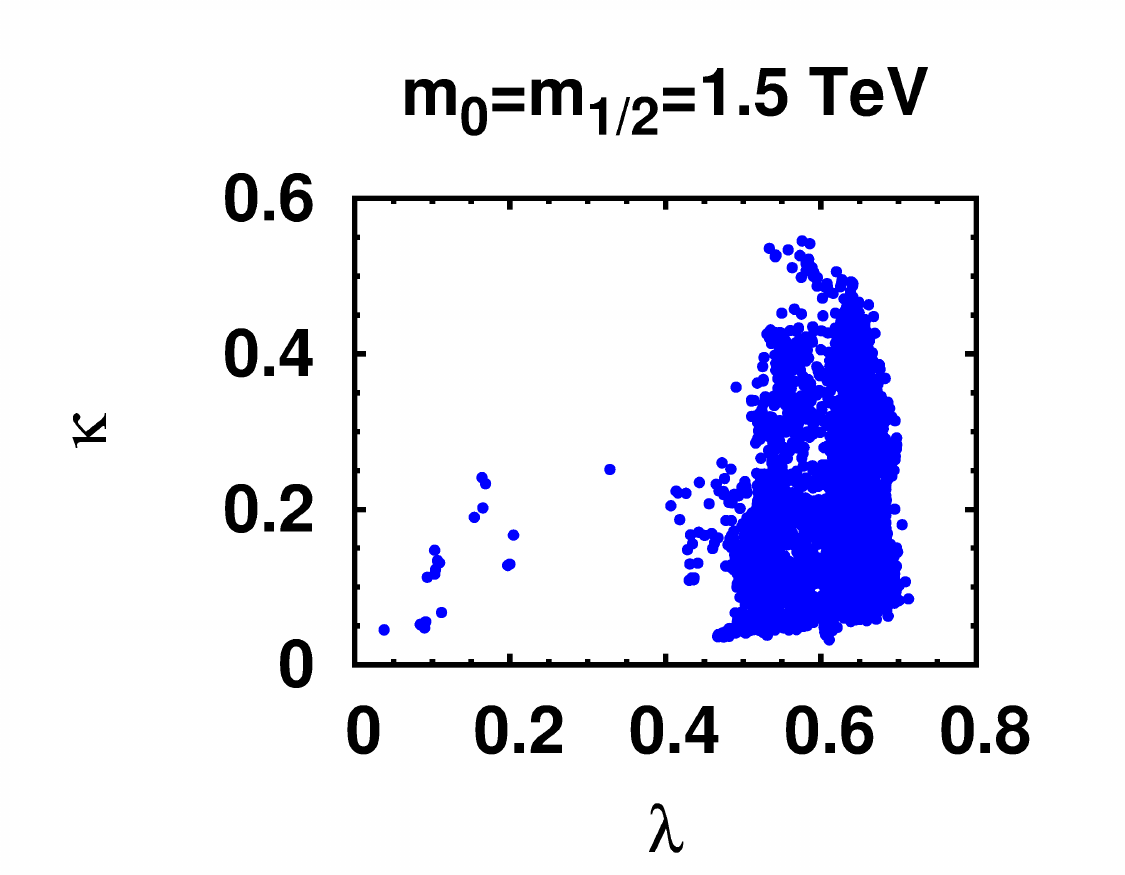}
\includegraphics[width=0.32\textwidth]{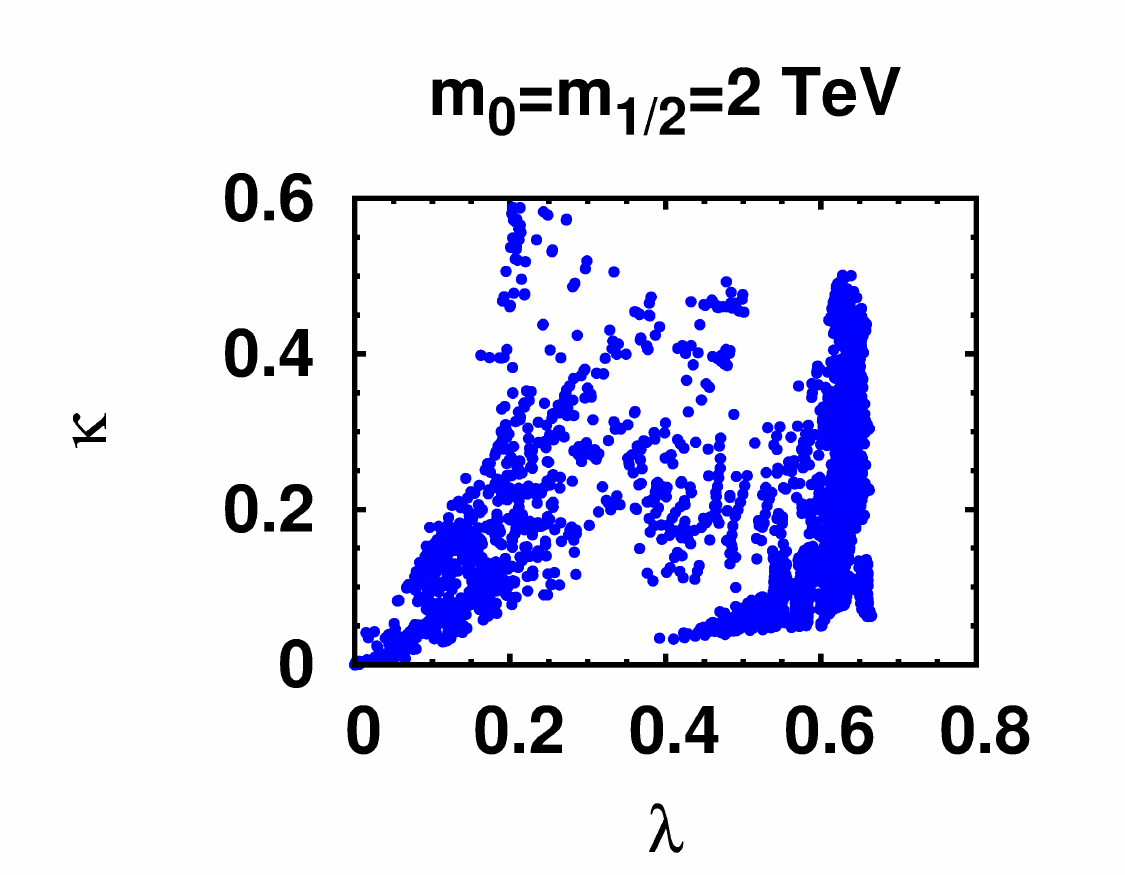}
\caption[]{
Allowed range for $\lambda-\kappa$ for $m_0=m_{1/2}=1,1.5$ and 2 TeV, respectively (left to right).
}
\label{f5}
\end{center}
\end{figure}

The  allowed range of $\lambda-\kappa$  depends on the choice of $m_0=m_{1/2}=1$, as demonstrated in FIg. \ref{f5}. One can identify two  preferred regions, which can be understood if one considers the approximate expression \cite{Ellwanger:2009dp} for a 125 GeV Higgs: 
\beq\label{eq4}
M_{H}^2\approx M_Z^2\cos^2 2\beta+ \Delta_{\tilde{t}} + \lambda^2 v^2 \sin^2 2\beta - \frac{\lambda^2}{\kappa^2}(\lambda-\kappa \sin 2 \beta)^2.  
\eeq 
The first tree level term can become at most $M_Z^2$ for large $\tan\beta$. The diffe\-rence between $M_Z$ and 125 GeV has to originate mainly from the lo\-garithmic stop mass corrections $\Delta_{\tilde{t}}$. The two remaining terms originate from the mixing between the 125 GeV Higgs boson and  the singlet of the NMSSM, which becomes large for large values of the couplings $\lambda$ and $\kappa$ and small $\tan\beta$. Note that these two terms contribute at tree level, so there is no logarithmic dependence, as is the case for the stop loop contribution. The allowed  region with large values of the couplings $\lambda$ and $\kappa$ is called \textit{Region I} in the following.  However, there exists a second solution to Eq. \ref{eq4} with small values of $\lambda,\kappa$ and larger values of $\tan\beta$, which can be obtained by a trade-off between the first two terms and the last two terms, which we call \textit{Region II}. Region II with its small couplings $\lambda$ and $\kappa$ is in some sense closer to the MSSM, although the singlet-like Higgs and its corresponding singlino-like LSP yield additional physics, like the possibility of double Higgs production and an LSP hardly coupling to matter.
If the values of $m_0$ and $m_{1/2}$ are enhanced the SUSY contribution to the tree level Higgs mass becomes larger, so  the contribution from the mixing with the singlet can be smaller. Then there is more freedom for the values of the NMSSM parameters and the regions I and II start to merge, as can be observed from the right panel in  Fig. \ref{f5}.  For larger values of $m_0,m_{1/2}$ the deviations of the signal values from the SM expectation start to decrease as well, as shown in Appendix \ref{signalstrengths}.

Intermediate values for $\lambda$ and $\kappa$ in Fig. \ref{f5} are disfavored by the fit because of  higher $\chi^2$ values, mainly from the signal strengths. This is discussed in more detail in  Appendix  \ref{lambda-kappa-scan}.

\section{Summary}
In this paper a new technique to sample the NMSSM parameter space is introduced, which allows an efficient sampling with complete coverage. This is obtained by sampling the 3D Higgs mass space instead of  the large 7D NMSSM parameter space. 
The reduction in the dimensionality is possible by assuming that the heavy Higgs masses $m_{H_3}$, $m_{A_2}$ and $m_{H^\pm}$ are approximately equal, which is true in the decoupling limit. Furthermore, one of the lighter Higgs bosons has to correspond to the observed 125 GeV Higgs boson. So instead of the 6 Higgs boson masses one has only 3 independent Higgs masses. This 3D mass space can be subdivided  in cells in the $m_{H_3}$, $m_{H_1}$, $m_{A_1}$ mass space, as shown in Fig. \ref{f2} on the left hand side. For each cell the corresponding NMSSM parameters can be determined by a Minuit fit to the Higgs masses with the NMSSM parameters as free parameters, a procedure which one would follow if all Higgs masses would have been measured. Here we assumed that the second lightest boson corresponds to the observed boson, i.e. $m_{H_2}=125$ GeV.  In case  $m_{H_1}=125$ GeV the 3D mass space would be spanned by $m_{H_3}$, $m_{H_2}$ and $m_{A_1}$, an option which was investigated as well.  The reduced 3D dimensionality of the Higgs mass space  allows to perform the fit for each cell, so one does not need a random random sampling of the parameter space. 
As discussed in the introduction, a random scan of highly correlated parameters (see  Fig. \ref{f4} for part of the correlations) will need a correlation matrix to efficiently scan the parameter space and guarantee complete coverage.
For example, one can observe from Fig. \ref{f4} immediately, that if $\mu_{eff}$ is scanned up to 1 TeV and $\tan\beta$ up to 60 that  many points for random values of $\mu_{eff}$ and $\tan\beta$ are disfavored by  higher $\chi^2$ values.  In the $\lambda,\kappa$ plane only two regions are preferred: Region I(II) at large(small) values of $\lambda,\kappa$, as shown in Fig. \ref{f5}. Why the $\chi^2$ values increase in the intermediate regions has been discussed in Appendix \ref{lambda-kappa-scan}.
The size of the regions depends on the values of the SUSY parameters, as is obvious from a comparison of the panels in Fig. \ref{f5}. This can be understood as follows:
the 125 GeV Higgs mass has at tree level contributions from the mixing with the singlet Higgs boson (largely determined by the Higgs couplings $\kappa,\lambda$) and the stop loops (largely determined by the stop mass). For light stops the corrections from stop loops are small and the values of $\lambda,\kappa$ have to be precisely tuned to reach the 125 GeV mass. 
For larger values of the stop masses one has more freedom in these NMSSM parameters and the allowed regions grow and start to overlap.

In summary, the novel scanning technique allows to study the NMSSM parameter space in a non-random way, which guarantees complete coverage and reveals many interesting features, like the high correlations of the NMSSM parameters and possible deviations of the signal strengths from the SM expectation for processes with SUSY particles contributing in the loops.

\clearpage
\providecommand{\href}[2]{#2}\begingroup\raggedright\endgroup

\newpage

\appendix

\section{Impact of the common SUSY masses}
\label{SUSYmasses}

\begin{figure}
\begin{center}
\hspace{-1cm}
\includegraphics[width=0.5\textwidth]{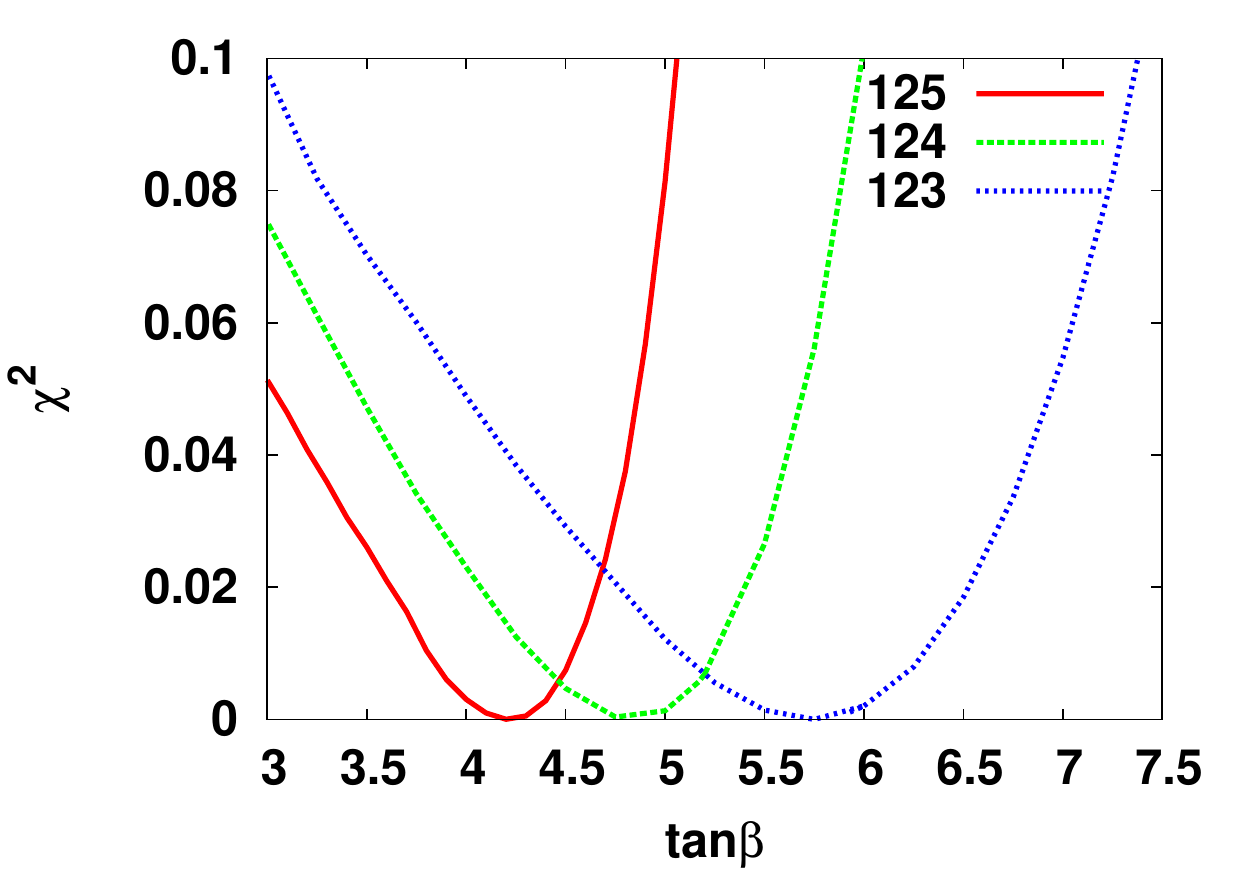}
\caption[]{ $\chi^2$ distribution for $m_{H1}=90$ GeV,$m_{H3}=1000$ GeV,$m_{A1}=200$ GeV and $m_0=m_{1/2}=$ 0.7 TeV. If hypothetically smaller values of the Higgs mass are considered, the $\chi^2$ distribution shifts and gets broader because the NMSSM parameters need not to be so finely tuned anymore to reach the 125 GeV. 
}
\label{f6}
\end{center}
\end{figure}

The broad $\chi^2$ distribution shown in Fig. \ref{f3} results from the SM Higgs mass constraint of 125 GeV.
A smaller Higgs mass requires smaller radiative stop corrections leading to more freedom in the choice of the NMSSM parameters and hence to a broader $\chi^2$ distribution for the same common SUSY masses $m_0,m_{1/2}$. 
The same effect is observed if smaller values of the SM Higgs mass are considered for small common SUSY masses. In this case a larger region of the NMSSM parameter space is compatible with the Higgs boson mass leading to a broader $\chi^2$ distribution. This is shown in Fig. \ref{f6} for 123-125 GeV. The Higgs mass combination was chosen to be: $m_{H1}=90$ GeV, $m_{H3}=1000$ GeV, $m_{A1}=200$ GeV, while $m_0=m_{1/2}=$ 0.7 TeV.  

\section{Correlation matrix example}
\label{matrix}

Example of a Minuit output of the correlation coefficients for the global minimum for the Higgs mass combination  $m_{H1}=90$ GeV, $m_{H3}=1000$ GeV, $m_{A1}=200$ GeV and $m_0=m_{1/2}=$ 1 TeV. The highest correlation of 0.755 is observed for the parameters $\kappa$ and $\mu_{eff}$, but this is still far from degeneracy, so the fit finds well defined minima for all parameters, as demonstrated in Fig. \ref{f3}. Note that these are the correlations between the parameter values of a single fit leading to single values of the parameters. If summed over all fits the fitted parameters vary in a correlated way, as shown  exemplary in Fig. \ref{f4}.\vspace*{5mm}

$ \bordermatrix{~ & \tan\beta & A_0 & A_\kappa & A_\lambda & \lambda & \kappa & \mu_{eff} \cr
                   \tan\beta & 1.000 & -0.350 & 0.197 & -0.182 & -0.396 & 0.528 & -0.685 \cr
                   A_0 & -0.350 & 1.000 & -0.157 & 0.680 & -0.312 & -0.186 & 0.135 \cr
                   A_\kappa & 0.197 & -0.157 & 1.000 & 0.345 & 0.218 & 0.306 & -0.225 \cr
                   A_\lambda & -0.182 & 0.680 & 0.345 & 1.000 & -0.218 & -0.374 & 0.273 \cr
                   \lambda & -0.396 & -0.312 & 0.218 & -0.218 & 1.000 & -0.406 & 0.385 \cr
                   \kappa & 0.528 & -0.186 & 0.306 & -0.374 & -0.406 & 1.000 & -0.755 \cr
                  \mu_{eff} & -0.685 & 0.135 & -0.225 & 0.273 & 0.385 & -0.755 & 1.000 \cr}$

\section{Running of the trilinear couplings}
\label{running}

\begin{figure}
\begin{center}
\includegraphics[width=0.3\textwidth]{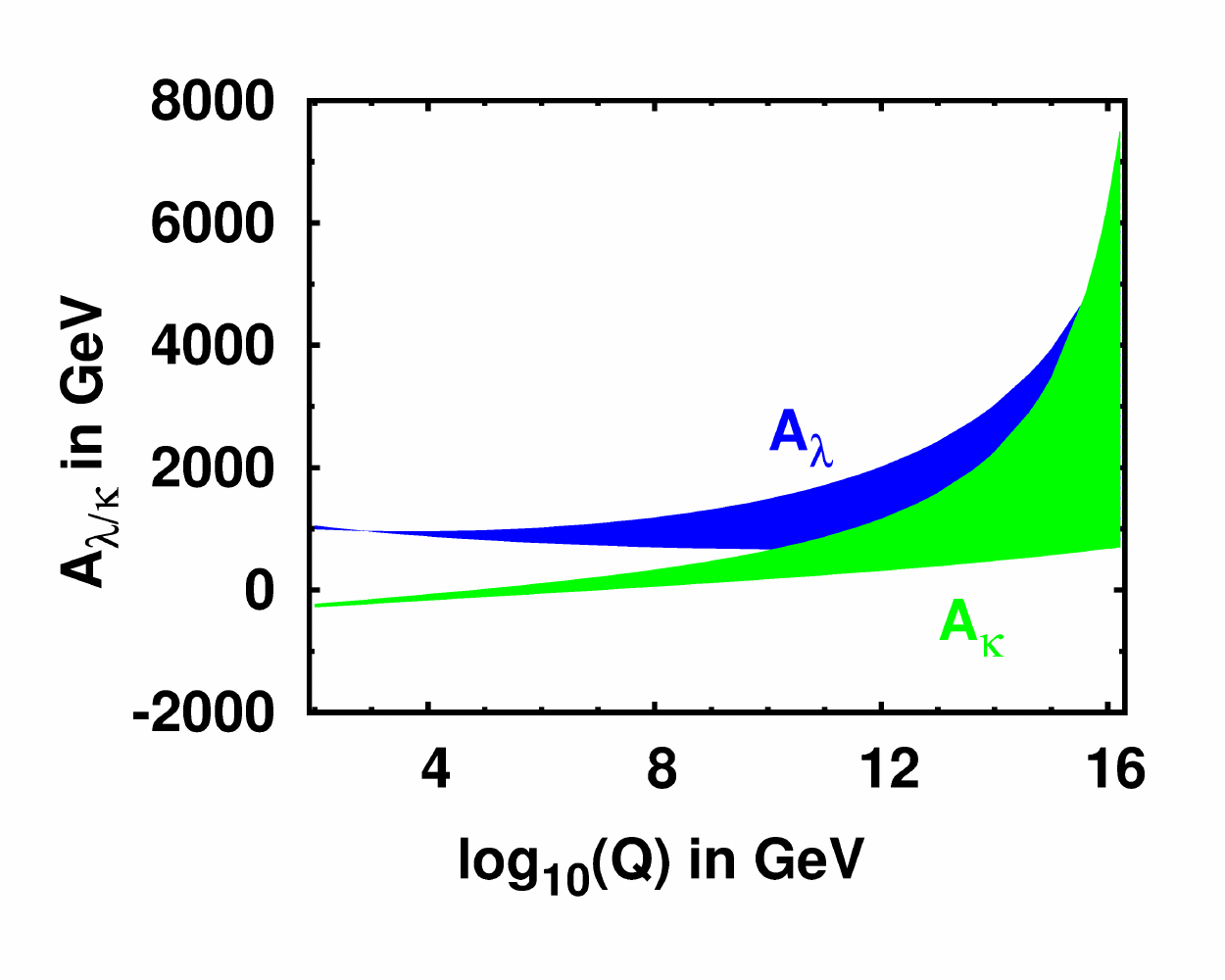}
\includegraphics[width=0.3\textwidth]{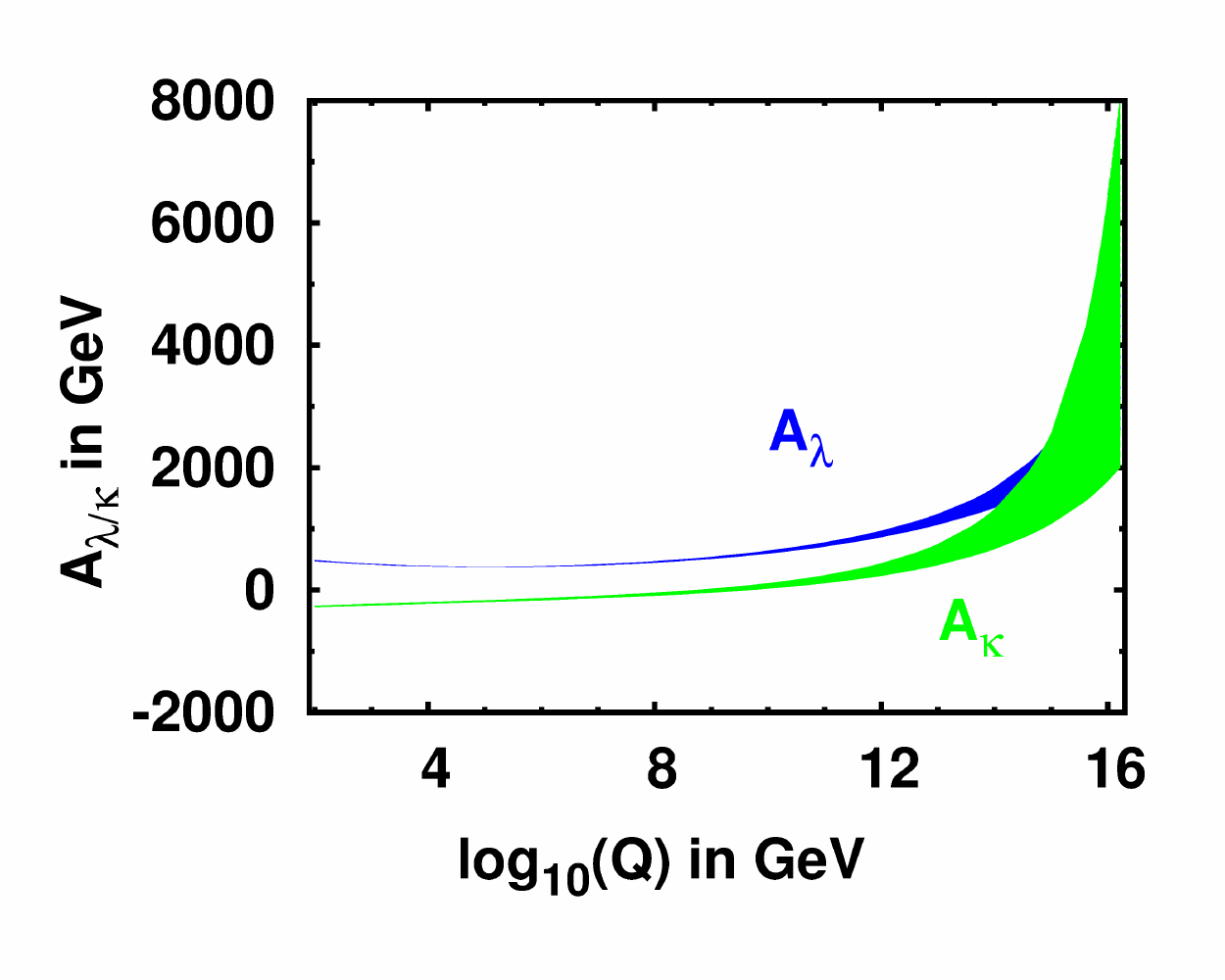}
\includegraphics[width=0.3\textwidth]{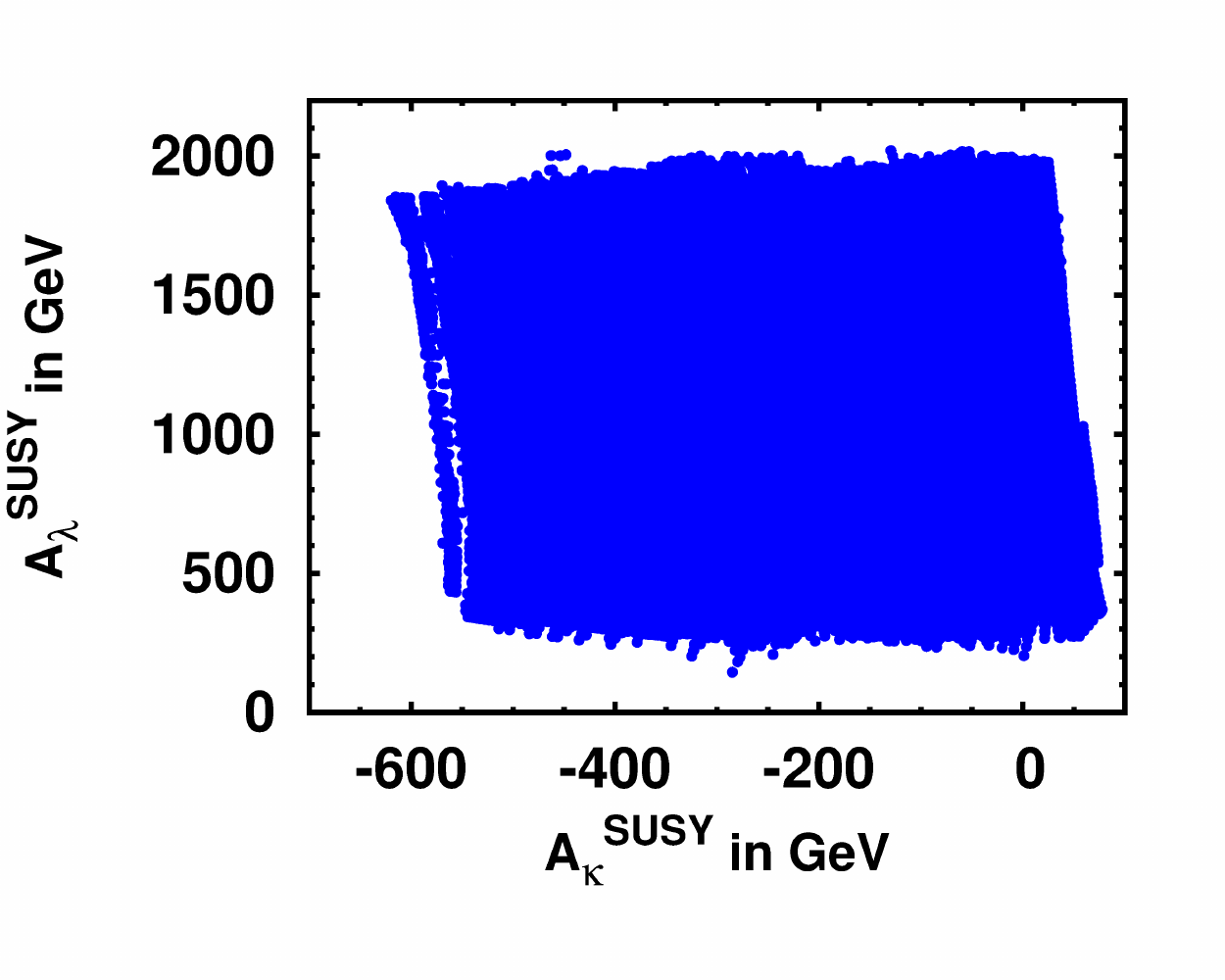}
\caption[]{ The left two panels show the running of the trilinear couplings $A_{\lambda}$ (blue region) and $A_{\kappa}$ (green region) from the GUT to the SUSY scale for two different Higgs mass combinatons.  One observes a fixed-point solution, i.e. the low energy values are largely independent of the choice of the GUT scale value. By a comparison of the fixed-point solutions in the left and middle panel one observes that the fixed-point solutions are different, i.e. they depend on the chosen mass combination.
If summed over all mass combinations the spread in the SUSY scale values (the fixed-point solutions) is still not so small, as shown in the panel on the right.
}
\label{f7}
\end{center}
\end{figure}

The GUT scale values of the trilinear couplings $A_{\lambda}$ and $A_{\kappa}$ have a large allowed range, as shown in Fig. \ref{f4} in the text.  However, the values at the SUSY scale have only small variations, corresponding to the so-called fixed-point solutions from the RGEs, as shown  for two mass combinations on the  two left panels of Fig. \ref{f7}. From a comparison of the two left panels one observes that the fixed-point is different for different Higgs mass combinations, so the SUSY scale values still show  variation, if summed over all mass combinations,  as shown on the right panel of Fig. \ref{f7}.

\begin{figure}[]
\vspace{-0.7cm}
\begin{center}
\begin{minipage}{\textwidth}
\begin{tabular}{cccc}
& \small{\boldmath$m_0=m_{1/2}=1.0$ \textbf{TeV}} &\small{\boldmath$m_0=m_{1/2}=1.5$ \textbf{TeV}} & \small{\boldmath$m_0=m_{1/2}=2.0$ \textbf{TeV}} \\
&  \multirow{7}{*}{\includegraphics[width=0.3\textwidth]{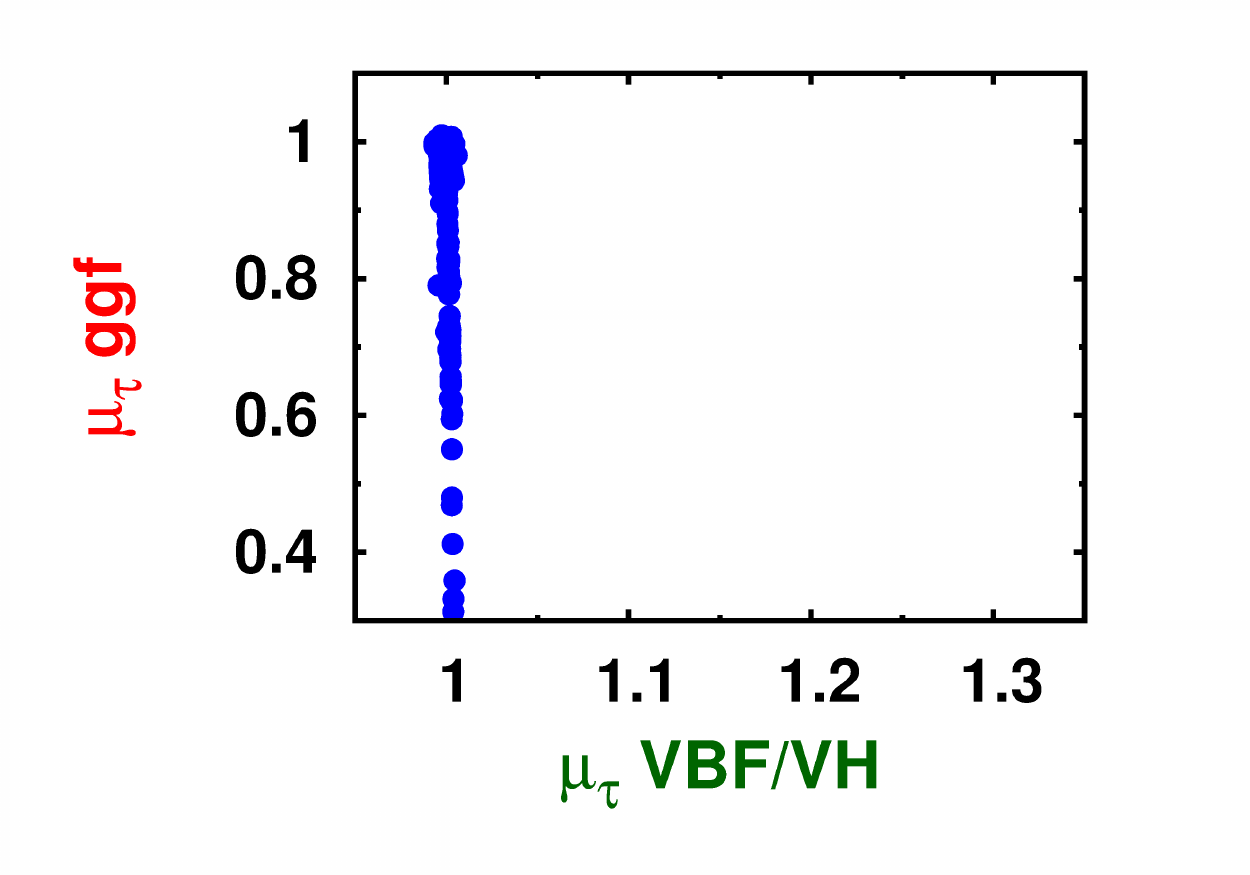}} 
& \multirow{7}{*}{\includegraphics[width=0.3\textwidth]{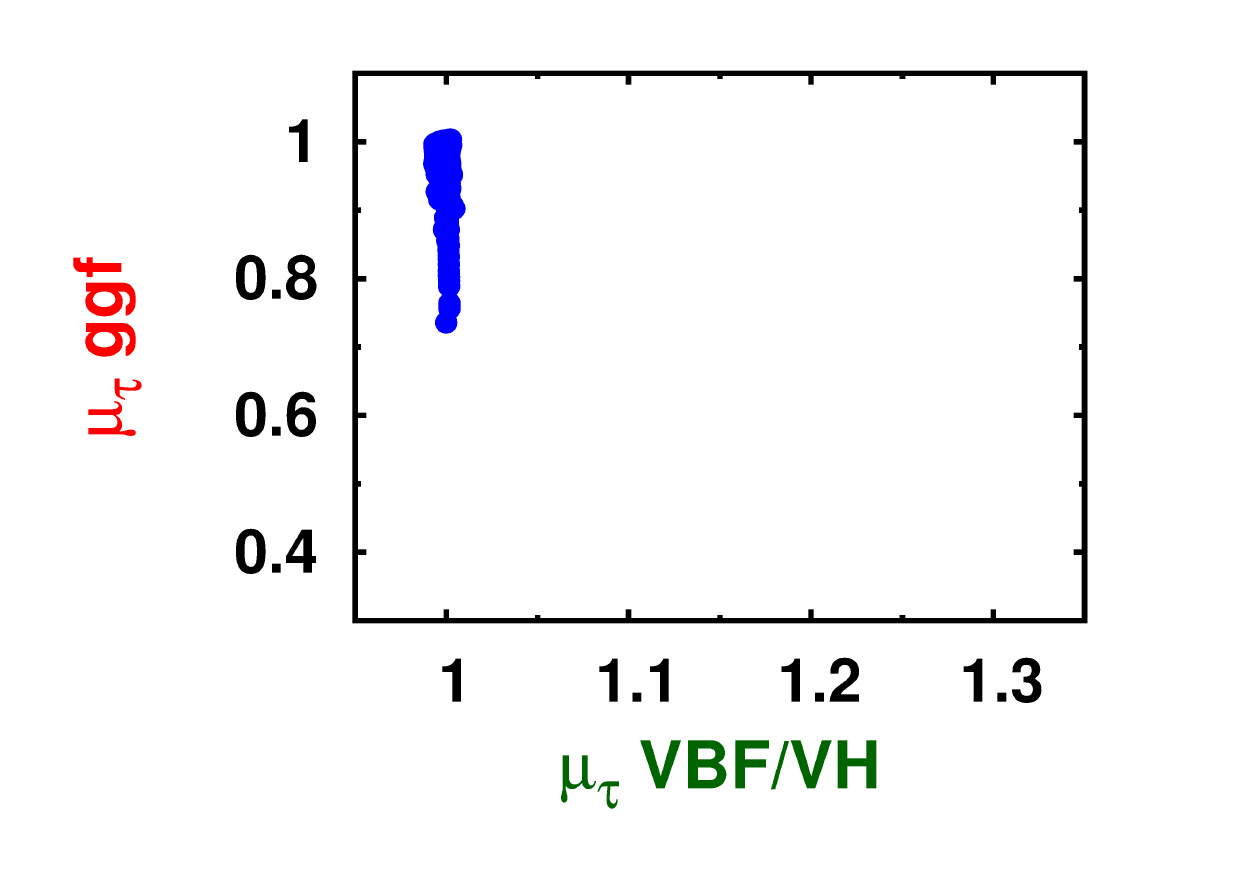}} 
& \multirow{7}{*}{\includegraphics[width=0.3\textwidth]{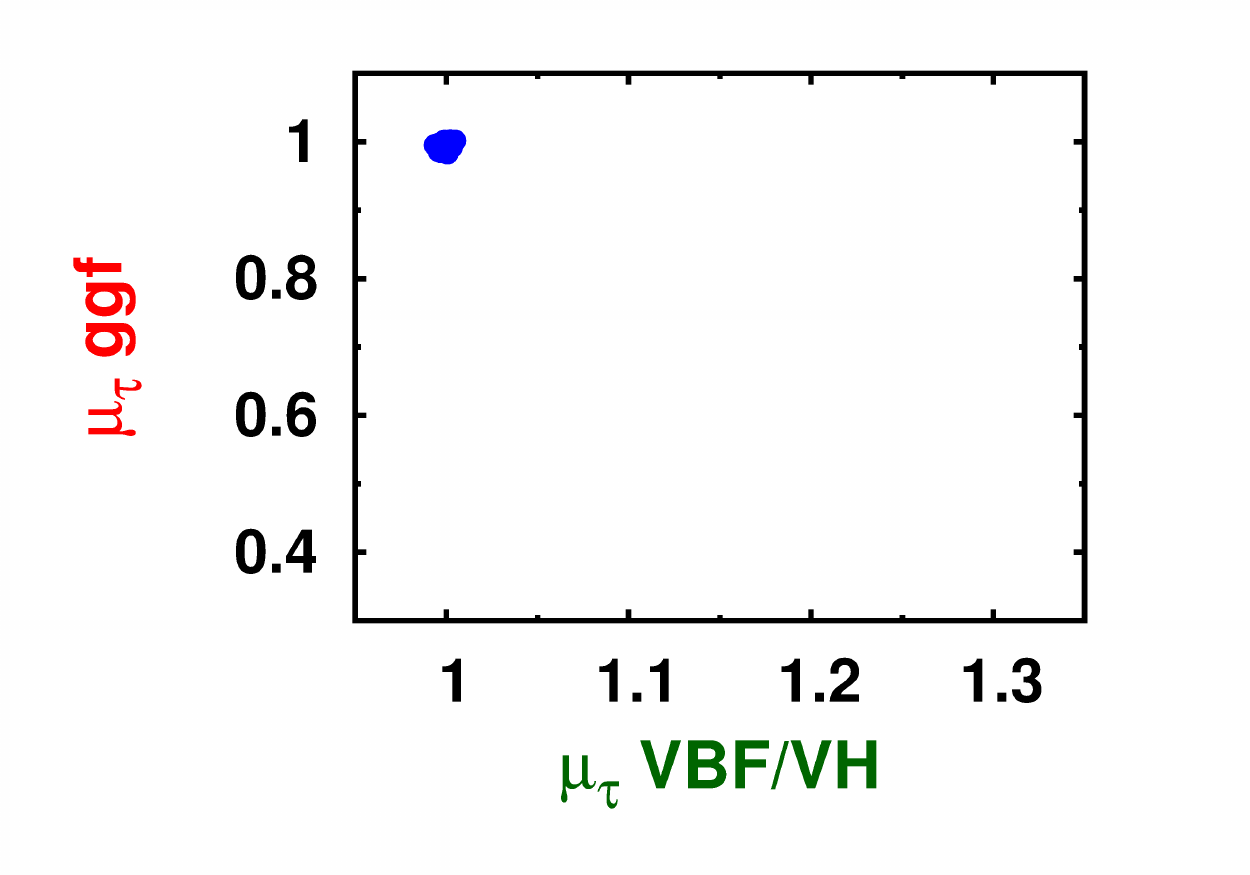}} \\[10mm]
 & &  & \\
 \small{\boldmath$\mu_{\tau\tau}$} & & & \\
 & &  & \\
&  &  & \\
 & &  & \\
& &  & \\
&   \multirow{7}{*}{\includegraphics[width=0.3\textwidth]{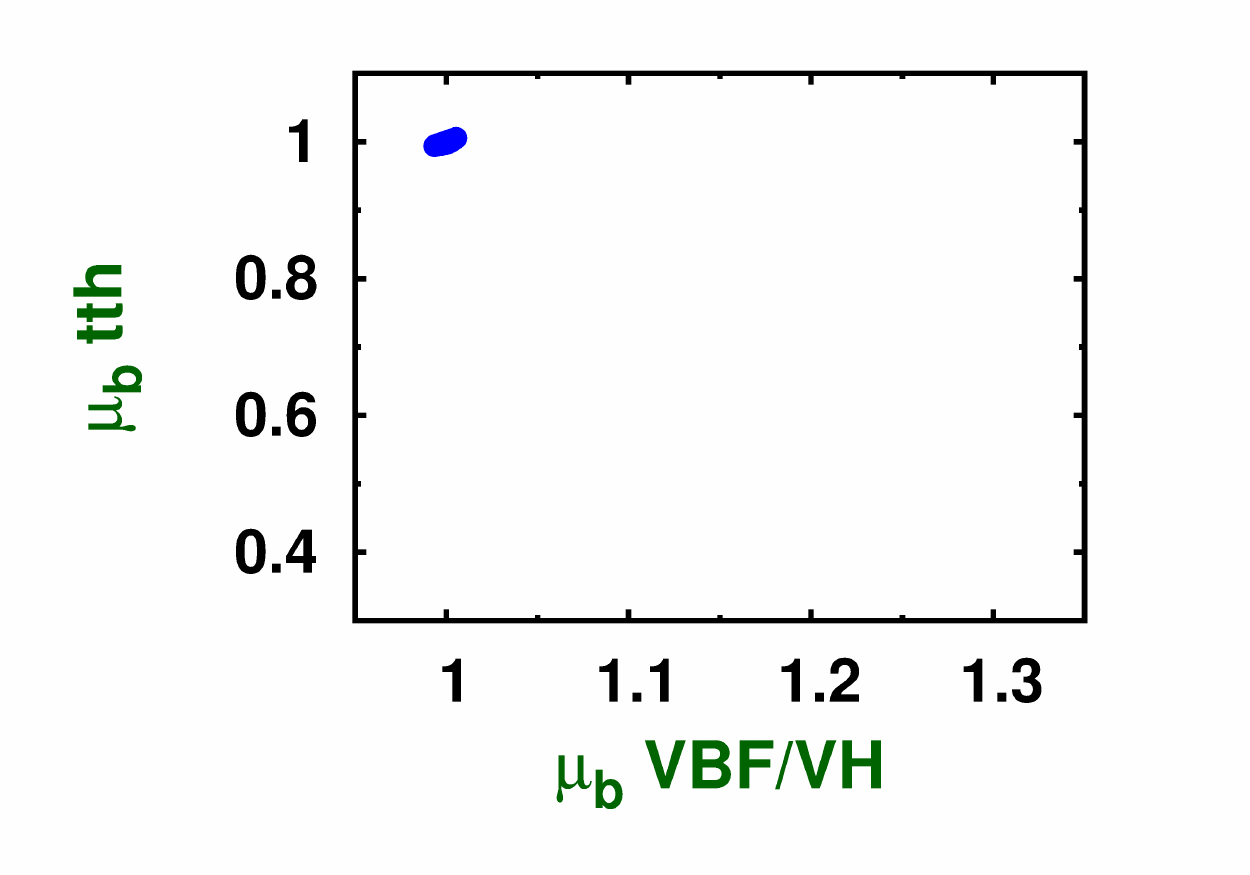}} 
& \multirow{7}{*}{\includegraphics[width=0.3\textwidth]{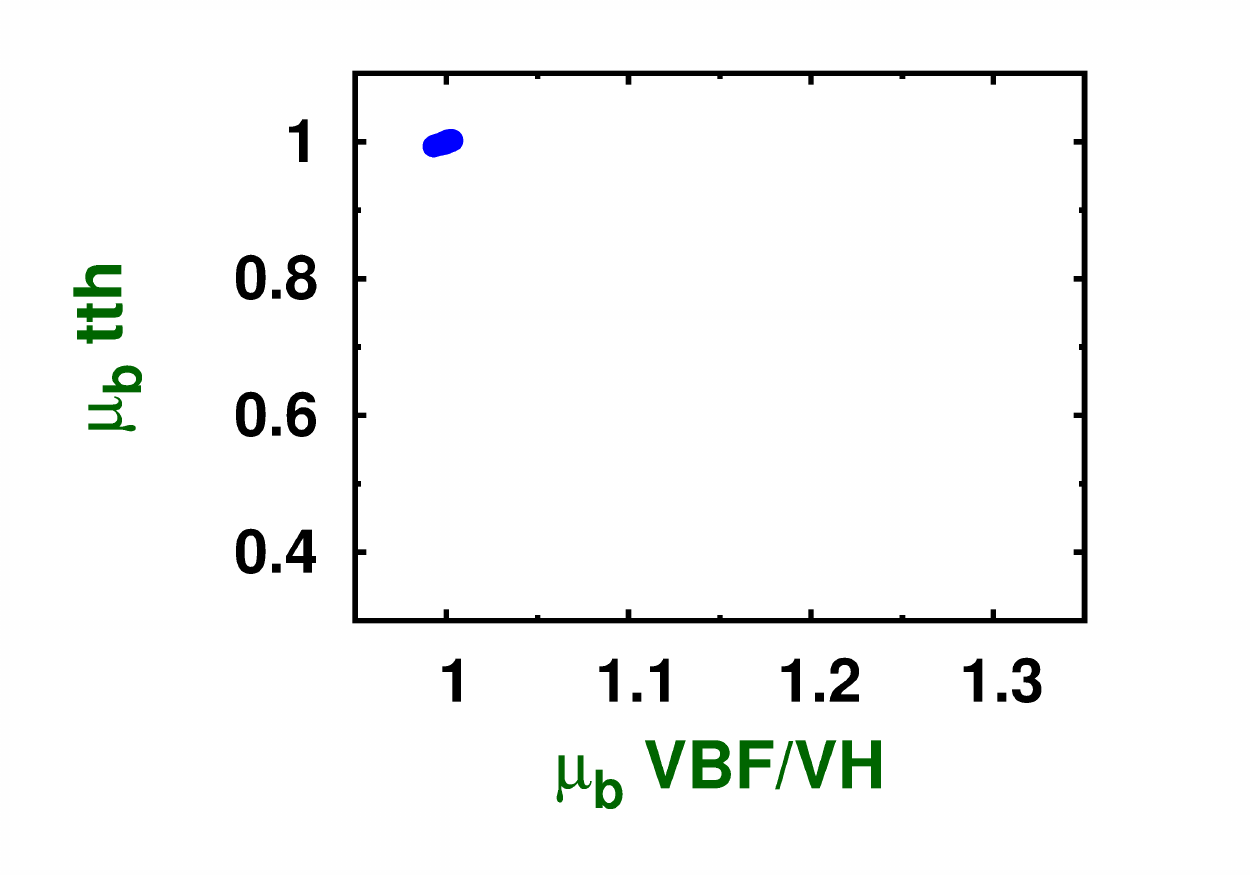}} 
&  \multirow{7}{*}{\includegraphics[width=0.3\textwidth]{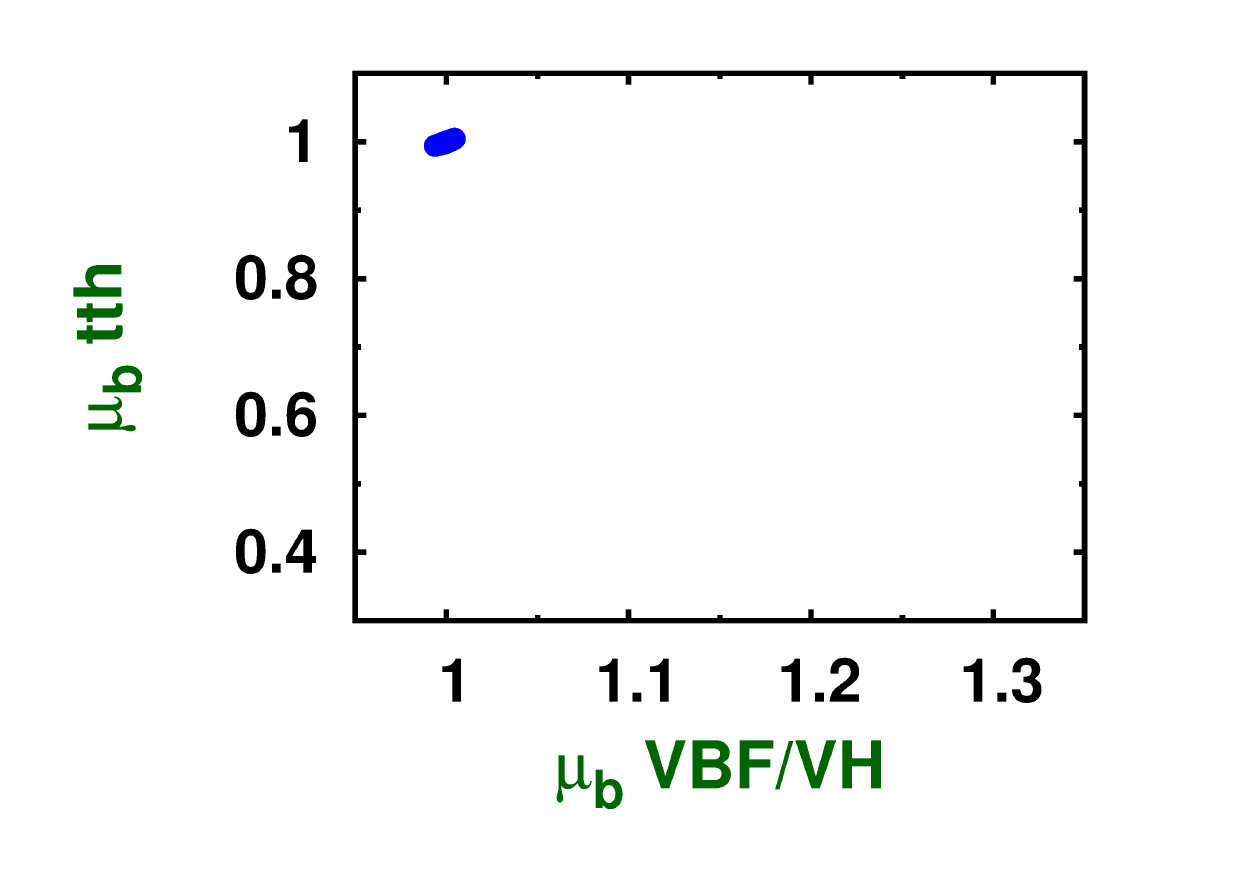}} \\[10mm]
 & &  & \\
 \small{\boldmath$\mu_{bb}$} & & & \\
 & &  & \\
 & &  & \\
& &  & \\
&  &  & \\
&   \multirow{7}{*}{\includegraphics[width=0.3\textwidth]{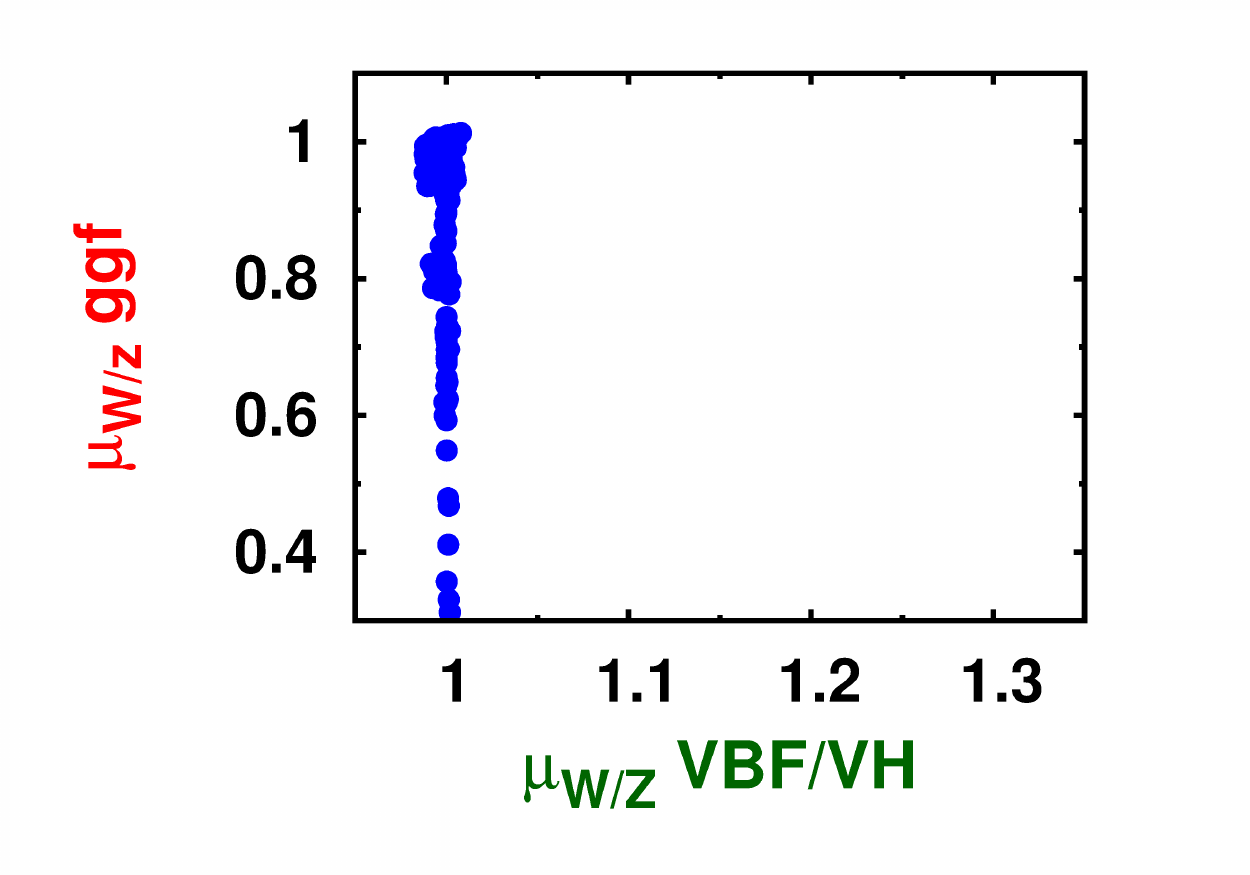}} 
& \multirow{7}{*}{\includegraphics[width=0.3\textwidth]{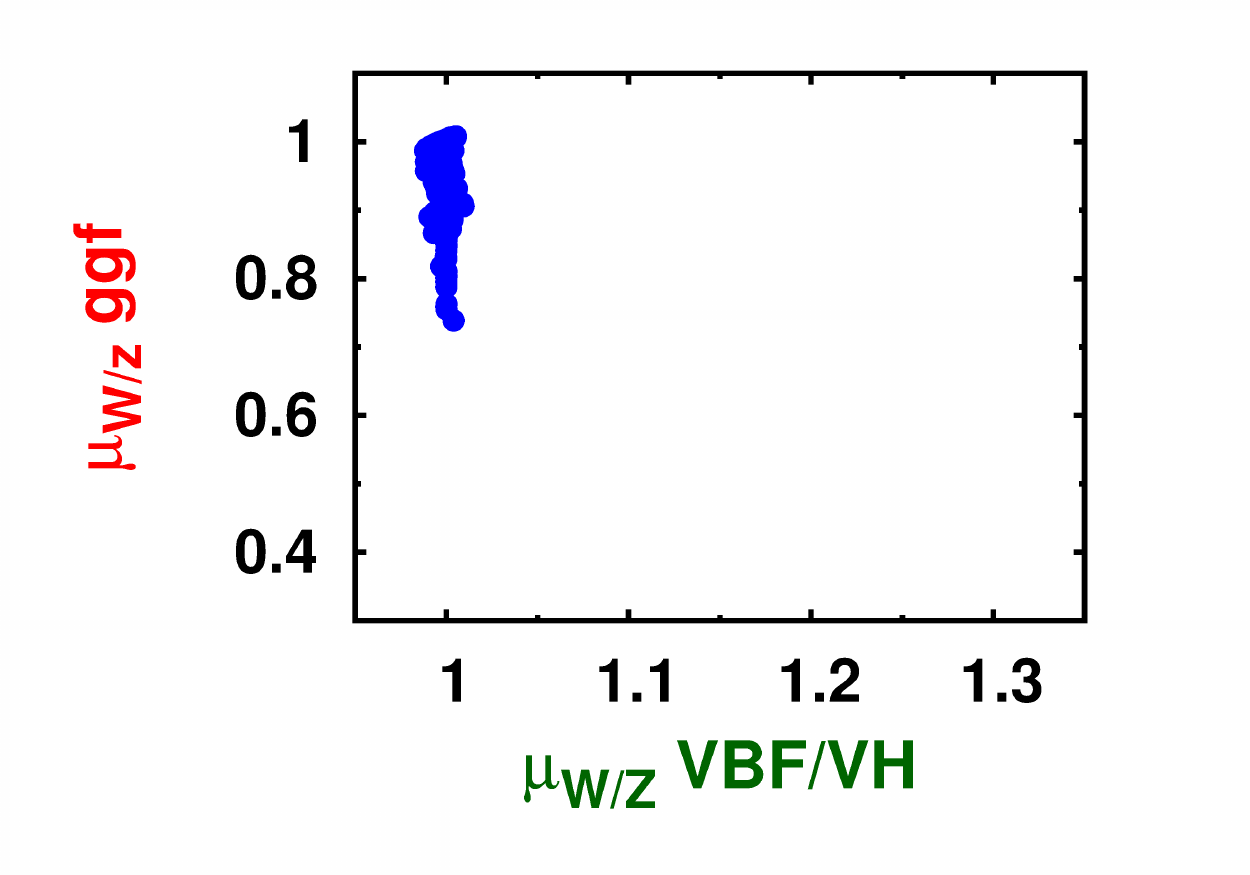}} 
&  \multirow{7}{*}{\includegraphics[width=0.3\textwidth]{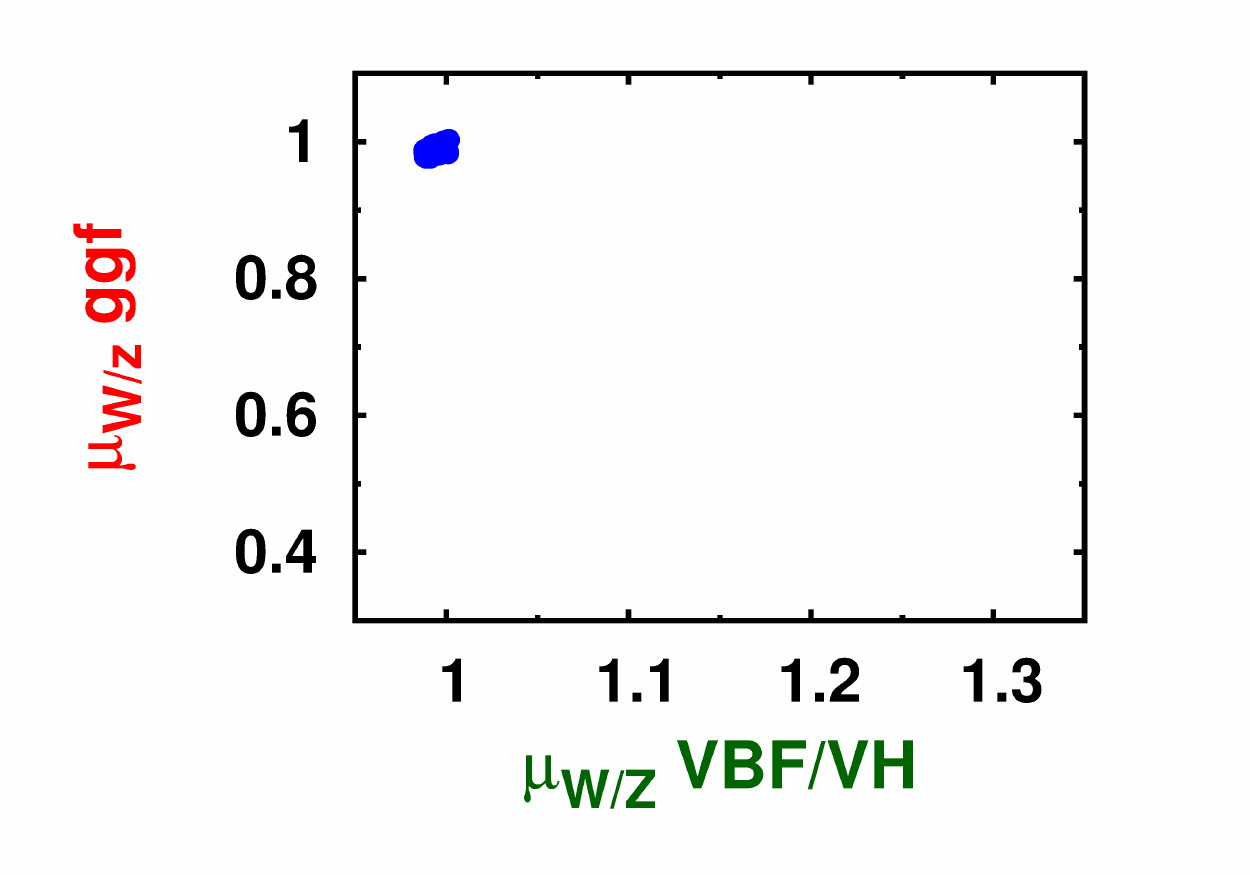}} \\[10mm]
 & &  & \\
 \small{\boldmath$\mu_{W/Z}$} & & & \\
 & &  & \\
 & &  & \\
& &  & \\
&  &  & \\
& \multirow{7}{*}{\includegraphics[width=0.3\textwidth]{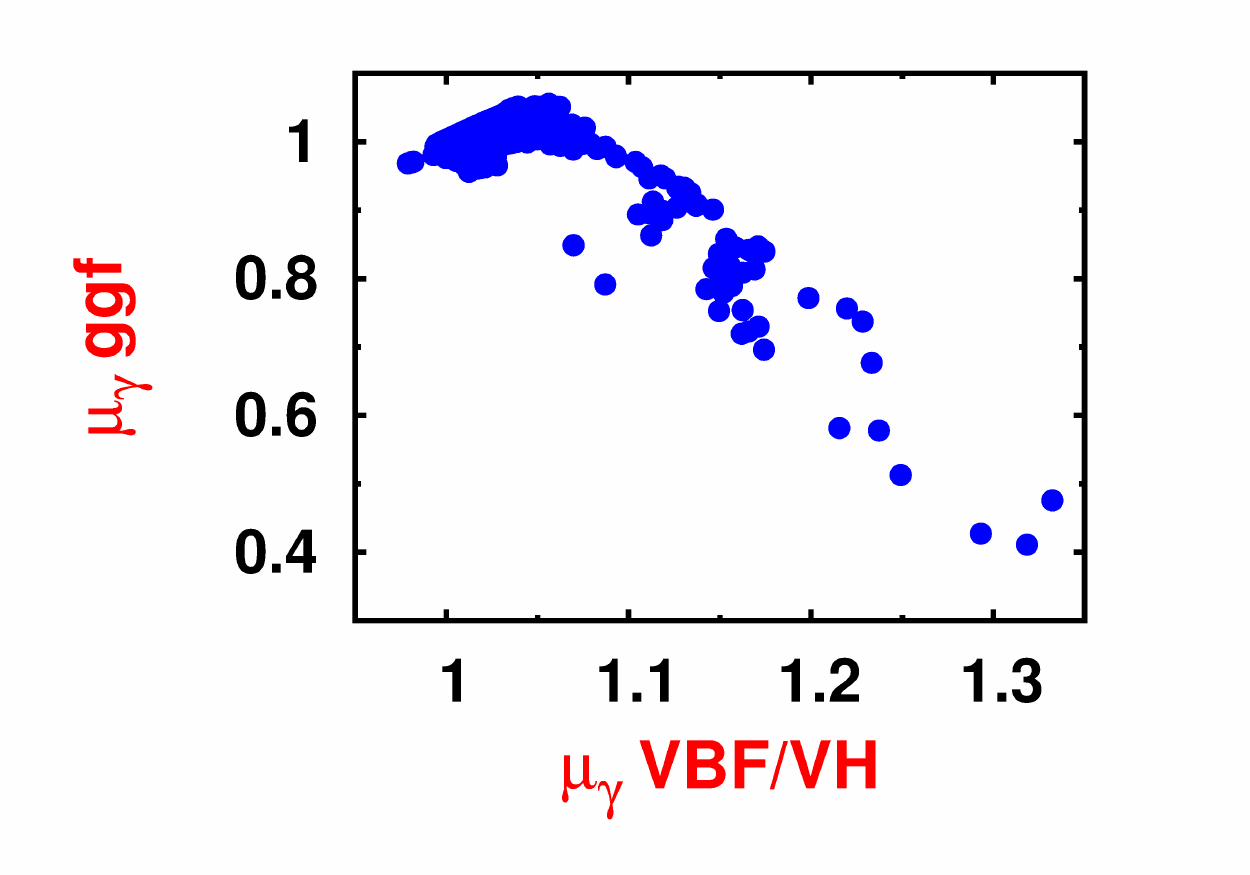}} 
& \multirow{7}{*}{\includegraphics[width=0.3\textwidth]{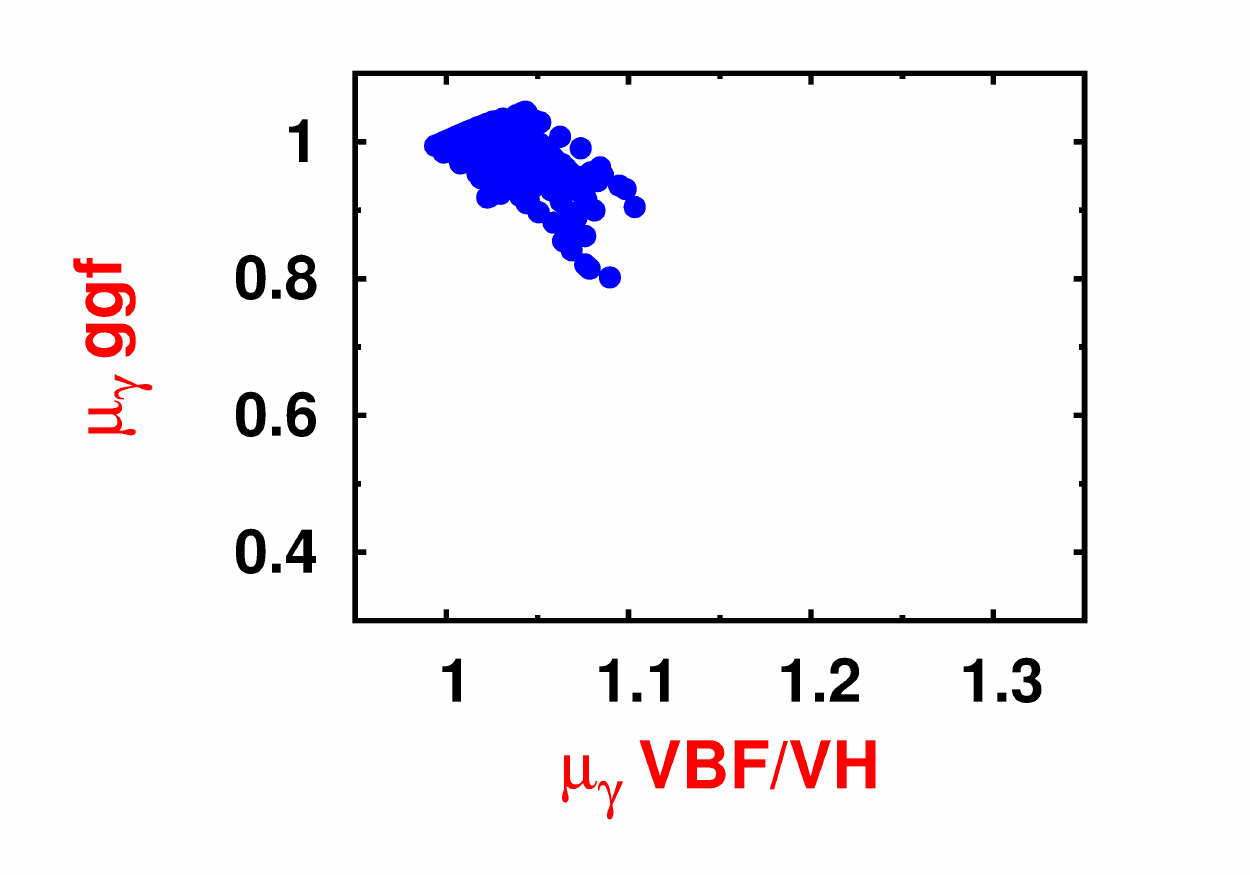}} 
&\multirow{7}{*}{\includegraphics[width=0.3\textwidth]{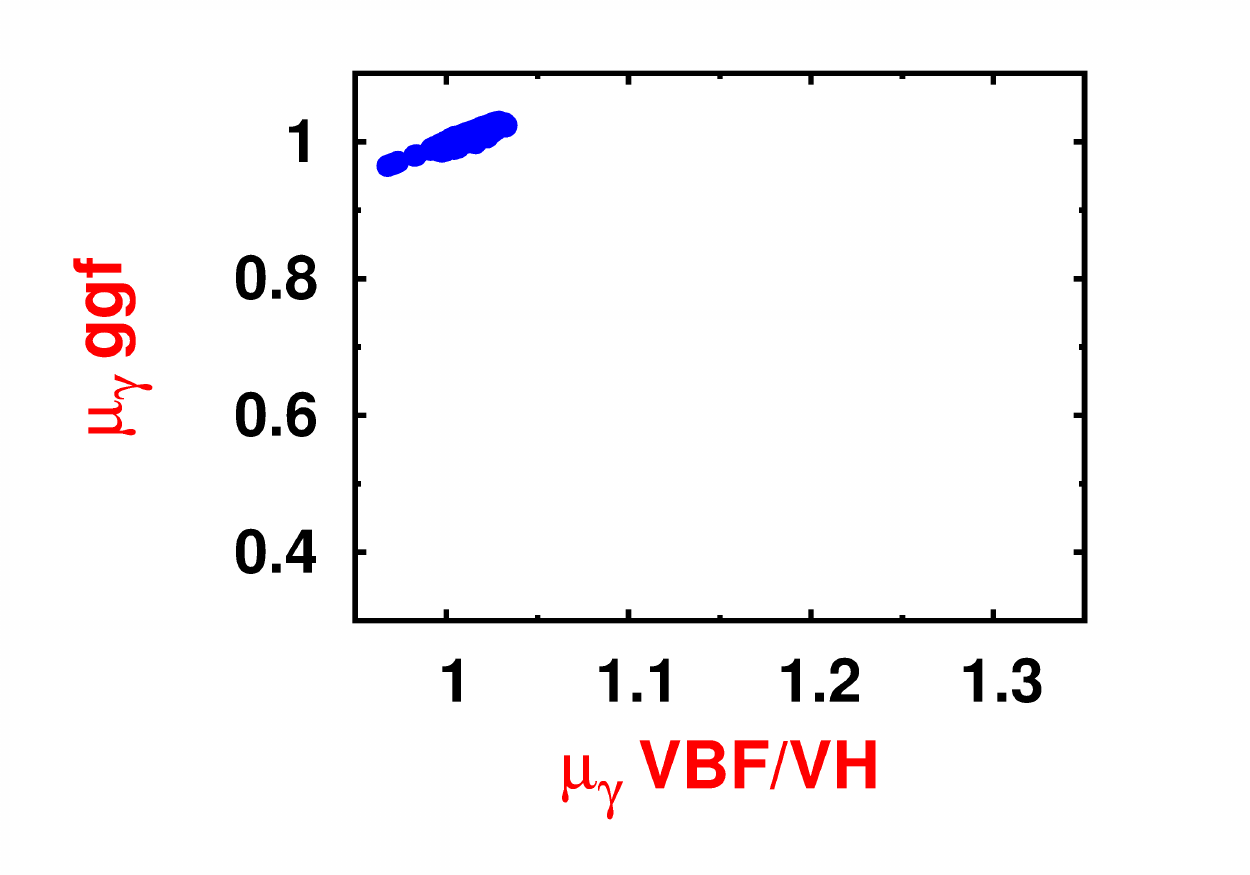}}\\[10mm]
 & &  & \\
 \small{\boldmath$\mu_{\gamma\gamma}$} & & & \\
 & &  & \\
 & &  & \\
& &  & \\
&  &  & \\
\end{tabular}
\end{minipage}
\caption[]{
Allowed range for all eight signal strengths from Eq. \ref{coupling5} for $m_0=m_{1/2}=1, 1.5, 2$ TeV, respectively (from left to right).
The axes labeled in green correspond to signal strengths without loops   and are always close to the SM-expectation of 1, while the axes labeled in red correspond to signal strengths with loops, which can deviate from 1, especially for low stop mass (left column) implying large SUSY contributions.
}
\label{f8}
\end{center}
\end{figure}
\section{Signal strengths versus SUSY masses}
\label{signalstrengths}
The results for the signal strengths from the Higgs mass sampling is shown in Fig. \ref{f8} by the  blue dots  for $m_0,m_{1/2}$ increasing from 1 to 2 TeV from left to right. The no-loop(loop) signal strengths are indicated by the green(red) colors on the axes. The fitted no-loop signal strengths are close to 1, but the signal strengths including loops differ from 1. One observes deviations of $\mu_{loop}$ up to 70\%, e.g. for $\mu_{\tau}^{ggf}$ for $m_0=m_{1/2}=1$ TeV in the left panel on the top. The tails of the distributed points represent points in the parameter space with low stop masses, which affect the loop signal strengths strongest. Larger common SUSY masses  $m_0,m_{1/2}$ for the panels in the two right columns lead to heavier stops (see Fig. \ref{f3}) and hence smaller  SUSY  contributions in the loops leading to smaller deviations of the signal strengths.

   \begin{figure}[]
\begin{center}
       \includegraphics[width=0.47\textwidth]{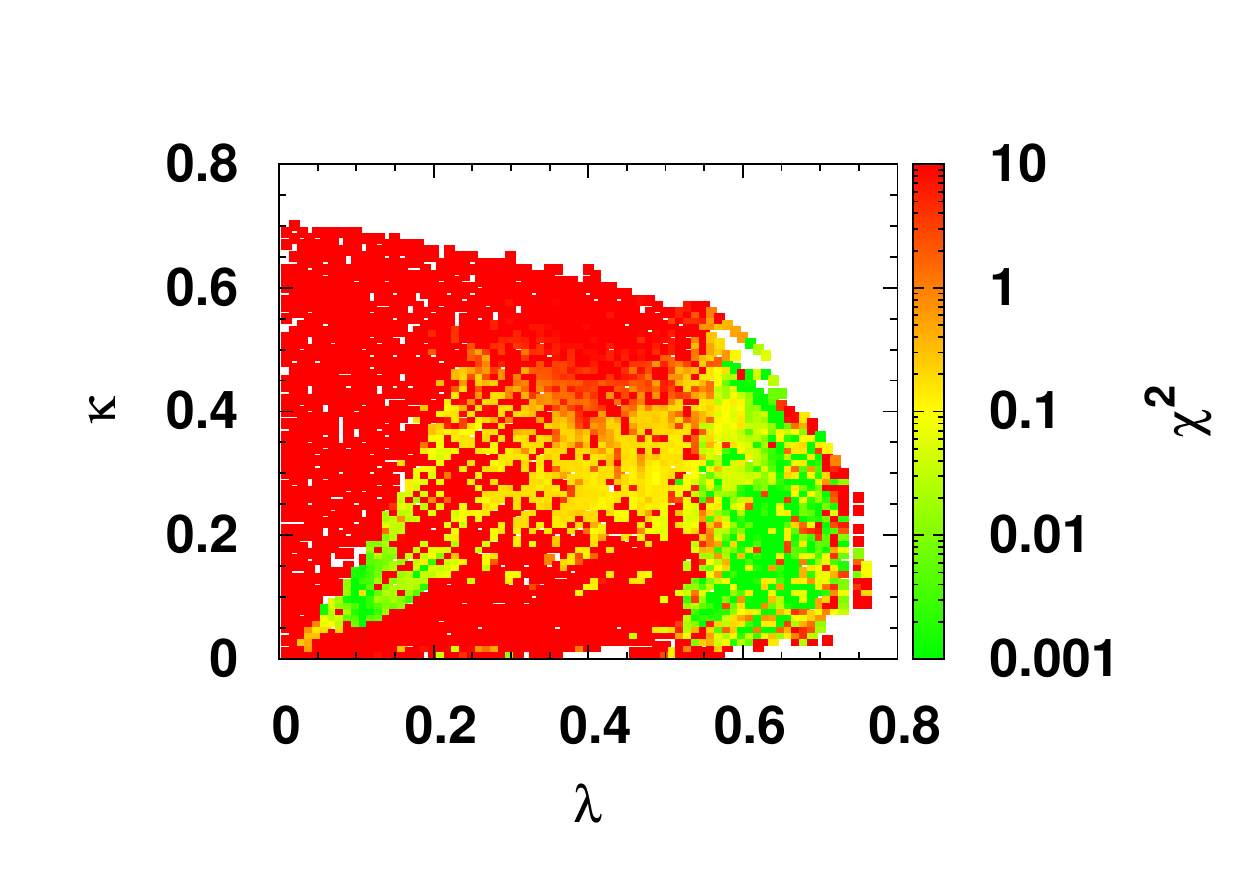}
       \includegraphics[width=0.47\textwidth]{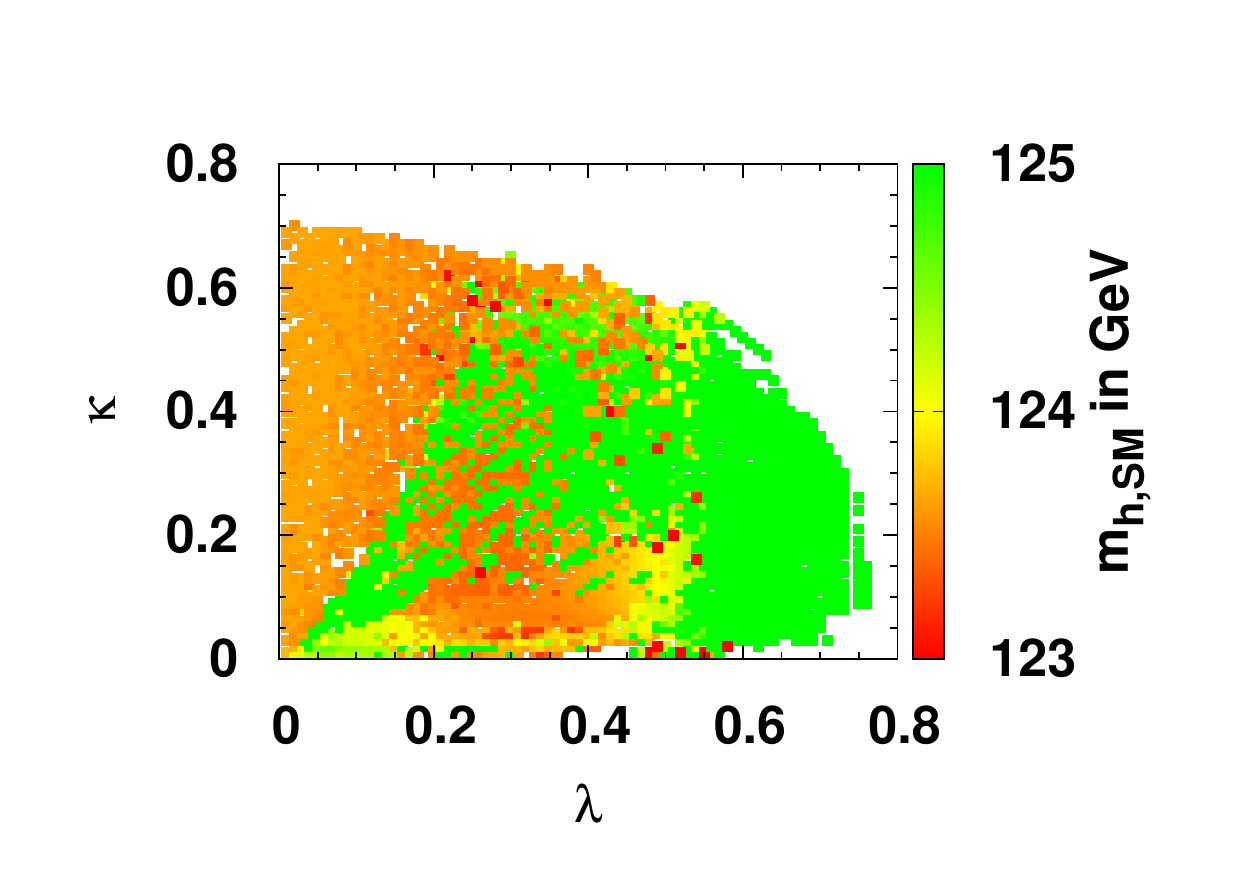}\vspace*{-3mm}
       \hspace*{0.02\textwidth}(a)\hspace*{0.46\textwidth} (b)\\[-4mm]
       \includegraphics[width=0.47\textwidth]{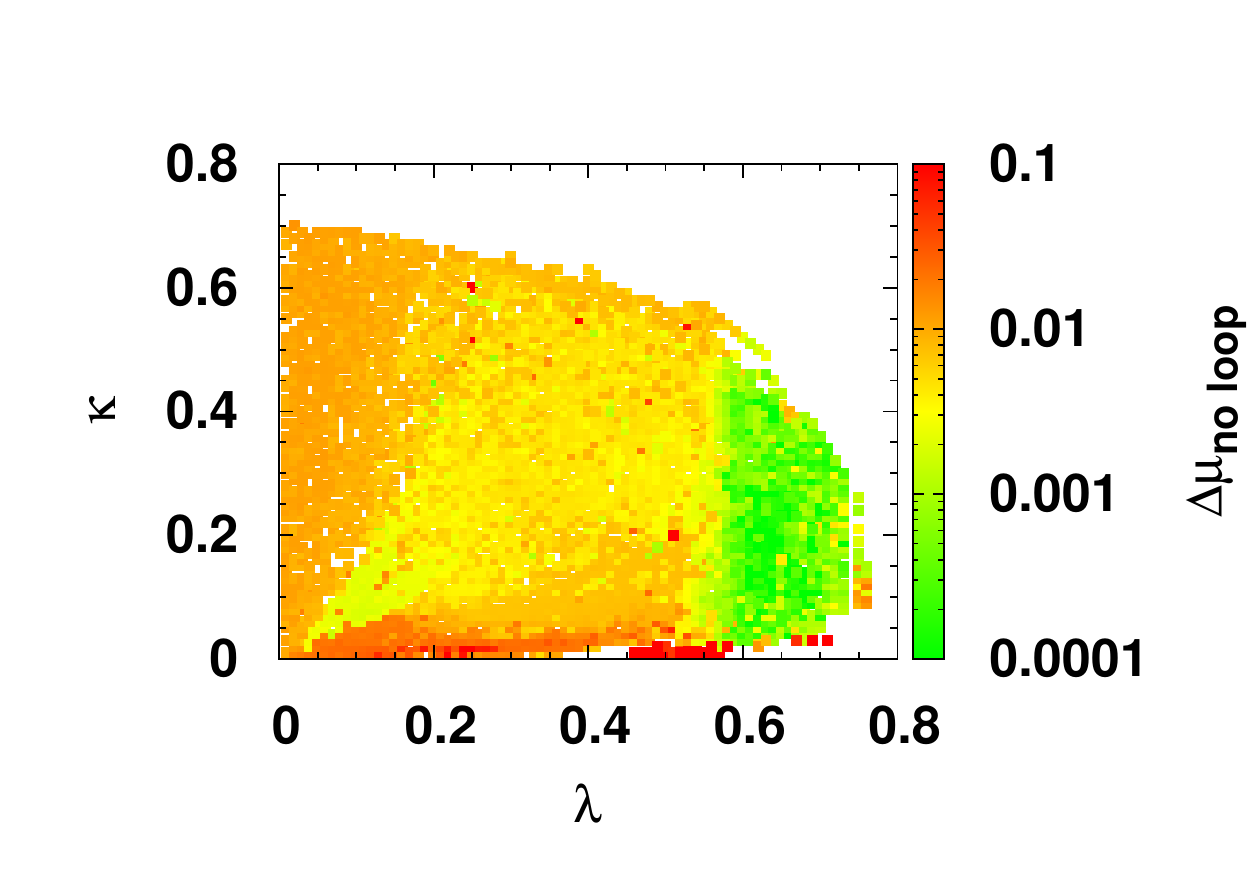}
       \includegraphics[width=0.47\textwidth]{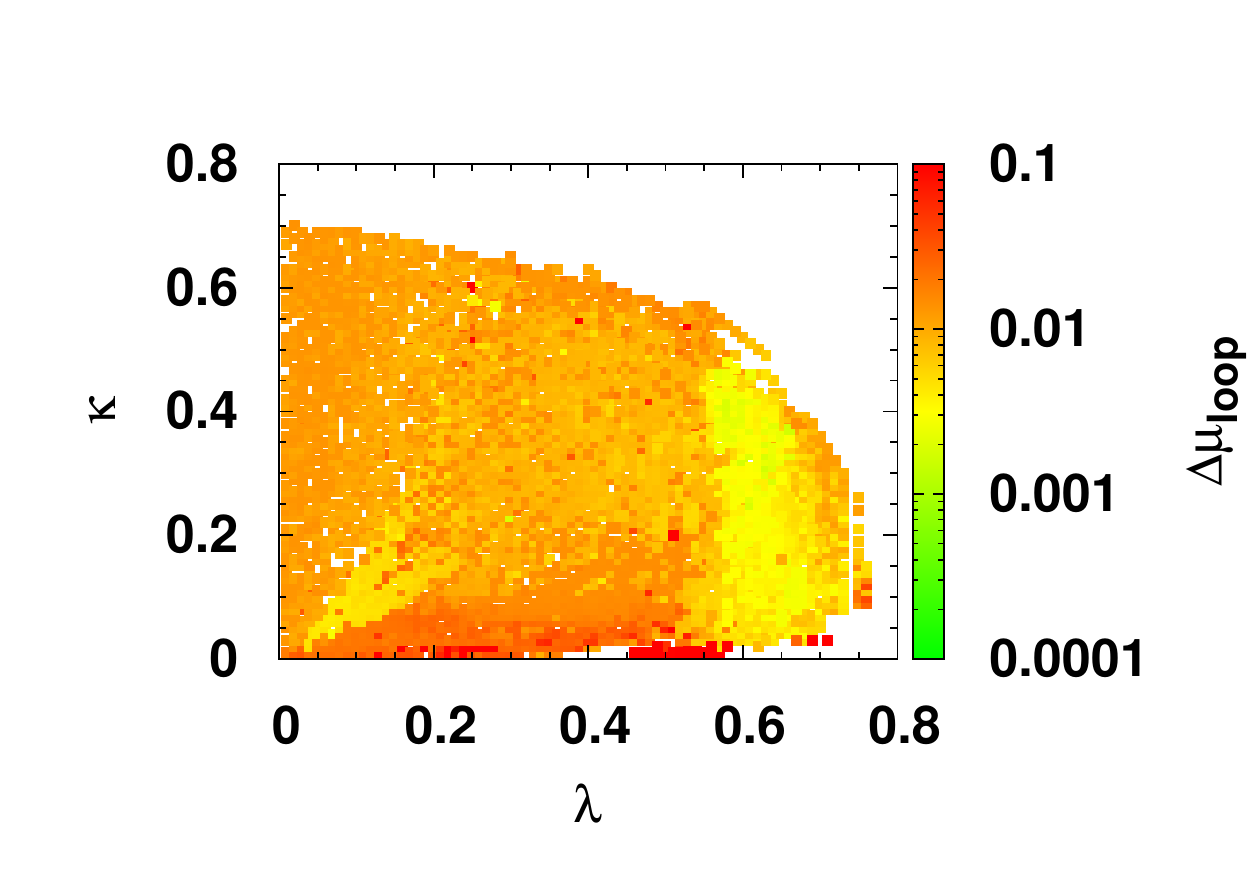}\vspace*{-3mm}
        \hspace*{0.02\textwidth}(c)\hspace*{0.46\textwidth} (d)\\[-1mm]
     \caption{ 
The  total $\chi^2$ function in the  $\lambda-\kappa$ plane,  as shown by the color coding in (a). The two green regions with the smallest  $\chi^2$ values   are called Region I (II) for the region on the right (left). here all constraints are fulfilled. Note that the absolute $\chi^2$ value depends on the choice of the error for the selected  Higgs mass combination, but the minimum of the $\chi^2$ distribution is always in the green regions. Different errors  only change the color coding.   For small values of $\lambda$ and large values of $\kappa$ the constraints are not fulfilled by  the Higgs mass, since the Higgs mass is too low in this region, as shown by the color coding in (b). In addition, the fitted signal strength $\mu_{no-loop}$ deviates from 1 as can be seen from (c), where the averaged difference from the SM value is shown. For intermediate values of $\lambda$ and $\kappa$ the main contribution to the total $\chi^2$ function is coming from the fermion signal strength.
The remaining signal strength $\mu_{loop}$, shown by the averaged difference from the SM value (d), deviates from 1 in almost the whole $\lambda-\kappa$ plane, except for Region I. 
}
     \label{f9}
\end{center}
   \end{figure}
\section{Results from the $\lambda-\kappa$ scan}
\label{lambda-kappa-scan}

\begin{table}
\footnotesize
\centering
\caption{ \label{t1}
Differences of the fitting procedures for the standard fit in Sect. \ref{sampling} (left column) and the fit parameters for the $\lambda-\kappa$ scan (right column).
}
\begin{tabular}{l|c|c}
	\hline\noalign{\smallskip}
	Procedure & standard & $\lambda-\kappa$ scan \\
	\noalign{\smallskip}\hline\noalign{\smallskip}
	Input & $A_1,H_{1/2},H_3$ & $\lambda,\kappa$\\
	Constraints & $H_{1/2}=$125 GeV, $\mu_{no-loop}=$1 & $H_{1/2}=$125 GeV, $\mu_{no-loop}=$1\\
	Output &  $\tan\beta,A_0,A_\kappa,A_\lambda,\mu_{eff},\lambda,\kappa$ & $\tan\beta,A_0,A_\kappa,A_\lambda,\mu_{eff}$ \\
	$\chi^2$ contribution & $\chi^2_{H_S},\chi^2_{H_3},\chi^2_{A_1},\chi^2_{H_{SM}},\chi^2_{\mu_{SM}},\chi^2_{LEP},\chi^2_{LHC}$ & $\chi^2_{H_{SM}},\chi^2_{\mu_{SM}},\chi^2_{LEP},\chi^2_{LHC}$ \\
	\noalign{\smallskip}\hline
\end{tabular}
\end{table}

   \begin{figure}[]
\begin{center}
       \includegraphics[width=0.42\textwidth]{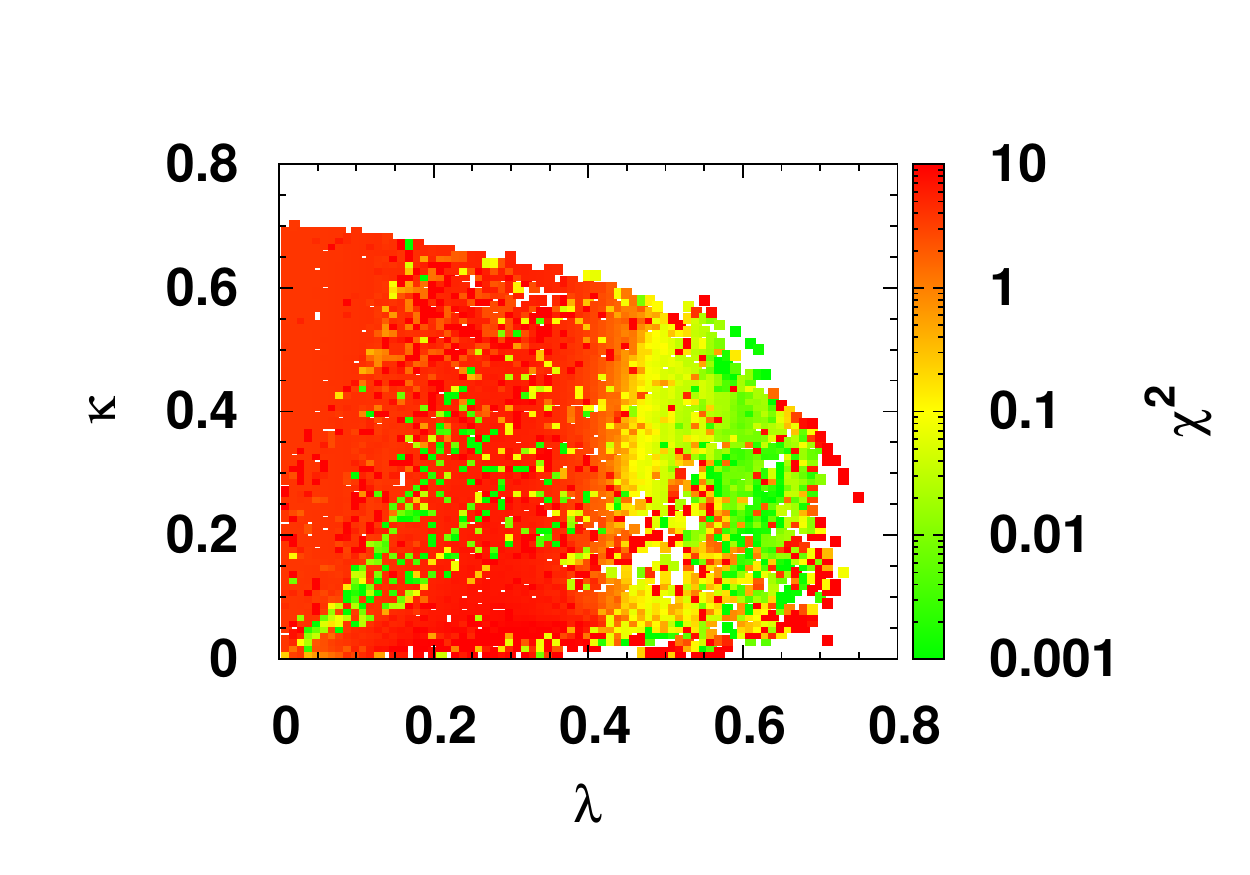}
       \includegraphics[width=0.42\textwidth]{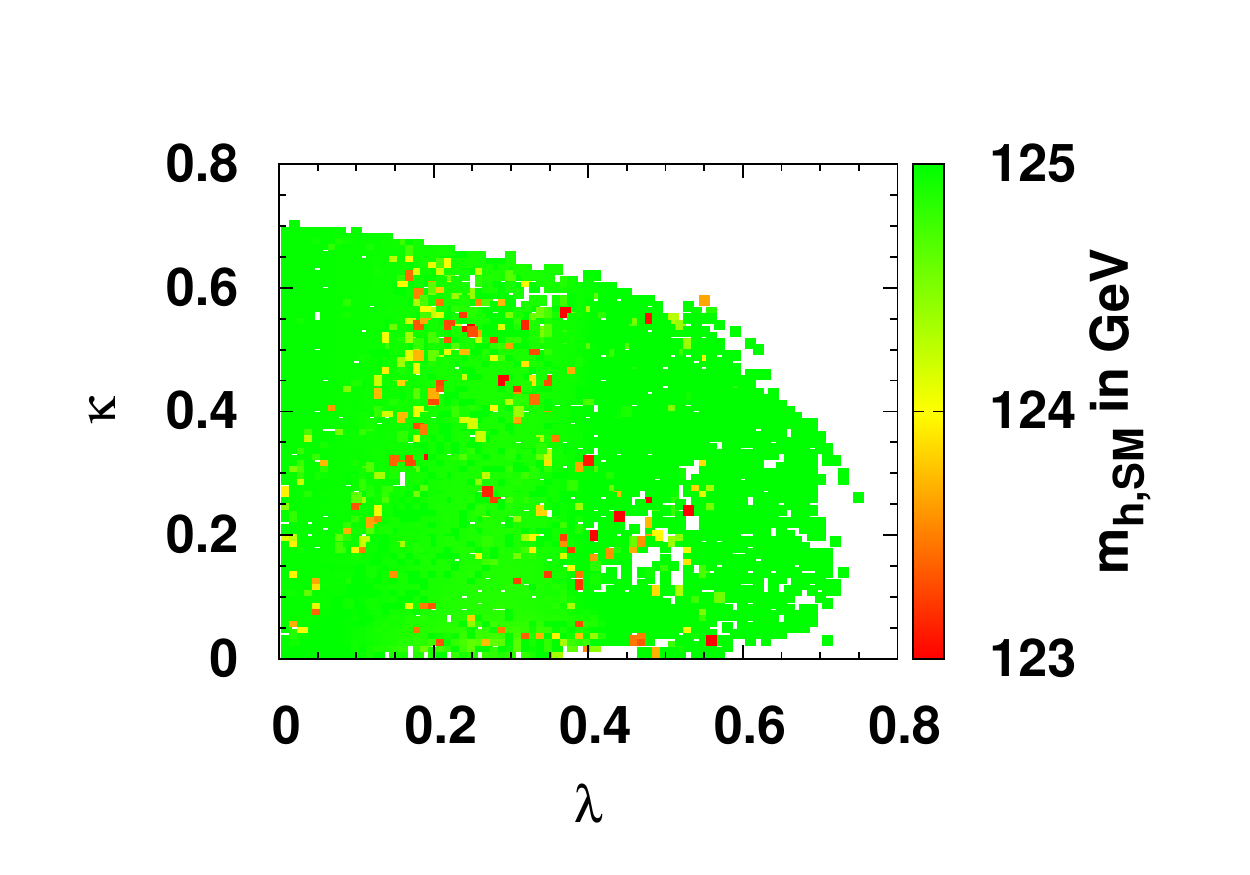}\vspace*{-3mm}
       \hspace*{0.02\textwidth}(a)\hspace*{0.46\textwidth} (b)\\[-4mm]
       \includegraphics[width=0.42\textwidth]{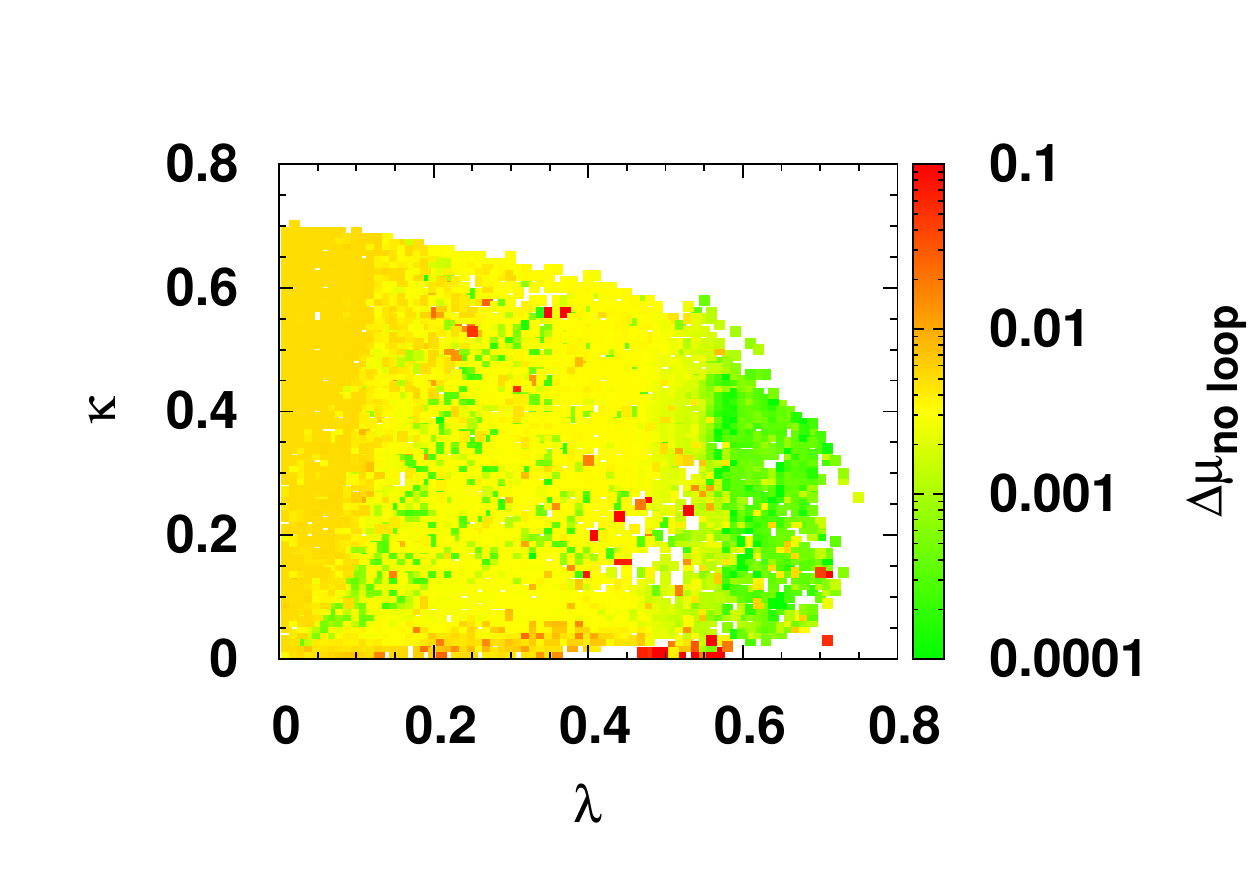}
       \includegraphics[width=0.42\textwidth]{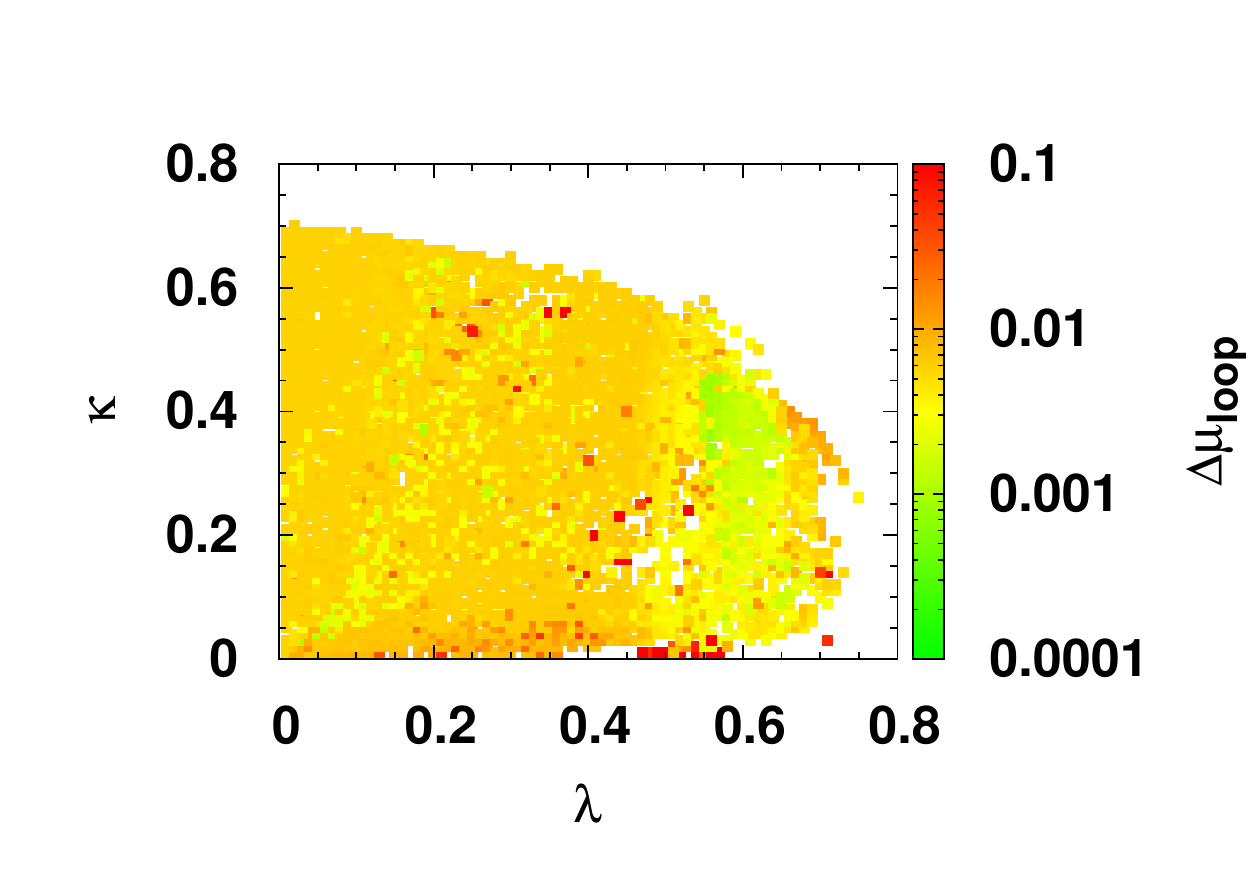}\vspace*{-3mm}
       \hspace*{0.02\textwidth}(c)\hspace*{0.46\textwidth} (d)\\[-1mm]
     \caption{ 
As Fig. \ref{f9}, but for $m_0=m_{1/2}=1.5$ TeV. 
}
     \label{f10}
\end{center}
   \end{figure}
   \begin{figure}[]
\begin{center}
       \includegraphics[width=0.42\textwidth]{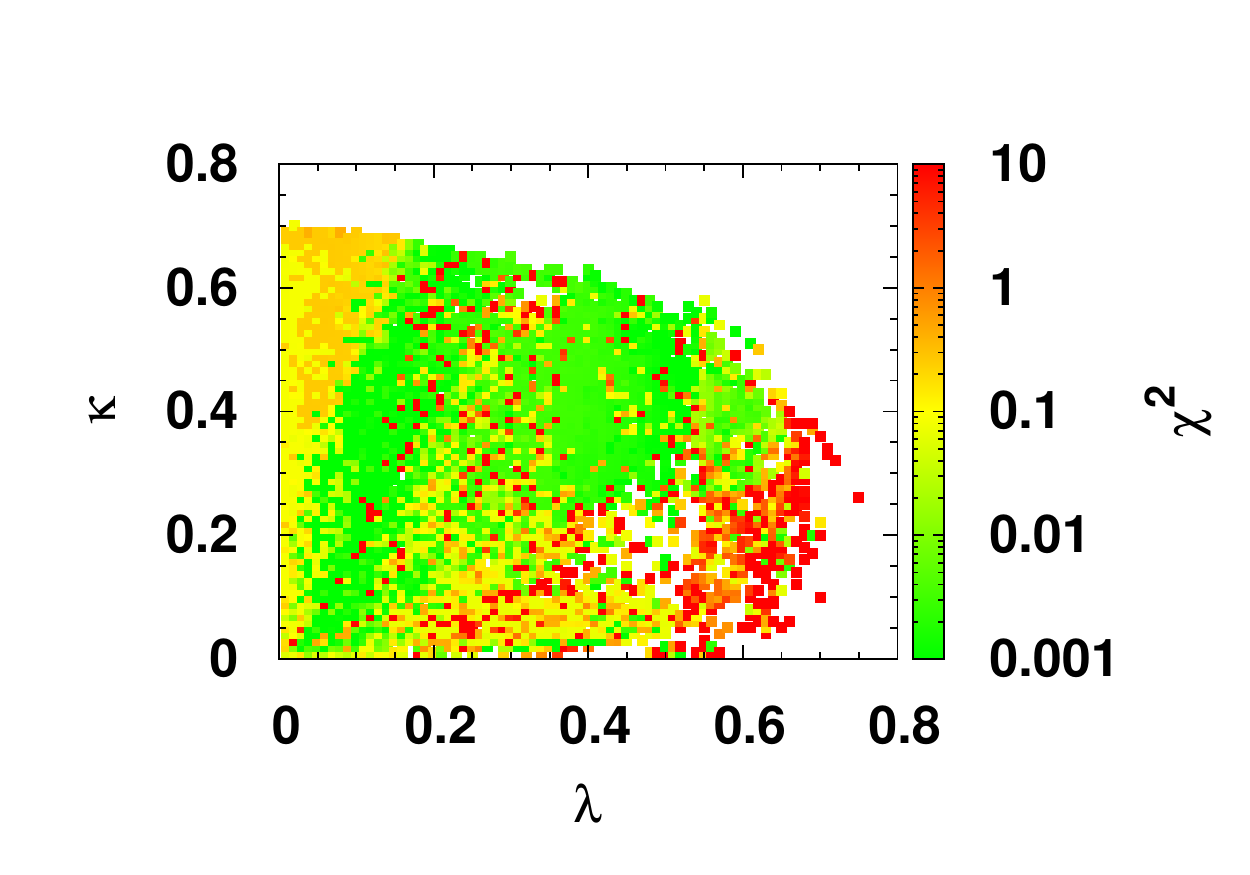}
       \includegraphics[width=0.42\textwidth]{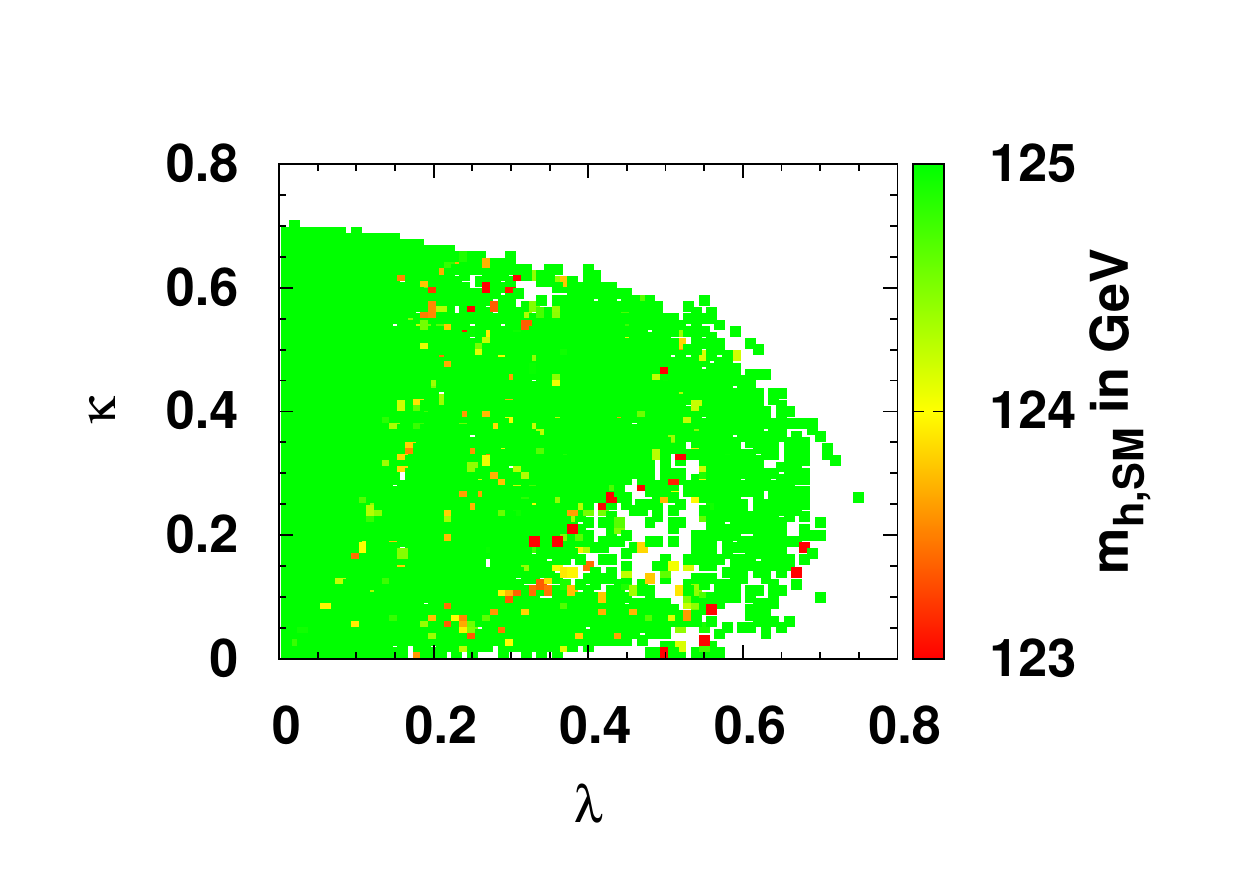}\vspace*{-3mm}
       \hspace*{0.02\textwidth}(a)\hspace*{0.46\textwidth} (b)\\[-8mm]
       \includegraphics[width=0.42\textwidth]{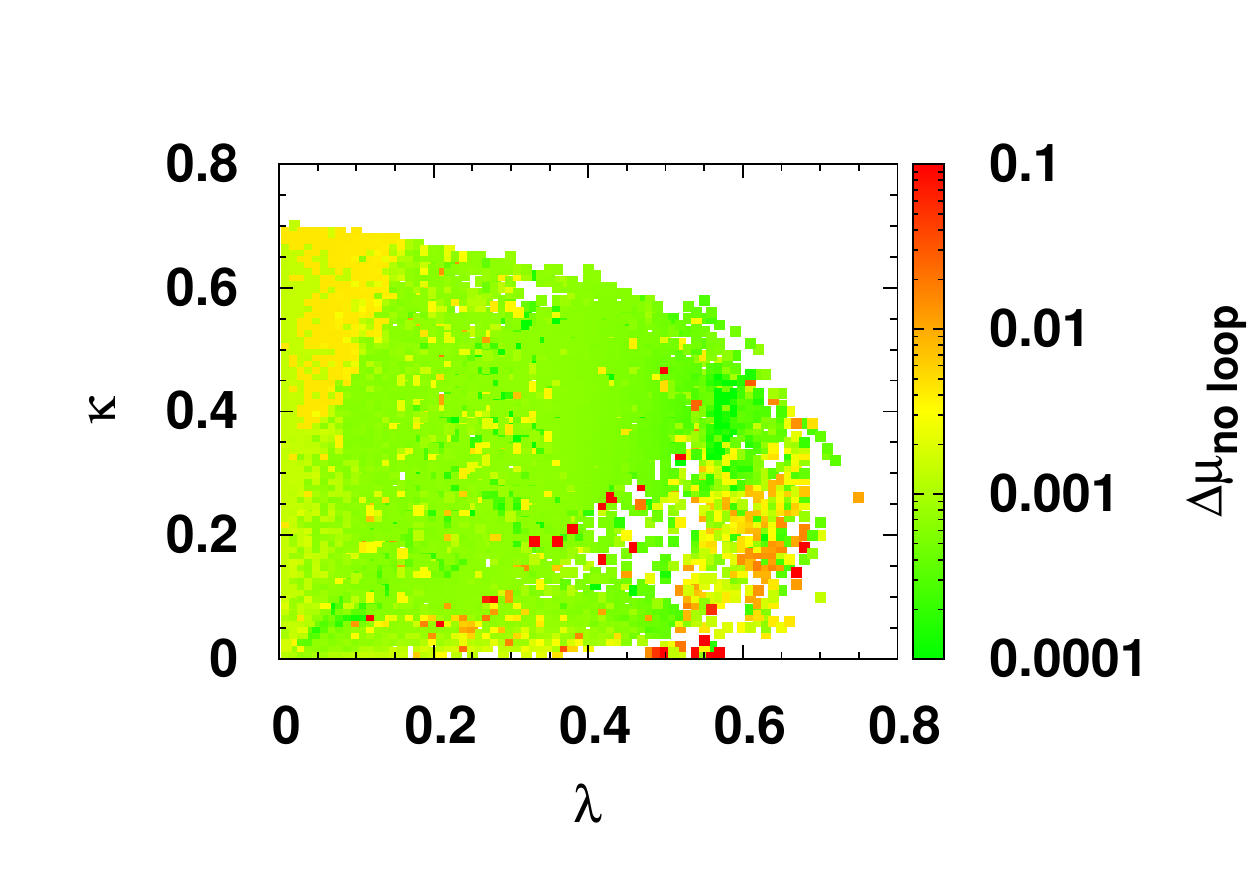}
       \includegraphics[width=0.42\textwidth]{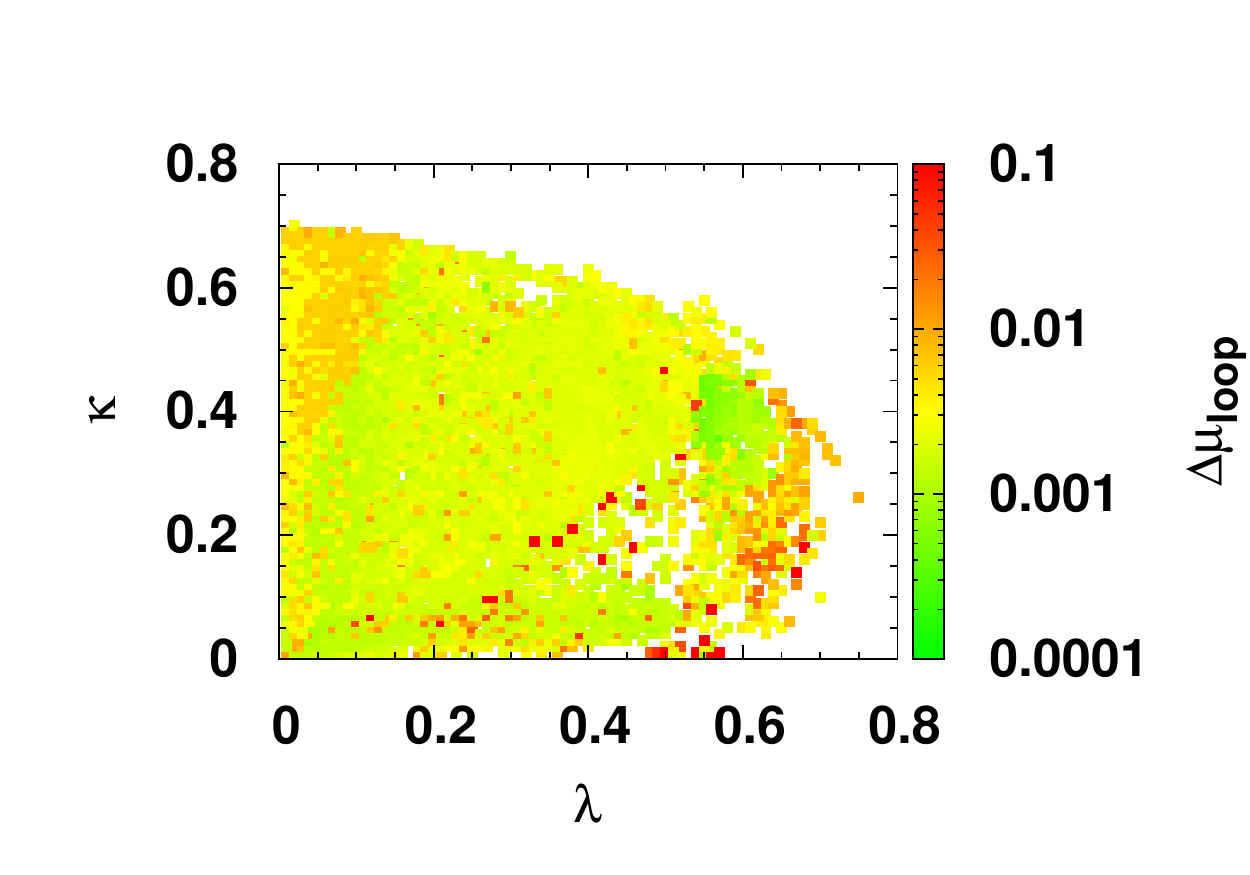}\vspace*{-3mm}
       \hspace*{0.02\textwidth}(c)\hspace*{0.46\textwidth} (d)\\[-1mm]
     \caption{ 
As Fig. \ref{f9}, but for $m_0=m_{1/2}=2$ TeV. 
}
     \label{f11}
\end{center}
   \end{figure}

   %
   %
As discussed in Sect. \ref{results} there are two preferred regions in the $\lambda,\kappa$ plane varying in size with the stop mass, as shown in Fig. \ref{f5}. One wonders what happens outside the preferred regions. In order to find out
the standard fit procedure was slightly modified by requiring the NMSSM parameters $\lambda$ and $\kappa$ to be fixed in the fit. This was repeated for all  $\kappa,\lambda$  combinations in the $\lambda$-$\kappa$ plane and check which constraints increase the $\chi^2$ of the fit outside the preferred regions.
Fixing the $\kappa,\lambda$ values in the fit reduces the number of free parameters to only 5, namely $\tan\beta,A_0,A_\kappa,A_\lambda$ and $\mu_{eff}$. The constraints are reduced  too since the free Higgs boson masses $A_1,H_1$ and $H_3$ are allowed to vary to minimize the $\chi^2$ function for a given combination of $\lambda$ and $\kappa$. The differences with the standard fit procedure, as discussed in Sect. \ref{sampling}, have been summarized in Table \ref{t1}. 

The two  regions with the smallest  $\chi^2$ values in the $\lambda$-$\kappa$ plane are  the green regions  in Fig. \ref{f9}(a). 
The remaining regions - called intermediate regions - have a larger $\chi^2$-value for various reasons: for small values of $\lambda$ and large values of $\kappa$ the Higgs mass of the observed Higgs boson is too low, as can be seen from Fig. \ref{f9}(b), where the color coding corresponds to the Higgs boson mass. The low  Higgs mass  (around 123 GeV) originates from the fact, that the stop corrections are too low for the chosen values of $m_0=m_{1/2}=$ 1 TeV.   In addition, the fitted signal strength $\mu_{no-loop}$ deviates from 1 as can be seen from Fig. \ref{f9}(c), where the color coding corresponds to the averaged difference from the SM value, i.e. $\Delta\mu_{no-loop}=\frac{1}{4}\sum_{i=1}^4 (\mu^i - \mu_{theo})^2/\sigma^2_{\mu}$ with $\mu^i=\mu_{\tau\tau}^{VBF/VH},\mu_{bb}^{ttH},\mu_{bb}^{VBF/VH}$ and $\mu_{ZZ/WW}^{VBF/VH}$. 
The averaged difference of the remaining signal strengths $\mu_{loop}$, i.e. $\Delta\mu_{loop}=\frac{1}{4}\sum_{i=1}^4 (\mu^i - \mu_{theo})^2/\sigma^2_{\mu}$ where $\mu_i$ includes $\mu_{\tau\tau}^{ggf},\mu_{ZZ/WW}^{ggf},\mu_{\gamma\gamma}^{VBF/VH}$ and $\mu_{\gamma\gamma}^{ggf}$ is shown in Fig. \ref{f9}(d). 
Here, the deviations of the signal strengths $\mu_{loop}$ of the order of a few percent  correspond to the orange region.

If one increases the $m_0=m_{1/2}$ values from 1 TeV in Fig. \ref{f9} to 1.5 (2) TeV in Figs. \ref{f10} (\{ref{f11}) one observes that  the $\chi^2$ values in panels (a) decrease, mainly because of the increased values of the Higgs mass (panels (b)) and the decrease in the deviations of the signal strengths from the SM expectations  (panels (c) and (d)). Note that the absolute values of the test statistic $\chi^2$ are somewhat arbitrary because of the choice of the errors for the assumed Higgs mass values of the 3 chosen Higgs masses on the grid in Fig. \ref{f2}, i.e. $m_{H_1}$, $m_{A_1}$ and $m_{H_3}$. But the regions with minimum  $\chi^2$ values stay the same, as well as the deviations of the signal strengths from the SM expectations.


\begin{thebibliography}{10}%
\makeatletter
\providecommand{\hrefCMSnoop }[0]{\@secondoftwo}%
\makeatother

\bibitem{Miller:2003ay}
\hrefCMSnoop {} {D.~Miller, R.~Nevzorov, and P.~Zerwas, ``{The Higgs sector of
  the next-to-minimal supersymmetric standard model}'',} \textit{ Nucl.Phys.}
  \textbf{ B681} (2004) 3--30,
\href{http://www.arXiv.org/abs/hep-ph/0304049}{\texttt{ arXiv:hep-ph/0304049}}.

\bibitem{Ellwanger:2009dp}
\hrefCMSnoop {} {U.~Ellwanger, C.~Hugonie, and A.~M. Teixeira, ``{The
  Next-to-Minimal Supersymmetric Standard Model}'',} \textit{ Phys.Rept.}
  \textbf{ 496} (2010) 1--77,
\href{http://www.arXiv.org/abs/0910.1785}{\texttt{ arXiv:0910.1785}}.

\bibitem{Aad:2012tfa}
\hrefCMSnoop {} {{ ATLAS Collaboration} Collaboration, ``{Observation of a new
  particle in the search for the Standard Model Higgs boson with the ATLAS
  detector at the LHC}'',} \textit{ Phys.Lett.} \textbf{ B716} (2012) 1--29,
\href{http://www.arXiv.org/abs/1207.7214}{\texttt{ arXiv:1207.7214}}.

\bibitem{Chatrchyan:2012xdj}
\hrefCMSnoop {} {{ CMS} Collaboration, ``{Observation of a new boson at a mass
  of 125 GeV with the CMS experiment at the LHC}'',} \textit{ Phys. Lett.}
  \textbf{ B716} (2012) 30--61,
\href{http://www.arXiv.org/abs/1207.7235}{\texttt{ arXiv:1207.7235}}.

\bibitem{Beskidt:2012sk}
\hrefCMSnoop {} {C.~Beskidt, W.~de~Boer, D.~Kazakov{ et~al.}, ``{Constraints on
  Supersymmetry from LHC data on SUSY searches and Higgs bosons combined with
  cosmology and direct dark matter searches}'',} \textit{ Eur.Phys.J.} \textbf{
  C72} (2012) 2166,
\href{http://www.arXiv.org/abs/1207.3185}{\texttt{ arXiv:1207.3185}}.

\bibitem{Buchmueller:2013rsa}
\hrefCMSnoop {} {O.~Buchmueller, R.~Cavanaugh, A.~De~Roeck{ et~al.}, ``{The
  CMSSM and NUHM1 after LHC Run 1}'',}
\href{http://www.arXiv.org/abs/1312.5250}{\texttt{ arXiv:1312.5250}}.

\bibitem{Fowlie:2012im}
\hrefCMSnoop {} {A.~Fowlie, M.~Kazana, K.~Kowalska{ et~al.}, ``{The CMSSM
  Favoring New Territories: The Impact of New LHC Limits and a 125 GeV
  Higgs}'',} \textit{ Phys.Rev.} \textbf{ D86} (2012) 075010,
\href{http://www.arXiv.org/abs/1206.0264}{\texttt{ arXiv:1206.0264}}.

\bibitem{Bechtle:2013mda}
\hrefCMSnoop {} {P.~Bechtle, K.~Desch, H.~K. Dreiner{ et~al.}, ``{Constrained
  Supersymmetry after the Higgs Boson Discovery: A global analysis with
  Fittino}'',}
\href{http://www.arXiv.org/abs/1310.3045}{\texttt{ arXiv:1310.3045}}.

\bibitem{Beskidt:2013gia}
\hrefCMSnoop {} {C.~Beskidt, W.~de~Boer, and D.~Kazakov, ``{A comparison of the
  Higgs sectors of the CMSSM and NMSSM for a 126 GeV Higgs boson}'',} \textit{
  Phys.Lett.} \textbf{ B726} (2013) 758--766,
\href{http://www.arXiv.org/abs/1308.1333}{\texttt{ arXiv:1308.1333}}.

\bibitem{Haber:1984rc}
\hrefCMSnoop {} {H.~E. Haber and G.~L. Kane, ``{The Search for Supersymmetry:
  Probing Physics Beyond the Standard Model}'',} \textit{ Phys.Rept.} \textbf{
  117} (1985)
75--263.

\bibitem{deBoer:1994dg}
\hrefCMSnoop {} {W.~de~Boer, ``{Grand unified theories and supersymmetry in
  particle physics and cosmology}'',} \textit{ Prog.Part.Nucl.Phys.} \textbf{
  33} (1994) 201--302,
\href{http://www.arXiv.org/abs/hep-ph/9402266}{\texttt{ arXiv:hep-ph/9402266}}.

\bibitem{Martin:1997ns}
\hrefCMSnoop {} {S.~P. Martin, ``{A Supersymmetry primer}'',} \textit{
  Perspectives on supersymmetry II, Ed. G. Kane} (1997)
\href{http://www.arXiv.org/abs/hep-ph/9709356}{\texttt{ arXiv:hep-ph/9709356}}.

\bibitem{Beskidt:2017xsd}
\hrefCMSnoop {} {C.~Beskidt, W.~de~Boer, D.~I. Kazakov{ et~al.},
  ``{Perspectives of direct Detection of supersymmetric Dark Matter in the
  NMSSM}'',} \textit{ Phys. Lett.} \textbf{ B771} (2017) 611--618,
\href{http://www.arXiv.org/abs/1703.01255}{\texttt{ arXiv:1703.01255}}.

\bibitem{Dermisek:2008uu}
\hrefCMSnoop {} {R.~Dermisek and J.~F. Gunion, ``{Many Light Higgs Bosons in
  the NMSSM}'',} \textit{ Phys. Rev.} \textbf{ D79} (2009) 055014,
\href{http://www.arXiv.org/abs/0811.3537}{\texttt{ arXiv:0811.3537}}.

\bibitem{King:2012is}
\hrefCMSnoop {} {S.~King, M.~M{\"u}hlleitner, and R.~Nevzorov, ``{NMSSM Higgs
  Benchmarks Near 125 GeV}'',} \textit{ Nucl.Phys.} \textbf{ B860} (2012)
  207--244,
\href{http://www.arXiv.org/abs/1201.2671}{\texttt{ arXiv:1201.2671}}.

\bibitem{Cao:2012fz}
\hrefCMSnoop {} {J.-J. Cao, Z.-X. Heng, J.~M. Yang{ et~al.}, ``{A SM-like Higgs
  near 125 GeV in low energy SUSY: a comparative study for MSSM and NMSSM}'',}
  \textit{ JHEP} \textbf{ 1203} (2012) 086,
\href{http://www.arXiv.org/abs/1202.5821}{\texttt{ arXiv:1202.5821}}.

\bibitem{Ellwanger:2012ke}
\hrefCMSnoop {} {U.~Ellwanger and C.~Hugonie, ``{Higgs bosons near 125 GeV in
  the NMSSM with constraints at the GUT scale}'',} \textit{ Adv.High Energy
  Phys.} \textbf{ 2012} (2012) 625389,
\href{http://www.arXiv.org/abs/1203.5048}{\texttt{ arXiv:1203.5048}}.

\bibitem{Gunion:2012zd}
\hrefCMSnoop {} {J.~F. Gunion, Y.~Jiang, and S.~Kraml, ``{The Constrained NMSSM
  and Higgs near 125 GeV}'',} \textit{ Phys.Lett.} \textbf{ B710} (2012)
  454--459,
\href{http://www.arXiv.org/abs/1201.0982}{\texttt{ arXiv:1201.0982}}.

\bibitem{Cao:2013gba}
\hrefCMSnoop {} {J.~Cao, F.~Ding, C.~Han{ et~al.}, ``{A light Higgs scalar in
  the NMSSM confronted with the latest LHC Higgs data}'',} \textit{ JHEP}
  \textbf{ 11} (2013) 018,
\href{http://www.arXiv.org/abs/1309.4939}{\texttt{ arXiv:1309.4939}}.

\bibitem{Badziak:2013bda}
\hrefCMSnoop {} {M.~Badziak, M.~Olechowski, and S.~Pokorski, ``{New Regions in
  the NMSSM with a 125 GeV Higgs}'',} \textit{ JHEP} \textbf{ 1306} (2013) 043,
\href{http://www.arXiv.org/abs/1304.5437}{\texttt{ arXiv:1304.5437}}.

\bibitem{Barbieri:2013nka}
\hrefCMSnoop {} {R.~Barbieri, D.~Buttazzo, K.~Kannike{ et~al.}, ``{One or more
  Higgs bosons?}'',} \textit{ Phys. Rev.} \textbf{ D88} (2013) 055011,
\href{http://www.arXiv.org/abs/1307.4937}{\texttt{ arXiv:1307.4937}}.

\bibitem{King:2014xwa}
\hrefCMSnoop {} {S.~F. King, M.~M{\"u}hlleitner, R.~Nevzorov{ et~al.},
  ``{Discovery Prospects for NMSSM Higgs Bosons at the High-Energy Large Hadron
  Collider}'',} \textit{ Phys. Rev.} \textbf{ D90} (2014), no.~9, 095014,
\href{http://www.arXiv.org/abs/1408.1120}{\texttt{ arXiv:1408.1120}}.

\bibitem{Bernon:2014nxa}
\hrefCMSnoop {} {J.~Bernon, J.~F. Gunion, Y.~Jiang{ et~al.}, ``{Light Higgs
  bosons in Two-Higgs-Doublet Models}'',} \textit{ Phys. Rev.} \textbf{ D91}
  (2015), no.~7, 075019,
\href{http://www.arXiv.org/abs/1412.3385}{\texttt{ arXiv:1412.3385}}.

\bibitem{Guchait:2015owa}
\hrefCMSnoop {} {M.~Guchait and J.~Kumar, ``{Light Higgs Bosons in NMSSM at the
  LHC}'',} \textit{ Int. J. Mod. Phys.} \textbf{ A31} (2016), no.~12, 1650069,
\href{http://www.arXiv.org/abs/1509.02452}{\texttt{ arXiv:1509.02452}}.

\bibitem{Potter:2015wsa}
\hrefCMSnoop {} {C.~T. Potter, ``{Natural NMSSM with a Light Singlet Higgs and
  Singlino LSP}'',} \textit{ Eur. Phys. J.} \textbf{ C76} (2016), no.~1, 44,
\href{http://www.arXiv.org/abs/1505.05554}{\texttt{ arXiv:1505.05554}}.

\bibitem{Bandyopadhyay:2015tva}
\hrefCMSnoop {} {P.~Bandyopadhyay, C.~Coriano, and A.~Costantini, ``{Probing
  the hidden Higgs bosons of the $Y = 0$ triplet- and singlet-extended
  Supersymmetric Standard Model at the LHC}'',} \textit{ JHEP} \textbf{ 12}
  (2015) 127,
\href{http://www.arXiv.org/abs/1510.06309}{\texttt{ arXiv:1510.06309}}.

\bibitem{Bomark:2015hia}
N.-E. Bomark, S.~Moretti, S.~Munir{ et~al.}, ``{A light NMSSM pseudoscalar
  Higgs boson at the LHC Run 2}'', in \textit{ {2nd Toyama International
  Workshop on Higgs as a Probe of New Physics (HPNP2015) Toyama, Japan,
  February 11-15, 2015}}.
\newblock 2015.
\newblock
\href{http://www.arXiv.org/abs/1502.05761}{\texttt{ arXiv:1502.05761}}.
\newblock

\bibitem{Cao:2016uwt}
\hrefCMSnoop {} {J.~Cao, X.~Guo, Y.~He{ et~al.}, ``{Diphoton signal of the
  light Higgs boson in natural NMSSM}'',} \textit{ Phys. Rev.} \textbf{ D95}
  (2017), no.~11, 116001,
\href{http://www.arXiv.org/abs/1612.08522}{\texttt{ arXiv:1612.08522}}.

\bibitem{Muhlleitner:2017dkd}
\hrefCMSnoop {} {M.~M{\"u}hlleitner, M.~O.~P. Sampaio, R.~Santos{ et~al.},
  ``{Phenomenological Comparison of Models with Extended Higgs Sectors}'',}
  \textit{ JHEP} \textbf{ 08} (2017) 132,
\href{http://www.arXiv.org/abs/1703.07750}{\texttt{ arXiv:1703.07750}}.

\bibitem{Das:2016eob}
\hrefCMSnoop {} {S.~P. Das and M.~Nowakowski, ``{Light neutral CP-even Higgs
  boson within Next-to-Minimal Supersymmetric Standard model (NMSSM) at the
  Large Hadron electron Collider (LHeC)}'',} \textit{ Phys. Rev.} \textbf{ D96}
  (2017), no.~5, 055014,
\href{http://www.arXiv.org/abs/1612.07241}{\texttt{ arXiv:1612.07241}}.

\bibitem{Mariotti:2017vtv}
\hrefCMSnoop {} {A.~Mariotti, D.~Redigolo, F.~Sala{ et~al.}, ``{New LHC bound
  on low-mass diphoton resonances}'',}
\href{http://www.arXiv.org/abs/1710.01743}{\texttt{ arXiv:1710.01743}}.

\bibitem{Baum:2017gbj}
\hrefCMSnoop {} {S.~Baum, K.~Freese, N.~R. Shah{ et~al.}, ``{NMSSM Higgs boson
  search strategies at the LHC and the mono-Higgs signature in particular}'',}
  \textit{ Phys. Rev.} \textbf{ D95} (2017), no.~11, 115036,
\href{http://www.arXiv.org/abs/1703.07800}{\texttt{ arXiv:1703.07800}}.

\bibitem{Beskidt:2016egy}
\hrefCMSnoop {} {C.~Beskidt, W.~de~Boer, D.~I. Kazakov{ et~al.}, ``{Higgs
  branching ratios in constrained minimal and next-to-minimal supersymmetry
  scenarios surveyed}'',} \textit{ Phys. Lett.} \textbf{ B759} (2016) 141--148,
\href{http://www.arXiv.org/abs/1602.08707}{\texttt{ arXiv:1602.08707}}.

\bibitem{Beskidt:2017dil}
\hrefCMSnoop {} {C.~Beskidt, W.~de~Boer, and D.~I. Kazakov, ``{Can we discover
  a light singlet-like NMSSM Higgs boson at the LHC?}'',} \textit{ Phys. Lett.}
  \textbf{ B782} (2018) 69--76,
\href{http://www.arXiv.org/abs/1712.02531}{\texttt{ arXiv:1712.02531}}.

\bibitem{Das:2011dg}
\hrefCMSnoop {} {D.~Das, U.~Ellwanger, and A.~M. Teixeira, ``{NMSDECAY: A
  Fortran Code for Supersymmetric Particle Decays in the Next-to-Minimal
  Supersymmetric Standard Model}'',} \textit{ Comput.Phys.Commun.} \textbf{
  183} (2012) 774--779,
\href{http://www.arXiv.org/abs/1106.5633}{\texttt{ arXiv:1106.5633}}.

\bibitem{Aad:2015iea}
\hrefCMSnoop {} {{ ATLAS} Collaboration, ``{Summary of the searches for squarks
  and gluinos using $ \sqrt{s}=8 $ TeV pp collisions with the ATLAS experiment
  at the LHC}'',} \textit{ JHEP} \textbf{ 10} (2015) 054,
\href{http://www.arXiv.org/abs/1507.05525}{\texttt{ arXiv:1507.05525}}.

\bibitem{Djouadi:2005gi}
\hrefCMSnoop {} {A.~Djouadi, ``{The Anatomy of electro-weak symmetry breaking.
  I: The Higgs boson in the standard model}'',} \textit{ Phys. Rept.} \textbf{
  457} (2008) 1--216,
\href{http://www.arXiv.org/abs/hep-ph/0503172}{\texttt{ arXiv:hep-ph/0503172}}.

\bibitem{Haber:1995be}
H.~E. Haber, ``{Challenges for nonminimal Higgs searches at future
  colliders}'', in \textit{ {Perspectives for electroweak interactions in e+ e-
  collisions. Proceedings, Ringberg Workshop, Tegernsee, Germany, February 5-8,
  1995}}, pp.~219--232.
\newblock 1996.
\newblock \href{http://www.arXiv.org/abs/hep-ph/9505240}{\texttt{
  arXiv:hep-ph/9505240}}.
\newblock
[,151(1995)].

\bibitem{Djouadi:2005gj}
\hrefCMSnoop {} {A.~Djouadi, ``{The Anatomy of electro-weak symmetry breaking.
  II. The Higgs bosons in the minimal supersymmetric model}'',} \textit{ Phys.
  Rept.} \textbf{ 459} (2008) 1--241,
\href{http://www.arXiv.org/abs/hep-ph/0503173}{\texttt{ arXiv:hep-ph/0503173}}.

\bibitem{James:1975dr}
\hrefCMSnoop {} {F.~James and M.~Roos, ``{Minuit: A System for Function
  Minimization and Analysis of the Parameter Errors and Correlations}'',}
  \textit{ Comput.Phys.Commun.} \textbf{ 10} (1975) 343--367.

\bibitem{Beskidt:2014kon}
C.~Beskidt, ``{Supersymmetry in the Light of Dark Matter and a 125 GeV Higgs
  Boson}''.
\newblock PhD thesis, KIT, Karlsruhe, EKP,
2014.
\newblock

\bibitem{Khachatryan:2015nba}
\hrefCMSnoop {} {{ CMS} Collaboration, ``{Search for a very light NMSSM Higgs
  boson produced in decays of the 125 GeV scalar boson and decaying into $\tau$
  leptons in pp collisions at $\sqrt{s}=8$ TeV}'',} \textit{ JHEP} \textbf{ 01}
  (2016) 079,
\href{http://www.arXiv.org/abs/1510.06534}{\texttt{ arXiv:1510.06534}}.

\bibitem{Aad:2015oqa}
\hrefCMSnoop {} {{ ATLAS} Collaboration, ``{Search for Higgs bosons decaying to
  $aa$ in the $\mu\mu\tau\tau$ final state in $pp$ collisions at $\sqrt{s} = $
  8 TeV with the ATLAS experiment}'',} \textit{ Phys. Rev.} \textbf{ D92}
  (2015), no.~5, 052002,
\href{http://www.arXiv.org/abs/1505.01609}{\texttt{ arXiv:1505.01609}}.

\bibitem{Khachatryan:2015wka}
\hrefCMSnoop {} {{ CMS} Collaboration, ``{A search for pair production of new
  light bosons decaying into muons}'',} \textit{ Phys. Lett.} \textbf{ B752}
  (2016) 146--168,
\href{http://www.arXiv.org/abs/1506.00424}{\texttt{ arXiv:1506.00424}}.

\end{thebibliography}
\end{document}